\documentclass[preprint,longtitle,12pt]{elsarticle}
\usepackage{bm}
\usepackage{epsfig}
\usepackage{graphicx}
\usepackage{longtable}
\usepackage[nolists,tablesfirst,nofighead,notabhead,nomarkers]{endfloat}
\usepackage{rotating}
\usepackage{amssymb,amsmath}
\biboptions{square,comma,sort&compress} 
\usepackage[bf,Large,tight,figtopcap]{subfigure}

\def\lazz{\mathrel{\mathchoice {\vcenter{\offinterlineskip\halign{\hfil
$\displaystyle##$\hfil\cr<\cr\sim\cr}}}
{\vcenter{\offinterlineskip\halign{\hfil$\textstyle##$\hfil\cr<\cr\sim\cr}}}
{\vcenter{\offinterlineskip\halign{
\hfil$\scriptstyle##$\hfil\cr<\cr\sim\cr}}}
{\vcenter{\offinterlineskip\halign{\hfil$\scriptscriptstyle##
$\hfil\cr<\cr\sim\cr}}}}}
\def\gazz{\mathrel{\mathchoice {\vcenter{\offinterlineskip\halign{\hfil
$\displaystyle##$\hfil\cr>\cr\sim\cr}}}
{\vcenter{\offinterlineskip\halign{\hfil$\textstyle##$\hfil\cr>\cr\sim\cr}}}
{\vcenter{\offinterlineskip\halign{
\hfil$\scriptstyle##$\hfil\cr>\cr\sim\cr}}}
{\vcenter{\offinterlineskip\halign{\hfil$\scriptscriptstyle##
$\hfil\cr>\cr\sim\cr}}}}}
\def\pr{\prime}
\def\be{\begin{equation}}
\def\lan{\left\langle}
\def\ran{\right\rangle}
\def\ee{\end{equation}}
\def\barr{\begin{array}}
\def\earr{\end{array}}

\def\nn8{\\}
\def\l{\left}
\def\r{\right}
\def\dis{\displaystyle}
\def\ed{\end{document}}

\def\dg{\dagger}

\def\cod{{\cal O}^\dagger}
\def\co{{\cal O}}
\def\cg{{\cal G}}

\def\wx{{\widehat {x}}}

\def\wy{{\widehat {y}}}

\def\wtM{{\widetilde {M}}}

\journal{Annals of Physics}
\begin{document}
\begin{frontmatter}
\title{Random Matrix Theory for Transition Strength  Densities in Finite
Quantum Systems: Results from Embedded Unitary Ensembles}

\author[prl]{V.K.B. Kota\corref{cor1}}

\ead{vkbkota@prl.res.in}

\cortext[cor1]{Corresponding author. Phone: +91-79-26314464, 
Fax: +91-79-26314460}

\author[mex]{Manan Vyas}

\address[prl]{Physical Research Laboratory, Ahmedabad 380 009, India}

\address[mex]{Instituto de Ciencias F\'{i}sicas, Universidad Nacional Aut{\'o}noma 
de M\'{e}xico, C.P. 62210 Cuernavaca, M\'{e}xico}

\ead{manan@fis.unam.mx}

\begin{abstract}

Embedded random matrix ensembles are generic models for describing statistical
properties of finite isolated interacting quantum many-particle systems. For the
simplest spinless fermion (or boson) systems, with say $m$ fermions (or bosons)
in $N$ single particle states and interacting via $k$-body interactions, we have
EGUE($k$) [embedded GUE of $k$-body interactions)  with GUE embedding and the
embedding algebra is $U(N)$. A finite quantum system, induced by a transition
operator, makes transitions from its states to the states of the same system or
to those of another system. Examples are electromagnetic transitions (then the
initial and final systems are same), nuclear beta and double beta decay (then
the initial and final systems are different), particle addition to or removal
from a given system and so on. Towards developing a complete
statistical theory for transition strength densities (transition strengths
multiplied by the density  of states at the initial and final energies), we have
derived formulas for the lower order bivariate moments of the strength densities
generated by a variety of transition operators. Firstly, for a spinless fermion
system, using EGUE($k$) representation for a Hamiltonian  that is $k$-body and
an independent EGUE($t$) representation for a transition  operator that is
$t$-body and employing the embedding $U(N)$  algebra, finite-$N$ formulas for
moments up to order four are derived, for the first time, for the transition
strength  densities. Secondly, formulas for the moments  up to order
four are also derived  for systems with two types of spinless fermions and a
transition operator similar to beta decay and neutrinoless beta decay operators.
In addition, moments formulas are also derived  for a transition operator that
removes $k_0$ number of particles from a system of $m$ spinless fermions. In the
dilute limit, these formulas are shown to reduce to those for the  EGOE version
derived using the asymptotic limit theory of Mon and French [Ann. Phys. (N.Y.)
95  (1975) 90]. Numerical results obtained using the exact formulas for two-body
($k=2$) Hamiltonians (in some examples for $k=3$ and $4$)  and the asymptotic
formulas clearly establish that in general the smoothed (with respect to energy)
form of the bivariate transition strength  densities take  bivariate Gaussian
form for isolated finite quantum systems.  Extensions of these results to
bosonic systems and EGUE ensembles with further symmetries are discussed. 

\end{abstract}

\begin{keyword}
Finite many-particle quantum systems \sep  Embedded ensembles \sep
Transition strengths \sep Bivariate moments \sep $U(N)$  Wigner-Racah algebra 
\sep Asymptotics \sep Bivariate Gaussian

\PACS  05.30.-d \sep 21.60.Fw \sep 24.60.-k
\end{keyword}

\end{frontmatter}

\section{Introduction}

Wigner introduced random matrix theory (RMT) in physics in 1955 primarily to
understand statistical properties of neutron resonances in heavy nuclei
\cite{Po-65,Weiden}.  Depending on the global symmetry properties of the
Hamiltonian of a quantum system, namely rotational symmetry and time-reversal
symmetry, we have Dyson's tripartite classification of random matrices giving
the classical random matrix ensembles, the Gaussian orthogonal (GOE), unitary
(GUE) and symplectic (GSE) ensembles. In the last three decades, RMT has found
applications not only in all branches of quantum physics but also in many other
disciplines such as econophysics, wireless communication, information theory,
multivariate statistics, number theory, neural and biological networks and so on
\cite{rmt1,rmt2,rmt3,rmt4,rmt5}.  However, in  the context of isolated finite
many-particle quantum systems, classical random matrix ensembles are too
unspecific to account for important features of the physical system at hand. One
refinement which retains the basic stochastic  approach but allows for such
features consists in the use of embedded random matrix ensembles 
\cite{Br-81,Ko-01,BW-JPA,PW,Go-12,kota}.

Finite quantum systems such as nuclei, atoms, quantum dots,  small metallic
grains, interacting spin systems modeling quantum computing core and ultracold
atoms, share one common property - their constituents predominantly interact
via  two-particle interactions. Therefore, it is more appropriate to represent
an isolated  finite interacting quantum system, say with $m$ particles (fermions
or bosons) in $N$ single particle (sp) states by random matrix models  generated by
random $k$-body (note that $k < m$ and most often we have $k=2$)  interactions
and propagate the information in the interactions to many ($m$) particle spaces. Then
we have random matrix ensembles in $m$-particle spaces - these ensembles are
defined by representing the $k$-particle Hamiltonian ($H$) by GOE/GUE/GSE and
then the $m$ particle $H$ matrix is generated by the $m$-particle Hilbert
space geometry. The key element here is the recognition that there is a Lie
algebra that transports the information in the two-particle spaces to the
many-particle spaces.  As a GOE/GUE/GSE random matrix ensemble in two-particle
spaces is embedded  in the $m$-particle $H$ matrix, these ensembles are more
generically called embedded ensembles (EE). 

Embedded ensembles have proved to be rich in their content and wide in their 
scope. A book giving detailed discussion of the various properties and
applications of a wide  variety of embedded matrix ensembles is now available
\cite{kota}. Significantly, the study of embedded random matrix ensembles is
still  developing. Partly this is due to the fact that deriving generic
properties of  these ensembles is not mathematically tractable and this is the
topic of the  present paper. A general formulation for deriving exact analytical
results is to use  the Wigner-Racah algebra of the embedding Lie algebra
\cite{kota}. Till now, this has has been applied only in the study of one- and
two-point functions in eigenvalues. The focus in the present paper is on
transition strengths which  on one hand probe the structure of the
eigenfunctions of a quantum many-body  system and on the other are important
ingredients in many applications (for example, beta decay transition strengths
are essential for nucleosynthesis studies). Also, as emphasized in
\cite{Br-81,Weiden}, there are many open questions in the random matrix theory
for transition strengths in finite interacting quantum many-particle systems.

For finite quantum many-particle systems, induced by a transition operator, a
given system makes transitions from its states to the states of the same system
or to the states of another system. Examples are electromagnetic transitions
(then the initial and final systems are same), nuclear beta and double beta
decay (then the initial and final systems are different), particle addition to
or removal from a given system and so on.  Given a transition operator $\co$
acting on the $m$ particle eigenstates $\l|\l.E\ran\r.$ of $H$ will give
transition matrix elements $\l|\lan E_f \mid \co \mid E_i\ran\r|^2$. Fig. 1
shows a schematic picture of transition strengths. In the statistical theories,
it is more useful to deal with the corresponding transition strength density
(this will take into account degeneracies in the eigenvalues) defined by
\be
I_\co(E_i,E_f) = I(E_f)\,\l|\lan E_f \mid \co \mid E_i\ran\r|^2\,I(E_i) \;.
\label{strn-eq1}
\ee
In Eq. (\ref{strn-eq1}), $I(E)$ are state densities normalized to the dimension
of the $m$-particle spaces. Note that $E_i$ and $E_f$ belong to the same 
$m$-particle system or different systems depending on the nature of the
transition operator $\co$. In the discussion ahead, we will consider both
situations. Random matrix theory has been used in the past to derive the form of
the smoothed $I(E)$. In particular, exact (finite $N$) formulas for lower order 
moments $\lan H^p\ran$, $p=2$ and $4$ of $I(E)$ are derived both for EGOE and
EGUE ensembles using group theoretical methods directly \cite{Ko-05} or
indirectly \cite{Be-01}. Let us add that some results valid only in the
asymptotic limit (essentially $N \rightarrow  \infty$, $m \rightarrow \infty$,
$m/N \rightarrow 0$) are also available in literature for the state  densities
\cite{Mo-75,Muller}.  Going beyond the eigenvalue densities, here we will apply
group theoretical methods and derive for the first time some exact formulas for
the ensemble averaged lower order moments of the transition  strength densities
$I_\co(E_i,E_f)$. These then will give the general form  of the smoothed (with
respect to both $E_i$ and $E_f$) transition strength densities.  We will focus
on fermionic systems and at the end, discuss the extension to bosonic systems.
It is important to mention that in the present paper fluctuations in transition
strengths are not considered and they will be addressed in a future publication.
For some discussion on strength fluctuations see \cite{Weiden,Br-81,Ko-01}. More
importantly, using smoothed transition strengths statistical spectroscopy
analysis of transition strengths (for example for particle transfer, 
electromagnetic transitions, beta decay and even double beta decay matrix
elements) in nuclei and also in atoms  is possible \cite{KH,Fl-99}. In addition,
smoothed form of the strength densities can be used to calculate transition
strengths that are needed for example in astrophysical applications
\cite{Fuller,KM} and for neutrinoless double beta  decay
\cite{KH,Iachello,Shounen}. These formed the main motivation for the present
study.  Let us add that in the past, besides deriving the dilute limit formulas
for the bivariate moments of the transition strength densities in some
situations \cite{FKPT,Ma-arx}, there are  suggestions of using a polynomial
expansion theory \cite{DFW} (later in  \cite{FKPT} it was shown that the
polynomial expansion starts with the GOE  result and hence in general
inappropriate) for transition strengths,  a specialized theory for one-body
transition operators  \cite{Fl-1,Fl-2,Fl-99,KS} and using the bivariate
$t$-distribution form for  transition strength densities \cite{KSC}.

In general, the Hamiltonian may have many symmetries with the fermions (or
bosons) carrying additional degrees of freedom such as spin, orbital angular
momentum, isospin and so on. Also, the system may comprise of different  types
of fermions (or bosons); for example, in atomic nuclei we have protons and
neutrons. In addition, a transition operator may preserve particle number and
other quantum numbers or it may change them. Of all these various situations,
here we will consider three different systems in detail. (i) Firstly, we will
consider a system of spinless fermions and a transition operator that preserves
particle number.  (ii) Secondly, we will consider a system with two types of
spinless fermions with the transition operator changing $k_0$ number of
particles of one type to $k_0$ number of particles of other type as in nuclear
beta decay and neutrinoless double beta decay. (iii) Third system considered is
transition operators that remove (add) say $k_0$ number of particles from (to) a
system of $m$ spinless fermions.  In all these we will restrict to EGUE. Before
giving a short preview, let us mention that some of the results in the present
paper are reported in three conferences earlier \cite{Bregenz,Ghent,Dubna} and
in a thesis \cite{Ma-arx}.

Section 2 gives some basic results for EGUE($k$) for spinless fermion systems as
derived in \cite{Ko-05} and some of their extensions. Using these results,
formulas for the lower order bivariate moments of the transition strength
densities for the situation (i) above are derived and they are presented in
detail in Section 3.  The corresponding asymptotic limit formulas are presented
in Section 4. Similarly, results for the situation (ii) above are  given in
detail in Section 5 and their asymptotics are given in Section 6. In addition,
in Section 7, results for the situation (iii) are given along with the
corresponding asymptotic formulas. Extensions to bosonic systems is discussed
briefly in Section 8. Finally, Section 9 gives conclusions and future outlook.

\section{Basic EGUE($k$) results for a spinless fermion system}

Let us consider $m$ spinless fermions in $N$ degenerate sp states with the 
Hamiltonian $\widehat{H}$ a $k$-body operator,
\be
\widehat{H} = \sum_{i,j} V_{ij}(k)\;A^\dagger_i(k)\,A_j(k)\;,\;\;\;V_{ij}(k)
=\lan k,i \mid \widehat{H} \mid k,j\ran\;.
\label{strn-eq2}
\ee
Here $A^\dagger_i(k)$ is a $k$ particle (normalized) creation operator and
$A_i(k)$ is the corresponding annihilation operator (a hermitian conjugate). 
Also, $i$ and $j$ are $k$-particle indices. Note that the $k$ and $m$ particle
space dimensions are $\binom{N}{k}$ and $\binom{N}{m}$ respectively. We will
consider $\widehat{H}$ to be EGUE($k$) in $m$-particle spaces. Then $V_{ij}$ form a
GUE with $V$ matrix being Hermitian. The real and imaginary parts of $V_{ij}$
are independent zero centered Gaussian random variables with variance 
satisfying,
\be
\overline{V_{ab}(k)\, V_{cd}(k)} = V^2_H\,\delta_{ad}\delta_{bc}\;.
\label{strn-eq3}
\ee
Here the `overline' indicates ensemble average. From now on we will drop the hat
over $H$ and denote, when needed, $H$ by $H(k)$. Let us add that in physical
systems, $k=2$ is of great interest and in some systems such as atomic nuclei 
it is possible to have $k=3$ and even $k=4$ \cite{Zel3,Fu-13}.

The  $U(N)$ algebra that generates the embedding, as shown in \cite{Ko-05},
gives formulas for the lower order moments of the one-point function, the
eigenvalue density $I(E)=\overline{\lan\lan \delta(H-E)\ran\ran}$ and also for
the two-point function in the eigenvalues. In particular,   explicit formulas
are given in \cite{Ko-05,kota}  for $\overline{\lan H^P\ran^m}$, $P=2,4$ and
$\overline{\lan H^P\ran^m\,\lan H^Q \ran^m}$, $P+Q=2,4$. Used here is the $U(N)$
tensorial decomposition of the  $H(k)$ operator giving $\nu=0,1,\ldots,k$
irreducible parts $B^{\nu , \omega_\nu}(k)$ and  then,
\be
H(k) = \dis\sum^{k}_{\nu =0;\omega_\nu \in \nu} 
W_{\nu , \omega_\nu}(k)\; B^{\nu , \omega_\nu}(k) \;.
\label{strn-eq5}
\ee
With the GUE($k$) representation for the $H(k)$ operator, the expansion 
coefficients W's will be independent zero centered Gaussian random variables 
with, by an extension of Eq. (\ref{strn-eq3}),
\be
\overline{W_ {\nu_1 , \omega_{\nu_1}}(k)\;W_ {\nu_2 , \omega_{\nu_2}}(k)}
= V^2_H\;\delta_{\nu_1 , \nu_2} \delta_{\omega_{\nu_1} \omega_{\nu_2}}\;.
\label{strn-eq6}
\ee
For deriving formulas for the various moments, the first step is to apply the
Wigner-Eckart theorem for the matrix elements of $B^{\nu , \omega_\nu} (k)$.
Given the $m$-fermion states $\l.\l|f_m v_i\r.\ran$, we have with respect to the
$U(N)$ algebra, $f_m=\{1^m\}$, the antisymmetric irreducible representation
(irrep) in Young tableaux notation and $v_i$ are additional labels. Note that 
$\nu$ introduced above corresponds to the Young tableaux $\{2^\nu 1^{N-2\nu}\}$
and $\omega_\nu$ are additional labels. Now, Wigner-Eckart theorem gives
\be
\lan f_m v_f \mid B^{\nu , \omega_\nu}(k) \mid f_m v_i\ran = \lan f_m 
\mid\mid B^{\nu}(k) \mid\mid f_m\ran\;C^{\nu , \omega_\nu}_{f_m v_f\,,\;
\overline{f_m} \overline{v_i}} \;.
\label{W-E}
\ee
Here, $\lan \dots || \dots || \dots \ran$ is the reduced matrix element and 
$C^{--}_{----}$ is a $U(N)$ Clebsch-Gordan (C-G) coefficient [note that we are
not making a distinction between $U(N)$ and $SU(N)$]. Also, if $\l.\l|f_m
v_i\r.\ran$ represents a $m$-particle state, then $\l.\l|
\overline{f_m} \overline{v_i}\r.\ran$ represents a $m$-hole state (see
\cite{Ko-05} for details). In Young tableaux notation $\overline{f_m} =
\{1^{N-m}\}$. It is important to mention that $\l.\l| \nu \omega_\nu \r.\ran = 
\l.\l| \overline{\nu}\, \overline{\omega_\nu}\r.\ran$. Fig. 2 shows some
typical Young tableaux and also the Young tableaux $\overline{f}$ that 
corresponds to a given $\{f\}$. Definition of  $B^{\nu , \omega_\nu}(k)$  and 
the $U(N)$ Wigner-Racah algebra will give,
\be
\barr{l}
\l|\lan f_m \mid\mid B^{\nu}(k) \mid\mid f_m\ran\r|^2 = 
\Lambda^{\nu}(N,m,m-k)\;,
\;\\
\\
\Lambda^{\mu}(N^\pr,m^\pr,r) = \dis\binom{m^\pr-\mu}{r}\,\dis\binom{
N^\pr-m^\pr+r-\mu}{r}\;.
\earr \label{reduced}
\ee
Note that $\Lambda^{\nu}(N,m,k)$ is nothing but, apart from a $N$ and $m$
dependent factor, a $U(N)$ Racah coefficient \cite{Ko-05}. This and the 
various properties of the $U(N)$ Wigner and Racah coefficients give two 
formulas for the ensemble average of a product of any two $m$-particle matrix 
elements of $H$,
\be
\barr{l}
\overline{\lan f_m v_1 \mid H(k) \mid f_m v_2\ran\, \lan f_m v_3 \mid H(k) 
\mid f_m v_4\ran} \\
= V^2_H\;\dis\sum_{\nu=0;\omega_\nu}^{k}\Lambda^{\nu}(N,m,m-k)\; C^{\nu , 
\omega_\nu}_{f_m v_1\,,\;\overline{f_m} \overline{v_2}}\; 
C^{\nu , \omega_\nu}_{f_m v_3\,,\;\overline{f_m} \overline{v_4}}\;,
\earr \label{matrix-ele-a}
\ee
and also
\be
\barr{l}
\overline{\lan f_m v_1 \mid H(k) \mid f_m v_2\ran\, \lan f_m v_3 \mid H(k) 
\mid f_m v_4\ran} \\
= V^2_H\;\dis\sum_{\nu=0;\omega_\nu}^{m-k}\Lambda^{\nu}(N,m,k)\; 
C^{\nu , \omega_\nu}_{f_m v_1\,,\;\overline{f_m} \overline{v_4}}\; 
C^{\nu , \omega_\nu}_{f_m v_3\,,\;\overline{f_m} \overline{v_2}} \;.
\earr \label{matrix-ele-b}
\ee
Eq. (\ref{matrix-ele-b}) follows by applying a Racah transform to the
product of the two C-G coefficients appearing in Eq. (\ref{matrix-ele-a}). 
Let us mention some properties of the $U(N)$ C-G coefficients that are quite 
useful in deriving the formulas given in Sections 3 and 5,
\be
\barr{l}
\dis\sum_{v_i} C^{\nu , \omega_\nu}_{f_m v_i\,,\,\overline{f_m} 
\overline{v_i}} = \dis\binom{N}{m}^{1/2} \; \delta_{\nu , 0}\;,\;\;\;
C^{0,0}_{f_m v_i\,,\;\overline{f_m}
\overline{v_j}}={\binom{N}{m}}^{-1/2}\;\delta_{v_i\,,v_j}\;, \\
\\
C^{f_{ab} v_{ab}}_{f_a v_a\,,\,f_b v_b}=(-1)^{\phi(f_a , f_b , f_{ab})} \;
C^{f_{ab} v_{ab}}_{f_b v_b\,,\,f_a v_a}\;,\\
\\
\dis\sum_{v_i,v_j} C^{\nu , \omega_\nu}_{f_m v_i\,,\,\overline{f_m}
\overline{v_j}}\;C^{\nu^\pr , \omega_{\nu^\pr}}_{f_m v_j\,,\,\overline{f_m}
\overline{v_i}} = \delta_{\nu \nu^\pr} \; \delta_{\omega_\nu \omega_{\nu^\pr}}
\;.
\earr \label{sum00}
\ee 
Here, $(-1)^{\phi}$ is a phase factor depending on the irreps $f_a$, $f_b$ and
$f_{ab}$. See \cite{Ko-05,He-75,Bu-1,Bu-2} for further details on the phase
relations for $U(N)$ C-G coefficients. From now on we will
use the symbol $f_m$ only in the C-G coefficients,  Racah coefficients and
reduced matrix elements. However, for the matrix elements of an operator we will
use $m$ implying totally antisymmetric state for fermions (symmetric state for
bosons).  

Starting with Eq. (\ref{strn-eq5}) and using Eqs. (\ref{strn-eq6}),
(\ref{matrix-ele-b}) and (\ref{sum00}) will immediately give the formula,  
\be
\overline{\lan H^2(k)\ran^m} = {\binom{N}{m}}^{-1}\;\dis\sum_{v_i}\,
\overline{\lan m v_i \mid H^2(k) \mid m v_i\ran} = V^2_H \;\Lambda^0(N,m,k)\;.
\label{strn-eq7a}
\ee
Similarly, for $\overline{\lan H^4\ran^m}$,  the ensemble average is
decomposed into three terms as 
\be
\barr{l}
\overline{\lan H^4(k) \ran^m} = {\dis\binom{N}{m}}^{-1}\;\dis\sum_{v_i}\,
\overline{\lan m v_i \mid H^4(k) \mid m v_i\ran} 
\\= 
\resizebox{0.95\hsize}{!}{$\dis\sum_{v_i, v_j, v_p , v_l} \; 
\overline{\lan m v_i \mid H(k) \mid m v_j\ran\,
\lan m v_j \mid H(k) \mid m v_p\ran \lan m v_p \mid H(k) \mid m v_l\ran 
\lan m v_l \mid H(k) \mid m v_i\ran} $} 
\\ \\
= \resizebox{0.95\hsize}{!}{$\dis\sum_{v_i, v_j, v_p , v_l} \; \l[
\overline{\lan m v_i \mid H(k) \mid m v_j\ran\,\lan m v_j 
\mid H(k) \mid m v_p\ran} \;\; \overline{\lan m v_p \mid H(k) \mid m 
v_l\ran \lan m v_l \mid H(k) \mid m v_i\ran} \r.$}
\\ \\ \l.
+ \,\resizebox{0.95\hsize}{!}{$\overline{\lan m v_i \mid H(k) \mid m v_j\ran\,
\lan m v_l \mid H(k) \mid m v_i\ran} \;\; \overline{\lan m v_j \mid H(k) \mid m 
v_p\ran \lan m v_p \mid H(k) \mid m v_l\ran}$} \r.
\\ \\ \l.
+ \,\resizebox{0.95\hsize}{!}{$\overline{\lan m v_i \mid H(k) \mid m v_j\ran\,
\lan m v_p \mid H(k) \mid m v_l\ran} \;\; \overline{\lan m v_j \mid H(k) \mid m 
v_p\ran \lan m v_l \mid H(k) \mid m v_i\ran}$}\,\r]  \;.
\earr \label{strn-eq7ab}
\ee
It is easy to see that the first two terms simplify to give $2\l[\overline{\lan
H^2\ran^m}\r]^2$ and the third term is simplified by applying Eq.
(\ref{matrix-ele-a}) to the first ensemble average and Eq. (\ref{matrix-ele-b})
to the second ensemble average. Then, the final result is
\be
\resizebox{0.95\hsize}{!}{$\overline{\lan H^4(k)\ran^m} = 
2 \l[ \overline{\lan H^2(k)\ran^m}\r]^2 
+ V^4_H \;{\dis\binom{N}{m}}^{-1} \; \dis\sum_{\nu=0}^{min(k,m-k)}\,
\Lambda^{\nu}(N,m,k)\, \Lambda^{\nu}(N,m,m-k)\,d(N:\nu)\;,$}
\label{strn-eq7b}
\ee
where
\be
d(N:\nu) = {\dis\binom{N}{\nu}}^2 -{\dis\binom{N}{\nu -1}}^2\;.
\label{eq-dnu}
\ee
Finally, a by-product of Eqs. (\ref{matrix-ele-b}) and (\ref{sum00}) is
\be
\dis\sum_{v_j}\,\overline{\lan m v_i \mid H(k) \mid m v_j\ran\, 
\lan m v_j \mid H(k) 
\mid m v_k\ran} = \overline{\lan H^2(k)\ran^m}\;\delta_{v_i , v_k} \;,
\label{sum-hh}
\ee
and we will use this in Section 3. Now we will discuss results for the
moments of the transition strength densities generated by a transition operator
$\co$.

\section{Lower-order moments of transition strength densities:
$H$ EGUE($k$) and $\co$ an independent EGUE($t$)}

For a spinless fermion system, similar to the $k$-body $H$ operator, we will consider 
a $t$-body transition operator $\co$ represented by EGUE($t$) in the $m$-particle 
spaces. Then, the matrix of $\co$ in $t$-particle space will be a GUE with the matrix 
elements $\co_{ab}(t)$ being zero centered independent Gaussian variables with the 
variance satisfying, 
\be
\overline{\co_{ab}(t)\, \co_{cd}(t)} = V^2_\co\,\delta_{ad} \; \delta_{bc}\;.
\label{strn-eq3a}
\ee
Further, we will assume that the GUE representing $H$ in $k$-particle spaces and
the GUE representing $\co$ in $t$ particle spaces are independent (this is
equivalent to the statement that $\co$ does not generate diagonal elements $\lan
E_i \mid \co \mid E_i\ran$; see \cite{FKPT}). It is important to mention that in
the past there were attempts to numerically study, in some nuclear $(2s1d)$
shell examples, transition strengths using the eigenvectors generated by a EGOE
(with symmetries) representation for $H$ but taking $\co$ to be a realistic
transition operator. The results are found to be in variance  with those
obtained using a $H$ defined by a realistic two-body interaction  and $\co$ a
realistic transition operator \cite{Zel}. As described in the present paper, a
proper random matrix theory for transition strengths has to employ ensemble
representation for both the Hamiltonian and the transition operator.

With EGUE representation, $\co$ is Hermitian and hence, $\cod=\co$. Moments of
the transition strength densities $I_\co(E_i,E_f) $ are defined by
\be
M_{PQ}(m) = \overline{\lan \cod(t) H^Q(k) \co(t) H^P(k) \ran^m} = \overline{\lan \co(t) H^Q(k) 
\co(t) H^P(k) \ran^m}\;.
\label{strn-eq4}
\ee
Here the ensemble average is w.r.t. both EGUE($k$) and EGUE($t$). Now we will
derive formulas for $M_{PQ}$ with $P+Q=2$ and $4$; the moments with odd value of 
($P+Q$) will vanish by definition. 

Firstly, the unitary decomposition of $\co(t)$ gives, 
\be 
\co(t) =
\dis\sum^{t}_{\nu =0;\omega_\nu \in \nu}  U_{\nu , \omega_\nu}(t)\; B^{\nu ,
\omega_\nu}(t) \;. 
\label{strn-eq5a} 
\ee 
The $U$'s satisfy a relation similar to Eq. (\ref{strn-eq6}). Now, for $P=Q=0$,
using Eq. (\ref{strn-eq7a}), we have
\be
\overline{\lan \co(t) \; \co(t) \ran^m} = V^2_\co \; \Lambda^0(N,m,t) \;. 
\label{strn-eq8}
\ee 
Moreover, we have the relations 
\be 
\overline{\lan \co(t) \co(t) H^P(k) \ran^m} =
\overline{\lan \co(t) \co(t) \ran^m}\; \overline{\lan H^P(k) \ran^m} = \overline{\lan \co(t)
H^P(k) \co(t) \ran^m}  
\label{strn-eq9} 
\ee 
and their proof is as follows. Let us consider $\overline{\lan \co(t) \co(t) H^P(k)
\ran^m}$. Then, 
\be 
\overline{\lan \co(t) \co(t) H^P(k) \ran^m} = {\binom{N}{m}}^{-1}\,
\dis\sum_{v_i,v_j} \overline{\lan m v_i \mid
\co(t) \co(t) \mid m v_j\ran}  \overline{\lan m v_j \mid H^P(k) \mid m v_i\ran} \;.
\label{strn-eq9a} 
\ee 
Now applying Eq. (\ref{sum-hh}) gives $\overline{\lan m v_i \mid \co(t) \,
\co(t) \mid m v_j\ran} =  \overline{\lan \co(t) \, \co(t) \ran^m}\,\delta_{v_i , v_j}$.
Substituting this into Eq. (\ref{strn-eq9a}) will give the first equality in 
Eq. (\ref{strn-eq9}). The second equality $\overline{\lan \co(t) \co(t) H^P(k) \ran^m}$ 
$=$ $\overline{\lan \co(t) H^P(k) \co(t) \ran^m}$ follows from the cyclic
invariance of the $m$-particle average. Equation (\ref{strn-eq9}) gives the moments, 
\be 
\barr{l} 
M_{20}(m) = M_{02}(m) = \overline{\lan \co(t) \; \co(t) \ran^m} 
\;\;\overline{ \lan H^2(k)\ran^m}\;, 
\\ \\ 
M_{40}(m) = M_{04}(m) = \overline{\lan \co(t) \; \co(t) \ran^m} \;\;
\overline{ \lan H^4(k)\ran^m}\;. 
\earr \label{strn-eq9c} 
\ee 
Formulas for $\overline{\lan \co(t) \, \co(t) \ran^m}$, $\overline{\lan H^2(k) \ran^m}$ and
$\overline{\lan H^4(k)\ran^m}$ follow from Eqs. (\ref{strn-eq8}), (\ref{strn-eq7a})
and (\ref{strn-eq7b}). Thus, the non-trivial moments $M_{PQ}$ for  $P+Q \leq 4$
are $M_{11}$, $M_{13}=M_{31}$ and $M_{22}$.

It is easy to recognize that the bivariate moment $M_{11}$ has same structure as
the third term in Eq. (\ref{strn-eq7ab}) for $\overline{\lan H^4(k) \ran^m}$ as the
$\co$ and $H$ ensembles are independent. Then, the formula for $M_{11}$ follows
directly from the second term in Eq. (\ref{strn-eq7b}). This gives,
\be
\barr{l}
M_{11}(m) =\overline{\lan \co(t)\, H(k)\,\co(t)\,H(k)\ran^m} 
\\ \\
= V^2_\co\; V^2_H\;{\dis\binom{N}{m}}^{-1}\; \dis\sum_{\nu=0}^{min(k,m-t)}\,
\Lambda^{\nu}(N,m,t)\, 
\Lambda^{\nu}(N,m,m-k)\,d(N:\nu)\;.
\earr \label{strn-eq10}
\ee
This equation has the correct $ t \leftrightarrow k$ symmetry. Using Eqs. 
(\ref{strn-eq10}) and (\ref{strn-eq7a}), we have formula for the bivariate 
correlation coefficient $\xi$,
\be
\resizebox{0.95\hsize}{!}{$
\xi(m) = \dis\frac{M_{11}(m)}{\dis\sqrt{M_{20}(m)\, M_{02}(m)}} = 
\dis\frac{\dis\sum_{\nu=0}^{min(t,m-k)}\,
\Lambda^{\nu}(N,m,k)\, \Lambda^{\nu}(N,m,m-t)\,d(N:\nu)}{
\dis\binom{N}{m}\;\Lambda^0(N,m,t)\,\Lambda^0(N,m,k)}\;.$}
\label{strn-eq11}
\ee
Turning to $M_{PQ}$ with $P+Q=4$, the first trivial moment is $M_{13}=M_{31}$. 
For $M_{31}$ we have,
\be
\barr{l}
M_{31}(m) = \overline{\lan \co(t) H(k) \co(t) H^3(k)\ran^m} \\ \\
= \resizebox{0.95\hsize}{!}{${\dis\binom{N}{m}}^{-1}\;
\dis\sum_{v_i, v_j, v_k, v_l} \, \overline{\lan m v_i \mid \co(t) 
\mid m v_j\ran\, \lan m v_j \mid H(k) \mid m v_k\ran \, 
\lan m v_k \mid \co(t) \mid m v_l\ran\,  \lan m v_l \mid 
H^3(k) \mid m v_i\ran}$} 
\\ \\
= \resizebox{0.95\hsize}{!}{${\dis\binom{N}{m}}^{-1}\;\dis\sum_{v_i, v_j,
v_k, v_l} \overline{\lan m v_i \mid \co(t) \mid m v_j\ran\,\lan m v_k \mid 
\co(t) \mid m v_l\ran} \; \overline{ \lan m v_j \mid H(k) \mid
m v_k\ran\,\lan m v_l \mid H^3(k) \mid m v_i\ran}$} \;. 
\earr \label{eq-m31a}
\ee
In Eq. \eqref{eq-m31a}, the last equality follows from the fact that the 
EGUE's representing $H$ and $\co$ are independent. The ensemble average 
of the product of two $\co$ matrix elements follows easily from Eq. 
(\ref{matrix-ele-b}) giving, 
\be
\resizebox{0.95\hsize}{!}{$\overline{\lan m v_i \mid \co(t) 
\mid m v_j\ran\,\lan m v_k \mid \co(t) \mid m 
v_l\ran} = V_\co^2 \, \dis\sum_{\nu=0}^{m-t} \Lambda^{\nu}(N,m,t) 
\;\dis\sum_{\omega_\nu} C^{\nu , \omega_\nu}_{f_m v_i\,,\;\overline{f_m} 
\overline{v_l}} \, C^{\nu , \omega_\nu}_{f_m v_k\,,\;\overline{f_m} 
\overline{v_j}}\;.$}
\label{eq-m31b}
\ee
The ensemble average of the product of a $H$ matrix element and $H^3$ matrix
element appearing in Eq. (\ref{eq-m31a}) is,
\be
\barr{l}
\overline{ \lan m v_j \mid H(k) \mid m v_k\ran\,\lan m v_l \mid H^3(k) 
\mid m v_i\ran} \\ \\
= \resizebox{0.95\hsize}{!}{$
\dis\sum_{v_p , v_q} \overline{\lan m v_j \mid H(k) \mid m v_k\ran\,
\lan m v_l \mid H(k) \mid m v_p\ran \lan m v_p \mid H(k) \mid m v_q\ran 
\lan m v_q \mid H(k) \mid m v_i\ran}$} 
\\ \\
= \resizebox{0.95\hsize}{!}{$\dis\sum_{v_p , v_q} \l[\overline{\lan m v_j 
\mid H(k) \mid m v_k\ran\, \lan m v_l \mid H(k) \mid m v_p\ran}\;\; 
\overline{\lan m v_p \mid H(k) \mid m v_q\ran \lan m v_q \mid H(k) 
\mid m v_i\ran}\r.$} 
\\ \\
+ \, \resizebox{0.95\hsize}{!}{$\overline{\lan m v_j \mid H(k) \mid m v_k\ran\,
\lan m v_q \mid H(k) \mid m v_i\ran}\;\; \overline{\lan m v_l \mid H(k) \mid m 
v_p\ran \lan m v_p \mid H(k) \mid m v_q\ran}$} 
\\ \\
+ \,\resizebox{0.95\hsize}{!}{$\l. \overline{\lan m v_j \mid H(k) \mid m v_k\ran\,
\lan m v_p \mid H(k) \mid m v_q\ran}\;\; \overline{\lan m v_l \mid H(k) \mid m 
v_p\ran \lan m v_q \mid H(k) \mid m v_i\ran} \r] \;.$}
\earr \label{eq-m31c}
\ee
The first two terms in Eq. (\ref{eq-m31c}) simplify to give $2\overline{\lan H^2(k) 
\ran^m} \, M_{11}(m)$, using Eq. (\ref{eq-m31b}). Simplifying the third term using 
Eqs. (\ref{matrix-ele-a}) and (\ref{matrix-ele-b}), we get
\be
\barr{l}
\resizebox{0.95\hsize}{!}{$\overline{\lan m v_j \mid H(k) \mid m v_k\ran\,
\lan m v_p \mid H(k) \mid m v_q\ran}\;\; \overline{\lan m v_l \mid H(k) 
\mid m v_p\ran \lan m v_q \mid H(k) \mid m v_i\ran}$} 
\\ \\
= \resizebox{0.95\hsize}{!}{$V_H^4 \dis\sum_{\nu_1=0;\omega_{\nu_1}}^{k} 
\dis\sum_{\nu_2=0;\omega_{\nu_2}}^{m-k} \Lambda^{\nu_1}(N,m,m-k)\;
\Lambda^{\nu_2}(N,m,k) \; C^{\nu_1 , \omega_{\nu_1}}_{f_m v_j\,,\;\overline{f_m} 
\overline{v_k}}\, 
C^{\nu_1 , \omega_{\nu_1}}_{f_m v_p\,,\;\overline{f_m} \overline{v_q}}\,
C^{\nu_2 , \omega_{\nu_2}}_{f_m v_l\,,\;\overline{f_m} \overline{v_i}}\, 
C^{\nu_2 , \omega_{\nu_2}}_{f_m v_q\,,\;\overline{f_m} \overline{v_p}}\;.$}
\earr \label{eq-m31d}
\ee
Combining this with Eq. (\ref{eq-m31b}) and applying the orthonormal properties
of the C-G coefficients will give the final formula for $M_{31}$,
\be
\barr{l}
M_{31}(m) =\overline{\lan \co(t)\, H(k)\,\co(t)\, H^3(k) \ran^m} \\ \\ 
= \resizebox{0.95\hsize}{!}{$2\;\overline{\lan [H(k)]^2\ran^m}\;M_{11}(m) 
+ \dis\frac{V^2_\co\; V^4_H}{\dis\binom{N}{m}}\; \dis\sum_{\nu=0}^{min(k,m-k,m-t)}
\,\Lambda^{\nu}(N,m,t)\, 
\Lambda^{\nu}(N,m,k)\,\Lambda^{\nu}(N,m,m-k)\,d(N:\nu) \;.$} 
\earr \label{strn-eq12}
\ee
Thus, $M_{31}(m)$ involves only the $\Lambda$ functions. 

To derive the formula for $M_{22}$, we will make use of the decompositions
similar to those in Eqs. (\ref{eq-m31a}) and (\ref{eq-m31c}). Then it is easy to
see that $M_{22}$ will have three terms and let us say they are $S_1$, $S_2$ 
and $S_3$. The first term is $S_1 = \overline{\lan \co(t) \, \co(t) \ran^m}\;
\l[\,\overline{\lan H^2(k)\ran^m}\;\r]^2$ and the second term $S_2$ is
\be
\barr{l}
S_2={\dis\binom{N}{m}}^{-1}\;\dis\sum_{v_i, v_j, v_r, v_l, v_p, v_Q} 
S_{21}(v_i, v_j, v_r, v_l) \, S_{22}(v_i, v_j, v_l, v_P, v_Q)\;;\\
\\
S_{21}(v_i, v_j, v_r, v_l)= \overline{\lan m,v_i \mid \co(t) \mid m,v_j \ran  
\lan m,v_r \mid \co(t) \mid m,v_l \ran}\;,\\   
\\
S_{22}(v_i, v_j, v_l, v_P, v_Q) = \overline{\lan m,v_j \mid H(k) \mid m,v_P\ran 
\lan m,v_Q \mid H(k) \mid m,v_i\ran} \\
\\ \;\;\;\;\;\;\;\;\;\;\;\;\;\;\;\;\;\;\;\;\;\;\;\;\;\;\;\;\;\;\;
\times\;\overline{\lan m,v_P \mid H(k) \mid 
m,v_r\ran \lan m,v_l \mid H(k) \mid m,v_Q\ran}\;.
\earr \label{strn-eq13x}
\ee
The sum involving $S_{22}$ is simplified by applying Eq. (\ref{matrix-ele-b})
to the two ensemble averages in $S_{22}$ and using the orthonormal properties of
the C.G coefficients. Now, applying Eq. (\ref{matrix-ele-a}) to the ensemble
average in $S_{21}$ and then multiplying with $S_{22}$ will (after using
again the orthonormal properties of the C-G coefficients) lead to the final 
result for $S_{2}$,
\be
S_2 = {\binom{N}{m}}^{-1} \;\dis\sum_{\nu=0}^{min(t,m-k)}\,
\Lambda^{\nu}(N,m,m-t)\, \l[\Lambda^{\nu}(N,m,k)\r]^2\,d(N:\nu) \;.
\label{strn-eq13y}
\ee
Similarly, the term $S_3$ is
\be
\barr{l}
S_3={\dis\binom{N}{m}}^{-1}\;\dis\sum_{v_i, v_j, v_r, v_l, v_p, v_Q} 
S_{31}(v_i, v_j, v_r, v_l, v_P, v_Q) \, S_{32}(v_i, v_r, v_P, v_Q)\;;\\
\\
S_{31}(v_i, v_j, v_r, v_l, v_P, v_Q)= \overline{\lan m,v_i \mid \co(t) 
\mid m,v_j \ran \lan m,v_r \mid \co(t) \mid m,v_l \ran} \\
\\ \;\;\;\;\;\;\;\;\;\;\;\;\;\;\;\;\;\;\;\;\;\;\;\;\;\;\;\;\;\;\;\;\;\;\;\;
\times\;\overline{\lan m,v_j \mid H(k) \mid 
m,v_P\ran \lan m,v_l \mid H(k) \mid m,v_Q\ran} \;,\\   
\\
S_{32}(v_i, v_r, v_P, v_Q) = \;\;\;\overline{\lan m,v_P \mid H(k) \mid m,v_r\ran 
\lan m,v_Q \mid H(k) \mid m,v_i\ran}\;.
\earr \label{strn-eq13xx}
\ee
The term $S_{31}$ is simplified by first using Eqs. (\ref{strn-eq5}) and 
(\ref{strn-eq5a}) and this gives, after carrying out the ensemble averages,
\be
\barr{rcl}
S_{31} & = & \resizebox{0.85\hsize}{!}{$V^2_\co \; V^2_H 
\;\dis\sum_{v_i, v_j, v_r, v_l, v_p, v_Q}\;
\dis\sum_{\nu_1=0;\omega_{\nu_1}}^t\;\dis\sum_{\nu_2=0;\omega_{\nu_2}}^k 
\lan m,v_i \mid B^{\nu_1 \omega_{\nu_1}}(t) \mid m,v_j \ran $}
\\ \\ & \times & 
\resizebox{0.85\hsize}{!}{$\lan m,v_r \mid B^{\nu_1 \omega_{\nu_1}}(t) \mid m,v_l \ran \;
\lan m,v_j \mid B^{\nu_2 \omega_{\nu_2}}(k) \mid m,v_P\ran \;
\lan m,v_l \mid B^{\nu_2 \omega_{\nu_2}}(k)  \mid m,v_Q\ran$} 
\\ \\
& = & V^2_\co \; V^2_H \;\dis\sum_{v_i, v_r, v_p, v_Q}\;
\dis\sum_{\nu_1=0;\omega_{\nu_1}}^t\;\dis\sum_{\nu_2=0;\omega_{\nu_2}}^k 
\lan m,v_i \mid B^{\nu_1 \omega_{\nu_1}}(t)\,B^{\nu_2 \omega_{\nu_2}}(k) 
\mid m,v_P \ran \\
\\
& \times & \lan m,v_r \mid B^{\nu_1 \omega_{\nu_1}}(t)\,B^{\nu_2 \omega_{\nu_2}}(k)  
\mid m,v_Q\ran \;.
\earr \label{strn-abcd}
\ee
Now, coupling the $B$'s in each matrix element in the last equality in Eq.
(\ref{strn-abcd}) applying the Wigner-Eckart theorem for the matrix elements of
the coupled tensor operator $[B^{\nu_1}(t) B^{\nu_2}(k)]^{\nu\,\omega_\nu}$
along with the application of Eq. (\ref{matrix-ele-b}) to the $S_{32}$ term 
will give the formula for the $S_3$ term. With this, the $M_{22}$ is given by,
\be
\barr{l}
M_{22}(m) = \overline{\lan \co(t) \, \co(t)\ran^m}\; \l[\,\overline{\lan H(k) \, H(k)
\ran^m}\,\r]^2 \\ \\
+ V^2_\co V^4_H \, {\dis\binom{N}{m}}^{-1} \; \l\{ \dis\sum_{\nu=0}^{
min(t,m-k)}\,\Lambda^{\nu}(N,m,m-t)\, 
\l[\Lambda^{\nu}(N,m,k)\r]^2\,d(N:\nu)\r.  \\
\\
+ \resizebox{0.95\hsize}{!}{$\dis\sum_{\nu=0}^{min(k+t,m-k)}
\Lambda^{\nu}(N,m,k)\,d(N:\nu)\;\;\dis\sum_{\nu_1=0}^{t} \; \l. \dis\sum_{\nu_2=0}^{k}\; \dis\sum_\rho 
\l|\lan f_m \mid\mid \l[B^{\nu_1}(t) B^{\nu_2}(k)\r]^{\nu\,,\,\rho} \mid\mid 
f_m\ran\r|^2\r\} \;.$}
\earr \label{strn-eq13a}
\ee
Here $\rho$ labels multiplicity of the irrep $\nu$ in the Kronecker product
$\nu_1 \times \nu_2 \rightarrow \nu$. We will see in Section 4 that Eq.
(\ref{strn-eq13a}) is useful in deriving  asymptotic results. In order to
evaluate $M_{22}$, we need a formula for the reduced matrix element in Eq.
(\ref{strn-eq13a}). This is obtained by considering the corresponding matrix
element, decoupling the coupled operator, using complete set of states between
the two operators, applying Wigner-Eckart theorem to the two matrix elements and
then simplifying the sums over all the C-G coefficients. This gives,
\be
\barr{l}
\lan f_m \mid\mid \l[B^{\nu_1}(t) B^{\nu_2}(k)\r]^{\nu\,,\,\rho} \mid\mid 
f_m\ran = \dis\sqrt{\dis\frac{d(N:\nu_1)\; d(N:\nu_2)}{d(N:\nu)\; 
\binom{N}{m}}} \\
\\
\times \lan f_m \mid\mid B^{\nu_1}(t) \mid\mid f_m\ran
\lan f_m \mid\mid B^{\nu_2}(k) \mid\mid f_m\ran
\;U(f_m\,\nu_1\,f_m\,\nu_2\,;\,f_m\,\nu)_{\rho}\;.
\earr \label{strn-eq13yy}
\ee
Eq. (\ref{strn-eq13yy}) gives the formula for $M_{22}$ in terms of $U(N)$ 
Racah coefficients,
\be
\barr{l}
M_{22}(m) =\overline{\lan \co(t)\, H^2(k)\,\co(t)\,H^2(k)\ran^m}  
= V^2_\co \; V^4_H\; \l\{\l[\Lambda^0(N,m,k)\r]^2\,\Lambda^0(N,m,t)\r. 
\\ \\
+ {\dis\binom{N}{m}}^{-1} \; \dis\sum_{\nu=0}^{min(t,m-k)}\,
\Lambda^{\nu}(N,m,m-t)\, 
\l[\Lambda^{\nu}(N,m,k)\r]^2\,d(N:\nu)  
\\ \\
+ {\dis\binom{N}{m}}^{-2} \; \dis\sum_{\nu=0}^{min(k+t,m-k)}
\;\dis\sum_{\nu_1=0}^{t}\;\; \dis\sum_{\nu_2=0}^{k}\,\Lambda^{\nu}(N,m,k)\, 
\Lambda^{\nu_1}(N,m,m-t)\, \\
\\
\times \l.\Lambda^{\nu_2}(N,m,m-k)\,d(N:\nu_1) d(N:\nu_2) 
\dis\sum_\rho \l[U(f_m\, \nu_1\,f_m\,\nu_2\;;\;f_m\,\nu)_\rho\r]^2\r\}\;. 
\earr \label{strn-eq13}
\ee
The $U$ or Racah coefficient in Eq. (\ref{strn-eq13}) is with
respect to $U(N)$ and a formula for this is not  available in closed form in the
literature. Deriving formulas for this  $U$-coefficient is an important open
problem. We will show ahead that it is possible to derive a formula valid in the
asymptotic ($N \rightarrow \infty$, $m \rightarrow \infty$, $m/N \rightarrow 0$
and $k,t$ fixed) limit.

\section{Asymptotic results for the bivariate moments for $H$ EGUE($k$) and 
$\co$ an independent EGUE($t$)}

Lowest order (sufficient for most purposes) shape parameters of the bivariate
strength density are the bivariate reduced cumulants of order four, i.e.
$k_{rs}(m)$, $r+s=4$. The $k_{rs}(m)$ can be written in terms  of the normalized
central moments  $\wtM_{PQ}(m)$ where $\wtM_{PQ}(m)=M_{PQ}(m)/M_{00}(m)$. 
Then, the scaled moments $\mu_{PQ}(m)$ are 
\be
\mu_{PQ}(m) = \dis\frac{\wtM_{PQ}(m)}{\l[\wtM_{20}(m)\r]^{P/2} 
\l[\wtM_{02}(m)\r]^{Q/2}}\;,\;\;\; P+Q \geq 2\;.
\label{eq.normom}
\ee
Note that $\mu_{20}(m)=\mu_{02}(m)=1$ and $\mu_{11}(m)=\xi(m)$. Now the 
fourth order cumulants are,
\be
\barr{c}
k_{40}(m) = \mu_{40}(m) - 3\,, \;\;k_{04}(m) = \mu_{04}(m) - 3\,,\\
k_{31}(m) = \mu_{31}(m) - 3\,\xi(m)\,,\;\;k_{13}(m) = \mu_{13}(m) - 3\,\xi(m)\,,\\
k_{22}(m) = \mu_{22}(m)- 2\;\xi^2(m) -1 \;. 
\earr \label{eq.cumu}
\ee
The $\l|k_{rs}(m)\r| \sim 0$ for $r+s \geq 3$ implies that the transition strength
density is close to a bivariate Gaussian (note that in our EGUE applications,
$k_{rs}(m)=0$ for $r+s$ odd by definition). Numerical results for $k_{rs}(m)$, 
$r+s=4$ and also for $\xi(m)$ for some typical values of $N$, $m$, $k$ and 
$t$ are shown in Table 1. It is seen that for sufficiently large values of $N$ and 
$m$ and $k+t$ relatively small, in  general the magnitude of the fourth order 
cumulants is very small ($< 0.3$) implying that for EGUE, transition strength 
densities approach a bivariate Gaussian. However, with increasing $k+t$ value 
it is seen that $\xi(m) \rightarrow 0$  and this is the GOE result. Also, in 
this limit the marginal densities approach semi-circle form giving 
$k_{40}(m) \rightarrow -1$. For a better understanding of these results,  
it is useful to derive expressions for $\mu_{PQ}(m)$ and thereby for $k_{PQ}(m)$, 
using Eq. (\ref{eq.cumu}), in the asymptotic (asymp) limit defined by  $N \rightarrow 
\infty$, $m \rightarrow \infty$, $m/N \rightarrow 0$  and fixed $k$ and $t$ 
with $t < k$. First we will consider $N \rightarrow \infty$ and $m$ fixed with 
$k,t << m$. Two relations we use are,
\be 
\dis\binom{N-p}{r} \stackrel{p/N \rightarrow 0}{\longrightarrow} 
\dis\frac{N^r}{r!}\;,\;\; d(N:\nu) \stackrel{\nu/N \rightarrow
0}{\longrightarrow} \dis\frac{N^{2\nu}}{ (\nu!)^2}\;. 
\label{asymp2} 
\ee 
Note that $d(N:\nu)$ is defined in Eq. (\ref{eq-dnu}). Let us start with the formula
for $\xi$ given  by Eq. (\ref{strn-eq11}). Applying Eq. (\ref{asymp2}), it is
easily seen that only the term with $\nu=t$ in the sum in Eq. (\ref{strn-eq11})
will contribute in the asymptotic limit. Using this and applying Eq.
(\ref{asymp2}) to the formula for $\Lambda^\nu$ given by Eq. (\ref{reduced}) will
lead to, 
\be \xi(m)  \longrightarrow  \dis\frac{m!\;N^{2t}}{N^m\;(t!)^2} 
\dis\frac{\binom{m-t}{k} \binom{N-m+k-t}{k}\binom{N-2t}{m-t}}{\binom{m}{k} 
\binom{N-m+k}{k} \binom{m}{t}\binom{N-m+t}{t}} 
\stackrel{\mbox{asymp}}{\longrightarrow}  {\binom{m}{k}}^{-1} \binom{m-t}{k}\;.
\label{asymp3} 
\ee 
Similarly, $\mu_{40}(m)$ will be, using Eqs. (\ref{strn-eq9c}), (\ref{strn-eq7b})
and (\ref{strn-eq7a}), 
\be 
\mu_{40}(m)  \longrightarrow  2 +
\dis\frac{m!\;N^{2k}}{N^m\;(k!)^2}  \dis\frac{\binom{m-k}{k}
\binom{N-m}{k}\binom{N-2k}{m-k}}{\binom{m}{k}  \binom{N-m+k}{k}
\binom{m}{k}\binom{N-m+k}{k}}  \stackrel{\mbox{asymp}}{\longrightarrow}  2+
{\binom{m}{k}}^{-1} \binom{m-k}{k}\;. 
\label{asymp4} 
\ee 
Turning to $\mu_{31}(m)$, it is easy to see from Eq. (\ref{strn-eq12}) that  
$M_{31}(m)$ has two terms. The first term is $2 \,\xi(m)\lan \co(t) \, \co(t) \ran^m 
[\lan H(k) H(k)\ran^m]^2$ and in the sum in the second term, only $\nu=k$ will 
survive in the asymptotic limit. These then will give, 
\be 
\barr{rcl} 
\mu_{31}(m) & = & \resizebox{0.75\hsize}{!}{$2 \, \xi(m) + {\dis\binom{N}{m}}^{-1}
\dis\frac{\Lambda^k(N,m,t)  \Lambda^k(N,m,k)
\Lambda^k(N,m,m-k)\,d(N:k)}{\Lambda^0(N,m,t)[ \Lambda^0(N,m,k)]^2}$} 
\\ \\
& \stackrel{\mbox{asymp}}{\longrightarrow} & \xi(m)\,\l[2 + {\dis\binom{m}{k}}^{-1}\,
\dis\binom{m-k}{k} \r] = \xi(m)\; \mu_{40}(m)\;. 
\earr 
\label{asymp5} 
\ee 
Here, Eq. (\ref{asymp2}) is used in the final simplifications. Finally, let us consider
$\mu_{22}(m)$.  Firstly, $M_{22}(m)$ has three terms as seen from Eqs.
(\ref{strn-eq13}) and (\ref{strn-eq13a}) and let us call them $X1$, $X2$ and
$X3$. Then, there will be corresponding three terms in $\mu_{22}(m)$ and we
call them $T1$, $T2$ and $T3$. It is seen that $T1=1$ and $T2$ is (in the
corresponding $X2$ sum, only $\nu=t$ term will contribute in the asymptotic
limit), 
\be 
\resizebox{0.95\hsize}{!}{$T2 \longrightarrow  {\dis\binom{N}{m}}^{-1}
\dis\frac{\Lambda^t(N,m,m-t) 
\l[\Lambda^t(N,m,k)\r]^2\,d(N:t)}{\Lambda^0(N,m,t)[\Lambda^0(N,m,k)]^2} 
\longrightarrow  {\binom{m}{k}}^{-2}\,{\binom{m-t}{k}}^2\;. $}
\label{asymp6} 
\ee
As a formula for the $U$-coefficient appearing in Eq. (\ref{strn-eq13}) 
is not available, we use  Eq. (\ref{strn-eq13a}) for simplifying $T3$. Then, in 
the $X_3$ sum only $\nu=t+k$ term will survive in the asymptotic limit giving
\be 
T3 \rightarrow {\dis\binom{N}{m}}^{-1} \dis\frac{\Lambda^{t+k}(N,m,k)\, 
d(N:t+k)}{\Lambda^0(N,m,t)[\Lambda^0(N,m,k)]^2}\; \l|\lan f_m \mid\mid
\l[B^{t}(t) B^{k}(k)\r]^{t+k} \mid\mid f_m\ran\r|^2 \;. 
\label{asymp7} 
\ee  
For further simplification of $T3$, we will use the relation 
\be 
\barr{l}
\resizebox{0.98\hsize}{!}{$\dis\binom{N}{m}\,\overline{\lan
\co(t) H(k) \co(t) H(k)\ran^m} = V^2_\co \;
V^2_H  \dis\sum_{\nu_1 =0}^t \dis\sum_{\nu_2=0}^k \dis\sum_{\nu ; \rho}  
\l|\lan f_m \mid\mid \l[B^{\nu_1}(t) B^{\nu_2}(k)\r]^{\nu ; \rho} \mid\mid 
f_m\ran\r|^2\;d(N:\nu)\;.$} 
\earr \label{asymp8} 
\ee 
Then, $T_3$ in the asymptotic limit will be 
\be 
T_3 \longrightarrow
\dis\frac{\Lambda^{t+k}(N,m,k)}{\Lambda^{0}(N,m,k)}\; \l[\xi(N \rightarrow
\infty)\r]  \rightarrow {\dis\binom{m}{k}}^{-2} \,\dis\binom{m-t-k}{k} 
\dis\binom{m-t}{k} \;. 
\label{asymp9} 
\ee 
Notice that, here we have used Eq. (\ref{asymp3}) for $\xi(m)$ in the $N
\rightarrow \infty$ limit. Now combining $T_1$, $T_2$ and $T_3$ we have 
\be
\mu_{22}(m) \stackrel{\mbox{asymp}}{\longrightarrow} 1 + 
{\dis\binom{m}{k}}^{-2}\,{\dis\binom{m-t}{k}}^2 + {\dis\binom{m}{k}}^{-2}\, 
\dis\binom{m-t-k}{k} \dis\binom{m-t}{k}\;. 
\label{asymp10} 
\ee  
Comparing Eq. (\ref{asymp9}) with the
formula for $T_3$ as given  by  Eq. (\ref{strn-eq13}), it is easy to see that in
the asymptotic limit 
\be 
\l[ U(f_m\; t\; f_m\; k\;;\; f_m\; t+k) \r]^2
\stackrel{\mbox{asymp}}{ \longrightarrow} {\binom{m}{k}}^{-1}\;\binom{m-t}{k}\;.
\label{asymp11} 
\ee  
The asymptotic formulas derived from the exact results and given by Eqs. 
(\ref{asymp3}), (\ref{asymp4}),  (\ref{asymp5}) and (\ref{asymp10}) are  
same as those given before in \cite{FKPT,kota} where the theory given by
Mon and French \cite{Mo-75}, that gives directly the asymptotic results, was 
used. This is a good test of the derivations in Section 3. It is important to
add that in the derivations given in Section 3 we have considered only the 
binary correlated terms in the various sums [the binary correlations come in the
ensemble average of the $W$'s and $U$'s defined by Eqs. (\ref{strn-eq5}) and 
(\ref{strn-eq5a})]. Within this constraint, the results in Section 3 are exact.
On the other hand, in the Mon and French theory, the binary correlated terms
are evaluated using formulas that are valid only in the asymptotic limit.

Asymptotic formulas for the bivariate moments will give the asymptotic formulas
for the fourth order bivariate cumulants  $k_{40}(m)$, $k_{31}(m)$ and $k_{22}(m)$ 
by applying Eqs. (\ref{eq.normom}) and (\ref{eq.cumu}). Then we have, 
\be 
\barr{rcl} 
k_{40}(m) = k_{04}(m) & = &
\dis\binom{m-k}{k}\,{\dis\binom{m}{k}}^{-1} -1\;, \\ 
\\ 
k_{31}(m) = k_{13}(m) & = & \xi(m)\,k_{40}(m)\;, \\ 
\\ 
k_{22}(m) & = & \xi^2(m) \;\l\{\dis\binom{m-k-t}{k}\,
{\dis\binom{m-t}{k}}^{-1} -1  
\r\}\;. 
\earr \label{eq-asympnew} 
\ee 
Remember that $\xi(m)$ is given by Eq. (\ref{asymp3}). Now,
the $1/m$ expansion of $k_{rs}(m)$ shows that $k_{rs}(m)=-k^2/m + O(1/m^2)$. 
This clearly demonstrates that we have for the transition strength densities bivariate
Gaussian form in general in the dilute limit defined by $N \rightarrow \infty$,
$m \rightarrow \infty$, $m/N \rightarrow 0$ with fixed $k$ and $t$. However, $\xi(m)
\rightarrow 1$ in the dilute limit and this gives a singular bivariate Gaussian that is
unphysical. Therefore, in practice the dilute limit condition will not be
realized and  there will be departures from the bivariate Gaussian form. Thus,
the corrections due to $k_{rs}(m)$, $r+s=4$ should be included and out of the many
ways to add the corrections, the Edgeworth expansion is considered to be the
best \cite{KH,Kend,Fell}. The Edgeworth corrected bivariate Gaussian form 
including $k_{rs}(m)$ with $r+s=4$ is given in Appendix A.

\section{EGUE results for moments of transition strength densities: two types of
spinless fermions with beta and double beta decay type transition operators}

Here we will consider a system with two types of spinless fermions with
$m_1$ fermions of type \#1 in $N_1$ sp states and $m_2$ fermions of type \#2 in
$N_2$ sp states with the system $H$ operator preserving $(m_1,m_2)$. This is
similar to protons ($p$) and neutrons ($n$) in a atomic nucleus. Thus, we have
two orbits with the first one having $N_1$ sp states and the second $N_2$ sp
states. Then the $H$ operator, assumed to be $k$-body, is given by,
\be
\barr{l}
H(k) = \dis\sum_{i+j=k} \dis\sum_{\alpha , \beta \in i} \dis\sum_{a,b \in j}\;
V_{\alpha a:\beta b}(i,j)\;
A^\dagger_{\alpha}(i)\, A_{\beta}(i)\, A^\dagger_a(j)\,A_b(j)\;; \\ \\
V_{\alpha a:\beta b}(i,j) = \lan i, \alpha : j, a \mid H \mid i, \beta : 
j,b\ran\;.
\earr \label{strn-eq14}
\ee
Here, we are using Greek labels $\alpha, \beta, \ldots$ to denote the many
particle states generated by fermions occupying the first orbit and the Roman
labels $a,b,\ldots$ for the many particle states generated by the fermions
occupying the second orbit. Note that $A^\dagger_{\alpha}(i)$  creates the state
$\l.\l|i,\alpha\r.\ran$ with $i$ number of fermions in the first orbit and
$A^\dagger_a(j)$ creates the state $\l.\l|j,a\r.\ran$ with $j$ number of
fermions in the second orbit. Similar is the action of the annihilation
operators $A_{\beta}(i)$ and $A_b(j)$. For example for a two-body Hamiltonian,
$(i,j)=(2,0)$, $(1,1)$ and $(0,2)$. This corresponds to $H = H_{pp} + H_{pn} + 
H_{nn}$ for atomic nuclei; with `p' denoting protons and `n'  denoting neutrons.
For each $(i,j)$ pair with $i+j=k$, we have a matrix $V(i,j)$ in the
$k$-particle space and $H$ matrix is a direct sum of these  matrices in the $k$
particle space. Their dimensions being $\binom{N_1}{2}$, $N_1 N_2$ and
$\binom{N_2}{2}$ respectively. Action of the $H$ operator on   the $\l.\l|m_1,
v_\alpha : m_2, v_a\r.\ran$ states of a $(m_1,m_2)$ system  generates
$(m_1,m_2)$ particle $H$ matrix; $v_\alpha$ and $v_a$ are respective additional
labels. The $H$ matrix dimension is $d(m_1,m_2) =\binom{N_1}{m_1}\,
\binom{N_2}{m_2}$. To proceed further, the $V(i,j)$ matrices are represented 
by  independent GUEs with matrix elements being zero centered Gaussian 
variables with variance,
\be
\overline{V_{\alpha a : \beta b}(i,j)\;V_{\alpha^\pr a^\pr : \beta^\pr b^\pr}
(i^\pr ,j^\pr)} 
=V^2_H(i,j)\;\delta_{i i^\pr} \; \delta_{j j^\pr} \; \delta_{\alpha \beta^\pr} \;
\delta_{a b^\pr} \; \delta_{\beta \alpha^\pr} \; \delta_{b a^\pr} \;.
\label{strn-eq15}
\ee
It is important to note that the embedding algebra for the EGUE generated by 
the action of the $V$ ensemble on $\l.\l|m_1, v_\alpha : m_2, v_a\r.\ran$ 
states is the direct sum algebra  $U(N_1) \oplus U(N_2)$. Thus we have 
EGUE($k$)-$[U(N_1) \oplus U(N_2)]$ ensemble. 

Going beyond the simple operators considered in Sections 2 and 3, here we will 
consider the following transition operator,
\be
\co(k_0) = \dis\sum_{\alpha , a} O_{\alpha a} A^\dagger_{\alpha}(k_0)\,
A_a(k_0)\;;\;\;\;O_{\alpha a} =
\lan k_0,\alpha \mid \co \mid k_0, a\ran\;.
\label{strn-eq16}
\ee
It is easy to see from Eq. (\ref{strn-eq16}) that the operator $\co(k_0)$
changes  $k_0$ number of  fermions in the second orbit to $k_0$ number of
fermions in  the first orbit. This is similar to the action of $\beta$ decay
operator (then $k_0=1$) and neutrinoless double beta decay operator ($k_0=2$). 
Eq. (\ref{strn-eq16}) gives the matrix elements of $\co$ for the matrix
representation of $\co$ in the defining space and this is in general a
rectangular ${\cal D}_A \times {\cal D}_B$ matrix where ${\cal D}_A=\binom{N_2}{k_0}$ and
${\cal D}_B=\binom{N_1}{k_0}$.  Just as in Section 3, we will assume GUE representation
for the $\co$ matrix in the defining space and then,
\be
\overline{O^\dagger_{\alpha a} \; O_{\beta b}}= V^2_\co\;\delta_{\alpha \beta} 
\;\delta_{a b}\;.
\label{strn-eq17}
\ee 
Note that the $O$ matrix in the defining space is a GUE in the sense that the
real and imaginary parts of the $O_{\alpha  a}$ are zero centered independent
Gaussian variables with variance given by Eq. (\ref{strn-eq17}). With the GUE
representation for the $O$ matrix in the defining space, we have EGUE  for $\co$
in many particle spaces. This matrix will be again a rectangular matrix with
matrix elements connecting $(m_1 ,m_2)$ states with $(m_1+k_0 ,  m_2-k_0)$
states by the $\co$ operator. Fig . 3 shows an example for the $H$ and $\co$
matrices in the defining and many-particle spaces.  With independent GUE 
representations for $H$ and $\co$ matrices in the defining spaces (then 
$V_{\alpha a:\beta b}(i,j)$ are independent of $O_{\alpha^\pr a^\pr}$), we will
derive formulas for the bivariate moments of the transition strength density 
\be 
\barr{l}
\resizebox{0.97\hsize}{!}{$I_{\co}^{(m_1, m_2)}(E_i , E_f) = I^{(m_1+k_0,
m_2-k_0)}(E_f)\,  \l|\lan (m_1+k_0, m_2-k_0), E_f \mid \co \mid  (m_1, m_2), E_i
\ran\r|^2\,  I^{(m_1, m_2)}(E_i) \;.$} 
\earr 
\label{strn-eq18} 
\ee 
Then, the ensemble averaged bivariate moments are 
\be  
M_{PQ}(m_1 , m_2)=\overline{\lan
\cod(k_0) H^Q(k) \co(k_0) H^P(k) \ran^{(m_1 , m_2)}}\;. 
\label{strn-eq19} 
\ee
Note that $\co$ takes $(m_1 , m_2)$ to $(m_1+k_0 , m_2-k_0)$ uniquely and
therefore the later is not specified explicitly in Eqs. (\ref{strn-eq19}) and 
(\ref{strn-eq18}). Also, it is important to note that $\cod \neq \co$.

Given one type of spinless fermions, we easily have
\be
\barr{c}
\lan m,v_1 \mid \dis\sum_\gamma A^\dagger_\gamma(k_0) A_\gamma(k_0) \mid
m,v_2 \ran^m = \dis\binom{m}{k_0}\,\delta_{v_1,v_2}\;,\\
\\
\lan m,v_1 \mid \dis\sum_\gamma A_\gamma(k_0) A^\dagger_\gamma(k_0) \mid 
m,v_2 \ran^m = \dis\binom{N-m}{k_0}\,\delta_{v_1,v_2}\;.
\earr \label{strn-eq20a}
\ee
Now, substituting complete set of state between the $A^\dagger$ and $A$
operators and applying the Wigner-Eckart theorem [note that 
$A^\dagger_{i}(k)$ and $A_j(k)$ transform as the $U(N)$ tensors 
$f_{k}=\{1^{k}\}$ and $\{\overline{f_{k}}\}= \{1^{N-k0}\}$] will give,
\be
\barr{l}
\lan m \mid\mid A^{\dagger}(k_0) \mid\mid m-k_0\ran \lan m-k_0 \mid\mid
A(k_0) \mid\mid m\ran = \dis\binom{N-k_0}{m-k_0} \;,\\
\\
\lan m \mid\mid A(k_0) \mid\mid m+k_0\ran \lan m+k_0 \mid\mid
A^\dagger(k_0) \mid\mid m\ran = \dis\binom{N-k_0}{m}\;.
\earr \label{strn-eq20z}
\ee 
For the derivations given ahead it is important to recognize that 
$A^\dagger_{\alpha}(k_0)$ and $A_\beta(k_0)$ transform as the $U(N_1)$ tensors 
$f_{k_0}=\{1^{k_0}\}$ and $\{\overline{f_{k_0}}\}=
\{1^{N_1-k_0}\}$ and similarly $A^\dagger_{a}(k_0)$ and $A_b(k_0)$ with respect
to $U(N_2)$. Moreover, using Eq. (\ref{strn-eq20a}), we have
\be
\barr{l}
\overline{\lan \cod(k_0) \, \co(k_0) \ran^{(m_1 , m_2)}} = 
V^2_\co\;\dis\binom{N_1 -m_1}{k_0}\,\dis\binom{m_2}{k_0}\;,\\
\\
\overline{\lan \co(k_0) \, \cod(k_0) \ran^{(m_1 , m_2)}} = 
V^2_\co\;\dis\binom{N_2 -m_2}{k_0}\,\dis\binom{m_1}{k_0} \;.
\earr \label{strn-eq20}
\ee

Before turning to the bivariate moments $M_{PQ}(m_1,m_2)$, let us consider the
second and the fourth moment of the state densities $I^{m_1,m_2}(E)$. 
For deriving formulas for these moments, unitary
decomposition of $H(k)$ needs to be carried out with respect to $U(N_1) \oplus U(N_2)$
algebra and this gives the  $B^{\nu , \omega_\nu}(k^\prime)$ tensors in $U(N_1)$
and $U(N_2)$ spaces. In order to distinguish the tensors, we call them $B$ and
$C$ in the two spaces respectively. Denoting these tensorial ranks by $\nu_1$
and $\nu_2$ respectively, we have
\be
H(k) = \dis\sum_{i+j=k} \dis\sum^{i}_{\nu_1=0\,;\,\omega_{\nu_1}}
\dis\sum^{j}_{\nu_2=0\,;\,\omega_{\nu_2}}
W_{ij}(\nu_1 , \omega_{\nu_1}\,;\,\nu_2 , \omega_{\nu_2})\;B^{\nu_1 , 
\omega_{\nu_1}}(i)\,C^{\nu_2 , \omega_{\nu_2}}(j)\;.
\label{strn-eq21}
\ee
Just as in Section 2, here also it can be proved that the $W$'s will be 
independent zero centered Gaussian variable with variances satisfying,
\be
\barr{l}
\resizebox{0.95\hsize}{!}{$\overline{W_{ij}(\nu_1 , \omega_{\nu_1}\,;\,\nu_2 , \omega_{\nu_2})
W_{rl}(\nu^{\pr}_1 , \omega_{\nu^{\pr}_1}\,;\,\nu^{\pr}_2 , 
\omega_{\nu^{\pr}_2})} = V^2_H \;\delta_{i,r} \, \delta_{j,l} \,
\delta_{\nu_1,\nu^{\pr}_1} \, \delta_{\nu_2, \nu^{\pr}_2} \,
\delta_{\omega_{\nu_1}, \omega_{\nu^{\pr}_1}}  \,
\delta_{\omega_{\nu_2}, \omega_{\nu^{\pr}_2}} \;.$}
\earr \label{strn-21a}
\ee
Using Eqs. (\ref{strn-eq21}) and (\ref{strn-21a}) and the results given 
in Section 2, it is straight forward to derive the formulas for 
$\overline{\lan H^P(k)\ran^{(m_1 , m_2)}}$ for $P=2$ and 4. We have for $P=2$,
\be
\overline{\lan H^2(k)\ran^{(m_1 , m_2)}} = \dis\sum_{i+j=k}\,V^2_H(i,j)\;
\Lambda^0(N_1, m_1, i) \,\Lambda^0(N_2, m_2, j)\;.
\label{strn-eq22}
\ee
Similarly for $P = 4$,
\be
\barr{l}
\resizebox{0.95\hsize}{!}{$\overline{\lan H^4(k)\ran^{(m_1 , m_2)}} = 
2 \l[ \; \overline{\lan H^2(k)\ran^{(m_1 , m_2)}} \;\r]^2
+ \dis\sum_{i+j=k} \,\dis\sum_{i^\pr + j^\pr =k} \;V^2_H(i,j)\, V^2_H(i^\pr , 
j^\pr)\;X(N_1, m_1 , i, i^\pr)\,Y(N_2, m_2 , j, j^\pr)\;;$}
\\ \\
\resizebox{0.95\hsize}{!}{$X(N_1, m_1 , i, i^\pr)= {\dis\binom{N_1}{m_1}}^{-1} \; \dis\sum_{\nu_1=0}^{
min(i, m_1-i^\pr)}\,\Lambda^{\nu_1}(N_1, m_1, m_1-i)\, \Lambda^{\nu_1}(
N_1, m_1, i^\pr)\,d(N_1 : \nu_1)\;,$}
\\ \\
\resizebox{0.95\hsize}{!}{$Y(N_2, m_2 , j, j^\pr)= \dis{\binom{N_2}{m_2}}^{-1} \dis\sum_{\nu_2=0}^{
min(j, m_2-j^\pr)}\,\Lambda^{\nu_2}(N_2, m_2, m_2-j)\, \Lambda^{\nu_2}(N_2, 
m_2, j^\pr)\,d(N_2 : \nu_2)\;.$}
\earr \label{strn-eq23}
\ee
It is also easy to show using Eqs. (\ref{strn-eq20a}) and (\ref{strn-eq17}) 
that,
\be
\barr{l}
\overline{\lan \cod(k_0) \co(k_0) H^P(k) \ran^{(m_1,m_2)}} = 
\overline{\lan \cod(k_0) \co(k_0) \ran^{(m_1,m_2)}}\;\;
\overline{\lan H^P(k) \ran^{(m_1,m_2)}}\;,\\
\\
\overline{\lan \cod(k_0) H^P(k) \co(k_0) \ran^{(m_1,m_2)}} = 
\overline{\lan \cod(k_0) \co(k_0) \ran^{(m_1,m_2)}}\;\;
\overline{\lan H^P(k) \ran^{(m_1+k_0, m_2-k_0)}}\;.
\earr \label{strn-eq24}
\ee
The bivariate moments $M_{00}$, $M_{20}$, $M_{02}$, $M_{40}$ and
$M_{04}$ follow directly by appropriately using Eqs. (\ref{strn-eq20}),
(\ref{strn-eq22})-(\ref{strn-eq24}) in Eq. \eqref{strn-eq19}.

Clearly, the first nontrivial bivariate moment is $M_{11}(m_1 , m_2)$ and it is
given by,
\be
\barr{l}
M_{11}(m_1,m_2) = \overline{\lan \cod(k_0) H(k) \co(k_0) H(k) \ran^{m_1,m_2}} 
= \l\{\dis\binom{N_1}{m_1}\;\dis\binom{N_2}{m_2}\r\}^{-1} 
\\ \\
\times \resizebox{0.95\hsize}{!}{$
\;\dis\sum_{\alpha_1, \alpha_2, \alpha_3, \alpha_4: a_1, a_2, a_3, a_4} 
\mbox{\bf Eavg}\l\{\lan m_1, \alpha_1 ; m_2, a_1 \mid \cod(k_0) \mid m_1+k_0, 
\alpha_2 ; m_2-k_0, a_2\ran \r.$}
\\ \\
\times \resizebox{0.8\hsize}{!}{$\l. \lan m_1+k_0, \alpha_2 ; m_2-k_0, a_2 \mid H(k) 
\mid m_1+k_0, \alpha_3 ; m_2-k_0, a_3\ran \r.$}
\\ \\
\times \resizebox{0.7\hsize}{!}{$\l. \lan m_1+k_0, \alpha_3 ; m_2-k_0, a_3 
\mid \co(k_0) \mid m_1, \alpha_4 ; m_2, a_4\ran \r.$}
\\ \\
\times \resizebox{0.55\hsize}{!}{$\l. \lan m_1, \alpha_4 ; m_2, a_4 
\mid H(k) \mid m_1, \alpha_1 ; m_2, a_1\ran \r\}$}
\earr
\nonumber
\ee
\be
\barr{l}
= \l\{\dis\binom{N_1}{m_1}\;\dis\binom{N_2}{m_2}\r\}^{-1}
\\ \\
\resizebox{0.95\hsize}{!}{$\times \dis\sum_{\alpha_1, \alpha_2, \alpha_3, 
\alpha_4: a_1, a_2, a_3, a_4} 
\mbox{\bf Eavg} \l\{
\lan m_1, \alpha_1 ; m_2, a_1 \mid \cod(k_0) \mid m_1+k_0, 
\alpha_2 ; m_2-k_0, a_2\ran \r.$} 
\\ \\
\times \;\l. \lan m_1+k_0, \alpha_3 ; m_2-k_0, a_3 
\mid \co(k_0) \mid m_1, \alpha_4 ; m_2, a_4\ran\r\} 
\\ \\
\resizebox{0.95\hsize}{!}{$
\times\; \mbox{\bf Eavg} \l\{\lan m_1+k_0, \alpha_2 ; m_2-k_0, a_2 \mid H(k) 
\mid m_1+k_0, \alpha_3 ; m_2-k_0, a_3\ran \r.$} 
\\ \\
\times\;\l. \lan m_1, \alpha_4 ; m_2, a_4 
\mid H(k) \mid m_1, \alpha_1 ; m_2, a_1\ran \r\} \\
\\
= V^2_\co \l\{\dis\binom{N_1}{m_1}\; \dis\binom{N_2}{m_2}\r\}^{-1} 
\dis\sum_{i+j=k}
\;\dis\sum_{\alpha_1, \alpha_2, \alpha_3, \alpha_4: a_1, a_2, a_3, 
a_4: \alpha, a} \;\;\dis\sum_{\nu, \omega_\nu, \nu^\pr, \omega_{\nu^\pr}} 
V^2_H(i,j)
\\ \\
\times \lan m_1, \alpha_1 ; m_2, a_1 \mid A^{\dagger}_a(k_0) A_{\alpha}(k_0) 
\mid m_1+k_0, \alpha_2 ; m_2-k_0, a_2\ran \\
\\
\times \lan m_1+k_0, \alpha_3 ; 
m_2-k_0, a_3 \mid A^{\dagger}_{\alpha}(k_0) A_{a}(k_0) \mid m_1, 
\alpha_4 ; m_2, a_4\ran \\
\\
\times \lan m_1+k_0, \alpha_2 ; m_2-k_0, a_2 \mid B^{\nu, \omega_\nu}(i) 
C^{\nu^\pr, \omega_{\nu^\pr}}(j) \mid m_1+k_0, \alpha_3 ; m_2-k_0, a_3\ran  \\
\\
\times \lan m_1, \alpha_4 ; m_2, a_4 \mid B^{\nu, \omega_\nu}(i) C^{\nu^\pr, 
\omega_{\nu^\pr}}(j) \mid m_1, \alpha_1 ; m_2, a_1\ran\;.
\earr \label{strn-eq25}
\ee
Here we have used `{\bf Eavg}' to denote ensemble average instead of using the
`overline'. We will use this notation at a few other places in the reminder of
this paper. In the second step above we have introduced complete set of states
between the operators to write the ensemble average of $\lan \cod(k_0) H(k) 
\co(k_0) H(k) \ran$ 
as the ensemble average of four matrix elements. In the next step, this ensemble
average is written as the product of the ensemble average of $\co$ matrix
elements and the ensemble average of $H$ matrix elements using the 
independence of the $\co$ and $H$ ensembles. Finally we have used 
Eq. (\ref{strn-eq21}) to get the final form in Eq. (\ref{strn-eq25}).
Now, it is easy to see that $M_{11}(m_1,m_2)$ factorizes into product of terms 
in $m_1$ and $m_2$ spaces giving, $M_{11}(m_1,m_2)$ to be of the form $V^2_\co\,
[\binom{N_1}{m_1}\;\binom{N_2}{m_2}]^{-1} 
\sum_{i+j=k} V^2_H(i,j) X(N_1, m_1, i, k_0) Y(N_2, m_2, j, k_0)$ where for 
example $X(N_1, m_1, i, k_0)$ is 
\be
\barr{l}
X(N_1, m_1, i, k_0) = \\
\\
\dis\sum_{\alpha_1, \alpha_2, \alpha_3, \alpha_4: \alpha, \nu, \omega_\nu} \;
\lan m_1, \alpha_1 \mid A_{\alpha}(k_0) \mid m_1+k_0, \alpha_2\ran 
\lan m_1+k_0, \alpha_3 \mid A^{\dagger}_{\alpha}(k_0)\mid m_1, \alpha_4\ran \\
\\
\times \lan m_1+k_0, \alpha_2 \mid B^{\nu, \omega_\nu}(i) \mid m_1+k_0, 
\alpha_3\ran \lan m_1, \alpha_4 \mid B^{\nu, \omega_\nu}(i) \mid m_1, 
\alpha_1\ran \;.
\earr \label{strn-eq26}
\ee 
Similarly $Y(N_2, m_2, j, k_0)$ can be written. Now, applying the Wigner-Eckart 
theorem along with Eqs. (\ref{reduced}) and (\ref{strn-eq20z}), we will be left 
with four C-G coefficients. These can be simplified to give a $U(N_1)$ $U-$ 
or Racah coefficient. Similarly, for the $Y$ function we will get a $U(N_2)$ Racah
coefficient. Putting these together will give the final formula for $M_{11}$ and
for later usage we will write it in the following form,  
\be
\barr{l}
M_{11}(m_1,m_2) = V^2_\co \; \l\{\dis\binom{N_1}{m_1}
\;\dis\binom{N_2}{m_2}\r\}^{-1} \, \dis\sum_{i+j=k} V^2_H(i,j) 
\;\dis\binom{N_1-k_0}{m_1}\,\dis\binom{N_2-k_0}{m_2-k_0} \\
\times \l[\dis\sum_{\nu_1=0}^i X_{11}(N_1, m_1, k_0, i, \nu_1)\r]\;
\l[\dis\sum_{\nu_2=0}^j Y_{11}(N_2, m_2, k_0, j, \nu_2)\r] \;;
\\ \\
X_{11}(N_1, m_1, k_0, i, \nu) = \l[\dis\binom{N_1}{k_0}\,d(N_1:\nu)\r]^{1/2} 
\\ \\
\times
\;\l[\Lambda^{\nu}(N_1, m_1, m_1-i)\,\Lambda^{\nu}(N_1, m_1+k_0, 
m_1+k_0-i)\r]^{1/2} \\
\\
\times (-1)^{\phi(f_{m_1+k_0}, \overline{f_{m_1}}, f_{k_0}) + \phi(f_{m_1}, 
\overline{f_{m_1}}, \nu)}\; U(f_{m_1+k_0}\,\overline{f_{m_1}}\,
f_{m_1+k_0}\,f_{m_1}\,;\,f_{k_0}\,\nu)\;,\\
\\
Y_{11}(N_2, m_2, k_0, j, \nu) = X_{11}(N_2, m_2-k_0, k_0, j, \nu) \\ \\
= \l[ \dis\binom{N_2}{k_0}\,d(N_2:\nu)\r]^{1/2}
\;\l[\Lambda^{\nu}(N_2, m_2, m_2-j)\,\Lambda^{\nu}(N_2, m_2-k_0, 
m_2-k_0-j)\r]^{1/2} \\
\\
\times (-1)^{\phi(f_{m_2}, \overline{f_{m_2-k_0}}, f_{k_0}) + \phi(f_{m_2-k_0}, 
\overline{f_{m_2-k_0}}, \nu)}\; U(f_{m_2}\,\overline{f_{m_2-k_0}}\,
f_{m_2}\,f_{m_2-k_0}\,;\,f_{k_0}\,\nu)\;.
\earr \label{strn-eq27}
\ee
Note that $f_r = \{1^r\}$ and $\nu=\{2^\nu , 1^{N-2\nu}\}$ with $N=N_1$ or $N_2$
as appropriate. Also, the $U$-coefficient in $X_{11}$ is with respect to
$U(N_1)$ while the one in $Y_{11}$ is with respect to $U(N_2)$. Formula for 
the $U$-coefficient appearing in Eq. (\ref{strn-eq27}) is available in 
\cite{He-75} and it is given by
\be
\barr{l}
U(\{1^n\}\,\{1^{N-p}\}\,\{1^n\}\,\{1^{p}\}\,;\;\{1^{n-p}\}\,
\{2^\nu 1^{N-2\nu}\}) = (-1)^{\phi(N,n,p,\nu)} \\
\\
\times \l[\dis\frac{(p!)^2\,(n-\nu)!\,(N+1)!\,[(N-n)!]^2\,(N-\nu -p)!\,
(N-2\nu +1)}{(\nu !)^2\,(p-\nu)!\,(n-p)!\,[(N+1-\nu)!]^2\,(N-n+p)!
\,(N-n-\nu)!}\r]^{1/2} \\
\earr \label{snt-8}
\ee
The phase factor $(-1)^\phi$ in Eq. (\ref{snt-8}) depends on the phase
convention \cite{He-75} and we will fix this phase later.
Eqs. (\ref{strn-eq25})-(\ref{strn-eq27}) show that in evaluating the bivariate
moment $\overline{\lan \cod(k_0) H^Q(k) \co(k_0) H^P(k) \ran^{m_1,m_2}}$ 
we can use $H(k)$ as $\sum_{i+j=k} H_1(i)H_2(j)$ and then the moment  
will be a sum of terms where each term is a product of two functions with 
one in the $m_1$ space [generated by $H_1(i)$ with body rank $i$] and other 
in the $m_2$ space [generated by $H_2(j)$ with body rank $j$]. Also, these 
functions follow from the results in Sections 2 and 3 by appropriate application. 
This will be seen in the formulas for the fourth order moments that are 
discussed below. 

Turning to the fourth order moments, we need $M_{13}$, $M_{31}$ and $M_{22}$ (we
have already discussed $M_{40}$ and $M_{04}$). As $\cod \neq \co$, here
$M_{13} \neq M_{31}$ and similarly $M_{40} \neq M_{04}$ (also $M_{20} \neq
M_{02}$). Following the procedure used for deriving the formula for
$M_{11}(m_1,m_2)$ and the results [Eqs. (\ref{eq-m31a}) and (\ref{eq-m31c})] 
given in Section 3 for $M_{31}(m)$, we have for $M_{31}(m_1,m_2)$,
\be
\barr{l}
M_{31}(m_1,m_2) = \overline{\lan \cod(k_0) H(k) \co(k_0) H^3(k) \ran^{m_1,m_2}} 
\\ \\
= 2\;\overline{\lan H^2(k)\ran^{m_1,m_2}}\;M_{11}(m_1,m_2) + V^2_\co\;
\l\{\dis\binom{N_1}{m_1}  \dis\binom{N_2}{m_2}\r\}^{-1}\; \\
\\
\resizebox{0.95\hsize}{!}{$
\times \dis\sum_{i_1+j_1=k} \dis\sum_{i_2+j_2=k} V^2_H(i_1,j_1)
V^2_H(i_2,j_2)\;X_{31}(N_1,m_1,i_1,i_2,k_0)\,Y_{31}(N_2,m_2,j_1,j_2,k_0)\;;$}
\earr
\nonumber
\ee
\be
\barr{l}
\resizebox{0.95\hsize}{!}{$X_{31}(N_1,m_1,i_1,i_2,k_0)=
\dis\sum_{\alpha_1, \alpha_2, 
\alpha_3, \alpha_4, \alpha_5, \alpha_6: \alpha, \nu_1, \omega_{\nu_1},
\nu_2, \omega_{\nu_2}}  
\lan m_1, \alpha_1 \mid A_{\alpha}(k_0) \mid m_1+k_0, \alpha_2\ran$}
\\ \\
\times 
\lan m_1+k_0, \alpha_3 \mid A^{\dagger}_{\alpha}(k_0)\mid m_1, \alpha_4\ran 
\lan m_1+k_0, \alpha_2 \mid B^{\nu_1, \omega_{\nu_1}}(i_1) \mid m_1+k_0, 
\alpha_3\ran
\\ \\
\times  \lan m_1, \alpha_5 \mid B^{\nu_1, \omega_{\nu_1}}(i_1) \mid m_1, 
\alpha_6\ran  \lan m_1, \alpha_4 \mid B^{\nu_2, \omega_{\nu_2}}(i_2) \mid m_1, 
\alpha_5\ran \\
\\
\times\; \lan m_1, \alpha_6 \mid B^{\nu_2, \omega_{\nu_2}}(i_2) \mid m_1, 
\alpha_1\ran\;, \\
\\
\resizebox{0.95\hsize}{!}{$Y_{31}(N_2,m_2,j_1,j_2,k_0)=\dis\sum_{b_1, b_2, 
b_3, b_4, b_5, b_6: a, \nu_3, \omega_{\nu_3}, \nu_4, \omega_{\nu_4}}  
\lan m_2, b_1 \mid A^{\dagger}_{a}(k_0) \mid m_2-k_0, b_2 \ran$} 
\\ \\
\times \; \lan m_2-k_0, b_3 \mid A_{a}(k_0)\mid m_2, b_4\ran 
\lan m_2-k_0, b_2 \mid C^{\nu_3, \omega_{\nu_3}}(j_1) \mid m_2-k_0, 
b_3\ran 
\\ \\
\times \; 
\lan m_2, b_5 \mid C^{\nu_3, \omega_{\nu_3}}(j_1) \mid m_2, b_6\ran 
\lan m_2, b_4 \mid C^{\nu_4, \omega_{\nu_4}}(j_2) \mid m_2, 
b_5\ran 
\\ \\ 
\times \lan m_2, b_6 \mid C^{\nu_4, \omega_{\nu_4}}(j_2) \mid m_2, b_1\ran\;.
\earr \label{strn-eq28-1}
\ee
The term $X_{31}$ (similarly $Y_{31}$) is simplified using Eqs. (\ref{W-E}), 
(\ref{reduced}), (\ref{matrix-ele-b}) and (\ref{strn-eq20z}) giving,
\be
\barr{l}
X_{31}(N_1,m_1,i_1,i_2,k_0)=\dis\binom{N_1-k_0}{m_1} \dis\sum_{\nu_1=0}^{i_1} 
\dis\sum_{\nu_2=0}^{m_1-i_2} \Lambda^{\nu_2}(N_1, m_1, i_2) 
\\ \\
\times \l[\Lambda^{\nu_1}(N_1, m_1, m_1-i_1)\,
\Lambda^{\nu_1}(N_1, m_1+k_0, m_1+k_0-i_1)\r]^{1/2} \; 
\\ \\
\times \dis\sum_{\alpha_1, \alpha_2, \alpha_3, \alpha_4, \alpha_5, \alpha_6: \alpha} 
C^{\nu_1 , \omega_{\nu_1}}_{f_{m_1+k_0} \alpha_2\,,\;\overline{f_{m_1+k_0}} 
\overline{\alpha_3}}\,
C^{\nu_1 , \omega_{\nu_1}}_{f_{m_1} \alpha_5\,,\;\overline{f_{m_1}} 
\overline{\alpha_6}}\,
C^{\nu_2 , \omega_{\nu_2}}_{f_{m_1} \alpha_4\,,\;\overline{f_{m_1}} 
\overline{\alpha_1}} \\
\\
\times\; 
C^{\nu_2 , \omega_{\nu_2}}_{f_{m_1} \alpha_6\,,\;\overline{f_{m_1}} 
\overline{\alpha_5}}\,
C^{\overline{k_0} , \overline{\alpha}}_{f_{m_1} \alpha_1\,,\;
\overline{f_{m_1+k_0} } \overline{\alpha_2}}\,
C^{k_0, \alpha}_{f_{m_1+k_0} \alpha_3\,,\;\overline{f_{m_1}} 
\overline{\alpha_4}}\,\;.
\earr \label{strn-eq28-2}
\ee
The sum over $\alpha_5$ and $\alpha_6$ of $C^{\nu_1 , \omega_{\nu_1}}_{f_{m_1} 
\alpha_5\,,\;\overline{f_{m_1}} \overline{\alpha_6}}\, C^{\nu_2 , \omega_{\nu_2}
}_{f_{m_1} \alpha_6\,,\;\overline{f_{m_1}} \overline{\alpha_5}}$ will give
$\delta_{\nu_1, \nu_2} \delta_{\omega_{\nu_1}, \omega_{\nu_2}}$. Now, the 
remaining four C-G coefficients sum up to give a Racah coefficient. Carrying out a
similar simplification of the C-G coefficients in $Y_{31}$, we finally obtain,
\be
\barr{l}
M_{31}(m_1,m_2) = 
\resizebox{0.75\hsize}{!}{$\overline{\lan \cod(k_0) H(k) \co(k_0) H^3(k) \ran^{m_1,m_2}} 
=  2\;\overline{\lan H^2(k)\ran^{m_1,m_2}}\;M_{11}(m_1,m_2)$} 
\\ \\
+ V^2_\co\l\{\dis\binom{N_1}{m_1}\; \dis\binom{N_2}{m_2}\r\}^{-1}\;
 \dis\sum_{i_1+j_1=k} \dis\sum_{i_2+j_2=k} V^2_H(i_1,j_1) \;
V^2_H(i_2,j_2) \\
\\
\times \l[\dis\binom{N_1-k_0}{m_1}\;  \dis\sum_{\nu_1=0}^{min(i_1,m_1-i_2)} 
\Lambda^{\nu_1}(N_1, m_1, i_2)\;X_{11}(N_1, m_1, k_0, i_1, \nu_1)\r] \\
\\
\times \l[\dis\binom{N_2-k_0}{m_2-k_0}\;  \dis\sum_{\nu_2=0}^{min(j_1,m_2-j_2)} 
\Lambda^{\nu_2}(N_2, m_2, j_2)\;Y_{11}(N_2, m_2, k_0, j_1, \nu_2)\r]\;.
\earr \label{strn-eq28}
\ee
The functions $X_{11}$ and $Y_{11}$ are defined in Eq. (\ref{strn-eq27}).
Following the same procedure as above, the formula for $M_{13}$ is,
\be
\barr{l}
M_{13}(m_1,m_2) = 
\resizebox{0.75\hsize}{!}{$\overline{\lan \cod(k_0) H^3(k) \co(k_0) H(k) \ran^{m_1,m_2}} = 
2\;\overline{\lan H^2(k)\ran^{m_1+k_0,m_2-k_0}}\;M_{11}(m_1,m_2)$} \\
\\
+ V^2_\co \l\{\dis\binom{N_1}{m_1}\; \dis\binom{N_2}{m_2}\r\}^{-1}\; 
\dis\sum_{i_1+j_1=k} \dis\sum_{i_2+j_2=k} V^2_H(i_1,j_1) \; V^2_H(i_2,j_2) \\
\\
\times \l[\dis\binom{N_1-k_0}{m_1}\;  \dis\sum_{\nu_1=0}^{min(i_2,m_1+k_0-i_1)} 
\Lambda^{\nu_1}(N_1, m_1+k_0, i_1)\;X_{11}(N_1, m_1, k_0, i_2, \nu_1)\r] \\
\\
\times \l[\dis\binom{N_2-k_0}{m_2-k_0}\;  \dis\sum_{\nu_2=0}^{min(j_2,m_2-k_0-j_1)} 
\Lambda^{\nu_2}(N_2, m_2-k_0, j_1)\;Y_{11}(N_2, m_2, k_0, j_2, \nu_2)\r]\;.
\earr \label{strn-eq29}
\ee   
Formula for $M_{22}(m_1,m_2)$ is more complicated and we will turn to this now.

Using the $H$ decomposition $H(k)=\sum_{i+j=k}\,H_1(i)H_2(j)$ mentioned just after
the $M_{11}(m_1,m_2)$ formula, the $M_{22}(m_1,m_2)$ can be written as,
\be
\barr{l}
M_{22}(m_1,m_2) = \overline{\lan \cod(k_0) H^2(k) \co(k_0) H^2(k) \ran^{m_1,m_2}} 
\\ \\
= \dis\sum_{i_1+j_1=k} \; \dis\sum_{i_2+j_2=k} \; \dis\sum_{i_3+j_3=k} \;
\dis\sum_{i_4+j_4=k} \;
\mbox{\bf Eavg}\l\{\lan \cod(k_0) H_1(i_1) H_2(j_1) H_1(i_2) \r.\r.
\\ \\
\l. \l. \times \; H_2(j_2) \co(k_0)
H_1(i_3) H_2(j_3) H_1(i_4) H_2(j_4)\ran^{m_1,m_2}\r\} \\
\\
= \resizebox{0.95\hsize}{!}{$\dis\sum_{i_1+j_1=k} \dis\sum_{i_2+j_2=k} \;
\l[\mbox{\bf Eavg}\l\{\lan \cod(k_0) H_1(i_1) 
H_2(j_1) H_1(i_1) H_2(j_1) \co(k_0) H_1(i_2) H_2(j_2) H_1(i_2) H_2(j_2)\ran^{m_1,m_2}
\r\} \r.$}
\\ \\
+ \;\resizebox{0.95\hsize}{!}{$\mbox{\bf Eavg}\l\{\lan \cod(k_0) H_1(i_1) H_2(j_1)
H_1(i_2) H_2(j_2) \co(k_0) H_1(i_2) H_2(j_2) H_1(i_1) H_2(j_1)\ran^{m_1,m_2}\r\}$} 
\\ \\
+ \; \resizebox{0.95\hsize}{!}{$\l. \mbox{\bf Eavg}\l\{\lan \cod(k_0) H_1(i_1) H_2(j_1)
H_1(i_2) H_2(j_2) \co(k_0) H_1(i_1) H_2(j_1) H_1(i_2) H_2(j_2)\ran^{m_1,m_2}\r\} \r]$}
\;.
\earr \label{eq-dumm1}
\ee
Thus, the ensemble averaged $M_{22}$ decomposes into three terms. The first term
is simple and the next two terms can be decomposed into averages in $m_1$ and
$m_2$ spaces. Then we have,
\be
\barr{l}
M_{22}(m_1,m_2) = \overline{\lan \cod(k_0) \co(k_0) \ran^{m_1,m_2}}\;\overline{\lan H^2(k) \ran^{m_1,m_2}}\;
\overline{\lan H^2(k) \ran^{m_1+k_0,m_2-k_0}} \\
\\
+ V^2_\co \l\{\dis\binom{N_1}{m_1}\;\dis\binom{N_2}{m_2}\r\}^{-1} \;
\dis\sum_{i_1+j_1=k}\; \dis\sum_{i_2+j_2=k} V^2_H(i_1,j_1) \; V^2_H(i_2,j_2) \\
\\
\times\;
X_{22:a}(N_1, m_1, i_1, i_2, k_0) \; Y_{22:a}(N_2, m_2, j_1, j_2, k_0) \\
\\
+ V^2_\co \; \l\{\dis\binom{N_1}{m_1}\;\dis\binom{N_2}{m_2}\r\}^{-1}\;
\dis\sum_{i_1+j_1=k}\; \dis\sum_{i_2+j_2=k} V^2_H(i_1,j_1)\; V^2_H(i_2,j_2) \\
\\
\times\;
X_{22:b}(N_1, m_1, i_1, i_2, k_0) \; Y_{22:b}(N_2, m_2, j_1, j_2, k_0) \;.
\earr \label{strn-eq30}
\ee
Note that the three terms in Eq. (\ref{eq-dumm1}) correspond directly to 
the three terms in Eq. (\ref{strn-eq30}). The second term involves $X_{22:a}$ 
and $Y_{22:a}$ functions. The $X_{22:a}$ function is
\be
\barr{l}
X_{22:a}(N_1, m_1, i_1, i_2, k_0) = \resizebox{0.6\hsize}{!}{$
\dis\sum_{\alpha_1, \alpha_2, 
\alpha_3, \alpha_4, \alpha_5, \alpha_6: \alpha, \nu_1, \omega_{\nu_1},\nu_2, 
\omega_{\nu_2}}  
\lan m_1, \alpha_1 \mid A_{\alpha}(k_0) \mid m_1+k_0, \alpha_2\ran$} 
\\ \\
\times\; \lan m_1+k_0, \alpha_4 \mid A^{\dagger}_{\alpha}(k_0)\mid m_1, \alpha_5\ran 
\lan m_1+k_0, \alpha_2 \mid B^{\nu_1, \omega_{\nu_1}}(i_1) \mid 
m_1+k_0, \alpha_3\ran 
\\ \\
\times\; \lan m_1, \alpha_6 \mid B^{\nu_1, \omega_{\nu_1}}(i_1)
\mid m_1, \alpha_1\ran 
\lan m_1+k_0, \alpha_3 \mid B^{\nu_2, \omega_{\nu_2}}(i_2) 
\mid m_1+k_0, \alpha_4\ran 
\\ \\
\times\; \lan m_1, \alpha_5 \mid B^{\nu_2, \omega_{\nu_2}}
(i_2) \mid m_1, \alpha_6\ran
\\ \\
=\dis\binom{N_1-k_0}{m_1} \dis\sum_{\nu_1=0}^{i_1} \dis\sum_{\nu_2=0}^{i_2} \;
\l[\Lambda^{\nu_1}(N_1, m_1+k_0, m_1+k_0-i_1)\,\Lambda^{\nu_1}(N_1, m_1, 
m_1-i_1)\r. \\
\\
\l. \times\;\Lambda^{\nu_2}(N_1, m_1+k_0, m_1+k_0-i_2)\,
\Lambda^{\nu_2}(N_1, m_1, m_1-i_2)\r]^{1/2} \\
\\
\times\; \dis\sum_{\alpha_1, \alpha_2, \alpha_3, \alpha_4, \alpha_5, \alpha_6: 
\alpha} 
C^{\nu_1 , \omega_{\nu_1}}_{f_{m_1+k_0} \alpha_2\,,\;\overline{f_{m_1+k_0}} 
\overline{\alpha_3}}\,
C^{\nu_1 , \omega_{\nu_1}}_{f_{m_1} \alpha_6\,,\;\overline{f_{m_1}} 
\overline{\alpha_1}}\,
C^{\nu_2 , \omega_{\nu_2}}_{f_{m_1+k_0} \alpha_3\,,\;\overline{f_{m_1+k_0}} 
\overline{\alpha_4}} \\
\\ 
\times \;C^{\nu_2 , \omega_{\nu_2}}_{f_{m_1} \alpha_5\,,\;\overline{f_{m_1}} 
\overline{\alpha_6}}\,
C^{\overline{k_0} , \overline{\alpha}}_{f_{m_1} \alpha_1\,,\;
\overline{f_{m_1+k_0}} \overline{\alpha_2}}\,
C^{k_0, \alpha}_{f_{m_1+k_0} \alpha_4\,,\;\overline{f_{m_1}} 
\overline{\alpha_5}}\,\;.
\earr \label{strn-eq31-1}
\ee
We have applied Eq. (\ref{reduced}) and the Wigner-Eckart theorem to get the second form in
Eq. (\ref{strn-eq31-1}). Simplification of the six C-G coefficients will
finally give a compact formula in terns of the $X_{11}$ functions introduced
earlier,
\be
\barr{l}
X_{22:a}(N_1, m_1, i_1, i_2, k_0) = \l\{\dis\binom{N_1}{k_0}\r\}^{-1}\;
{\dis\binom{N_1-k_0}{m_1}} \\
\\
\times\; \dis\sum_{\nu_1=0}^{i_1} \;X_{11}(N_1, m_1, k_0, i_1, \nu_1)
 \dis\sum_{\nu_2=0}^{i_2}\; X_{11}(N_1, m_1, k_0, i_2, \nu_2)\;.\\
\earr \label{strn-eqnn1}
\ee
Similarly, we have
\be
\barr{l}
Y_{22:a}(N_2, m_2, j_1, j_2, k_0) = \l\{\dis\binom{N_2}{k_0}\r\}^{-1}\;
{\dis\binom{N_2-k_0}{m_2-k_0}} \\
\\
\times\; \dis\sum_{\nu_1=0}^{j_1} \;Y_{11}(N_2, m_2, k_0, j_1, \nu_1)
\dis\sum_{\nu_2=0}^{j_2}\; Y_{11}(N_2, m_2, k_0, j_2, \nu_2)\;.\\
\earr \label{strn-eqnn2}
\ee
Finally, the third term in $M_{22}(m_1,m_2)$ involves the functions $X_{22:b}$ and 
$Y_{22:b}$. The expression for function $X_{22:b}$ is, 
\be
\barr{l}
X_{22:b}(N_1, m_1, i_1, i_2, k_0) = \dis\sum_{\alpha_1, \alpha_2, 
\alpha_4, \alpha_5, \alpha, \nu_1, \omega_{\nu_1},\nu_2, \omega_{\nu_2}} 
\lan m_1, \alpha_1 \mid A_{\alpha}(k_0) \mid m_1+k_0, \alpha_2\ran 
\\ \\
\times \; \resizebox{0.95\hsize}{!}{$\lan m_1+k_0, \alpha_4 \mid 
A^{\dagger}_{\alpha}(k_0)\mid m_1, \alpha_5\ran 
\lan m_1+k_0, \alpha_2 \mid B^{\nu_1, \omega_{\nu_1}}(i_1) B^{\nu_2, 
\omega_{\nu_2}}(i_2) \mid m_1+k_0, \alpha_4\ran$} 
\\ \\
\times\; \lan m_1, \alpha_5 \mid B^{\nu_1, \omega_{\nu_1}}(i_1) B^{\nu_2, 
\omega_{\nu_2}}(i_2) \mid m_1, \alpha_1\ran \;.
\earr \label{strn-eq31-2}
\ee
Now, simplifying Eq. (\ref{strn-eq31-2}) and similarly, $Y_{22:b}$, finally we get
\be
\barr{l}
X_{22:b}(N_1, m_1, i_1, i_2, k_0) =\dis\binom{N_1-k_0}{m_1} 
\l\{\dis\binom{N_1}{k_0}\;d(N_1 : \nu)\r\}^{1/2} 
\\ \\
\times \dis\sum_{\nu_1=0}^{i_1}
\dis\sum_{\nu_2=0}^{i_2} \dis\sum_{\nu=0}^{i_1+i_2} 
\dis\sum_\rho \lan m_1+k_0 \mid\mid \l[B^{\nu_1}(i_1) B^{\nu_2}(i_2)\r]^{
\nu:\rho} \mid\mid m_1+k_0\ran\;
\\ \\ 
\times \; \lan m_1 \mid\mid \l[B^{\nu_1}(i_1) 
B^{\nu_2}(i_2)\r]^{\nu:\rho} \mid\mid m_1\ran \; (-1)^{\phi(f_{m_1+k_0},\overline{f_{m_1}}, k_0) + \phi(f_{m_1},
\overline{f_{m_1}}, \nu)} 
\\ \\
\times\;U(f_{m_1+k_0}\,\overline{f_{m_1}}\,f_{m_1+k_0}\,f_{m_1}\,;\,
f_{k_0}\,\nu)\;,
\\ \\
Y_{22:b}(N_2, m_2, j_1, j_2, k_0) = \dis\binom{N_2-k_0}{m_2-k_0} 
\l\{\dis{\binom{N_2}{k_0}\;d(N_2 : \nu)}\r\}^{1/2}
\\ \\
\times \dis\sum_{\nu_1=0}^{j_1}
\dis\sum_{\nu_2=0}^{j_2} \dis\sum_{\nu=0}^{j_1+j_2} 
\dis\sum_\rho \lan m_2 \mid\mid \l[B^{\nu_1}(j_1) B^{\nu_2}(j_2)\r]^{
\nu:\rho} \mid\mid m_2\ran\;\\ \\
\times \; \lan m_2-k_0 \mid\mid \l[B^{\nu_1}(j_1) 
B^{\nu_2}(j_2)\r]^{\nu:\rho} \mid\mid m_2-k_0\ran 
\\ \\
\times \; (-1)^{\phi(f_{m_2},\overline{f_{m_2-k_0}}, k_0) + \phi(f_{m_2-k_0},
\overline{f_{m_2-k_0}}, \nu)}\; U(f_{m_2}\,\overline{f_{m_2-k_0}}\,f_{m_2}
\,f_{m_2-k_0}\,;\,f_{k_0}\,\nu)\;.\\
\earr \label{strn-eq31}
\ee
Combining Eqs. (\ref{strn-eq31}), (\ref{strn-eqnn1}), and (\ref{strn-eqnn2}) 
with (\ref{strn-eq30}) will give the formula for $M_{22}(m_1,m_2)$.
Note that the reduced matrix elements in Eq. (\ref{strn-eq31}) are given by 
Eq. (\ref{strn-eq13yy}) along with Eq. (\ref{reduced}).

\section{Asymptotic results for two types of spinless fermions with beta and
double beta decay type transition operators}

Employing the formulas derived in Section 5, for some typical values,
appropriate for atomic nuclei, for $N_1$, $N_2$, $m_1$ and $m_2$ with 
$k=2$ and $k_0$ taking values 1 and 2, numerical results for $\xi(m_1,m_2)$ 
and $k_{rs}(m_1,m_2)$ with $r+s=4$ are shown in Tables 2 and 3. In the 
examples shown in the Table, $N_1$ and $N_2$ are sufficiently large but not 
$m_1$ and $m_2$ for all the examples. For the situation with $m_1$ and $m_2$ 
sufficiently large (typically larger than say $6$), the $|k_{rs}(m_1,m_2)|$ are 
$\leq 0.3$ pointing that the bivariate transition strength density is close to a 
bivariate Gaussian. Also, typically $\xi \gazz 0.6$ shows that for the systems 
considered, EGUE is essential (as shown in \cite{FKPT}, $\xi =0$ for a GUE). For 
further understanding the smallness of $k_{rs}(m_1,m_2)$, we will derive 
asymptotic formulas first for $M_{rs}(m_1,m_2)$ and using them, for
$k_{rs}(m_1,m_2)$ with $r+s=4$ and also for $\xi(m_1,m_2)$.

Let us begin with the asymptotic limit defined by $N_1 \rightarrow \infty$, $N_2
\rightarrow \infty$, $m_1$, $m_2$ fixed with $k$ and $k_0$ much smaller than
$m_1$ and $m_2$. Note that in the dilute limit (or true asymptotic limit) we 
also have $m_1 \rightarrow \infty$, $m_2 \rightarrow \infty$, 
$m_1/N_1 \rightarrow 0$ and $m_2/N_2 \rightarrow 0$ and we will 
consider this in the later part of this section. First, we consider the 
following functions,
\be
\barr{l}
A_1(N,m,i,t)= \l[\dis\binom{N}{t}\,d(N:i)
\,\Lambda^i(N,m,m-i)\,\Lambda^i(N,m+t,m+t-i)\r]^{1/2} \\  
\\
\times \; {\dis\binom{N}{m}}^{-1}\dis\binom{N-t}{m}\;
\l| U(f_{m+t}, \overline{f_m}, f_{m+t}, f_m\,;\,f_t, i)\r|\;, \\
\\
A_2(N,m,i,t)= \l[\dis\binom{N}{t}\,d(N:i)
\,\Lambda^i(N,m,m-i)\,\Lambda^i(N,m-t,m-t-i)\r]^{1/2} \\  
\\
\times \;{\dis\binom{N}{m}}^{-1}\dis\binom{N-t}{m-t}\;
\l| U(f_{m}, \overline{f_{m-t}}, f_{m}, f_{m-t}\,;\,f_t, i)\r|\;, \\
\\
T(N,m,i)=\Lambda^0(N,m,i)\;,\\
\\
F(N,m,i,j)={\dis\binom{N}{m}}^{-1}\,\Lambda^i(N,m,m-i)\,\Lambda^i(N,m,j)\,d(N:i)\;.
\earr \label{eq.a12tf-def}
\ee
Using Eq. (\ref{reduced}) for $\Lambda$, Eq. (\ref{eq-dnu}) for $d(N:i)$ and Eq.
(\ref{snt-8}) for the $U$-coefficient, $A_1$ and $A_2$ in the
asymptotic limit simplify to
\be
\barr{rcl}
A_1(N,m,i,t) & = & \resizebox{0.65\hsize}{!}{$\dis\binom{m}{i}\,
\dis\binom{N+1}{i}\,\dis\binom{N-m}{i}\,
\dis\binom{N-m-i}{t}\,{\dis\binom{N-i+1}{i}}^{-1}$} 
\\ \\
& \stackrel{\mbox{asymp}}{\longrightarrow}  & \dis\binom{m}{i}\,
\dis\binom{N}{i}\,
\dis\binom{N-m-i}{t} \;,
\earr \label{eq-a1asymp}
\ee
and
\be
\barr{rcl}
A_2(N,m,i,t) & = & \resizebox{0.65\hsize}{!}{$\dis\binom{m}{i}\,
\dis\binom{N+1}{i}\,\dis\binom{N-m}{i}\,
\dis\binom{m-i}{t}\,{\dis\binom{N-i+1}{i}}^{-1}$} \\
\\
& \stackrel{\mbox{asymp}}{\longrightarrow}  & \dis\binom{m}{i}\,
\dis\binom{N}{i}\,\dis\binom{m-i}{t} \;.
\earr \label{eq-a2asymp}
\ee
Similarly $T$ and $F$ are given by,
\be
\barr{l}
T(N,m,i) \stackrel{\mbox{asymp}}{\longrightarrow}  \dis\binom{m}{i}\,
\dis\binom{N}{i}\;,\\
\\
F(N,m,i,j) \stackrel{\mbox{asymp}}{\longrightarrow} \dis\binom{m}{i}\,
\dis\binom{m-i}{j}\,\dis\binom{N}{i}\,\dis\binom{N}{j}\;.
\earr \label{eq-tfasymp}
\ee
Using Eqs. (\ref{strn-eq22}) and (\ref{eq-tfasymp}),
\be
\overline{\lan H^2(k) \ran^{m_1,m_2}} \; \stackrel{\mbox{asymp}}{\longrightarrow} \; 
\dis\sum_{i+j=k} V^2_H(i,j)\,T(N_1,m_1,i) T(N_2,m_2,j)\;.
\label{asym12-1}
\ee
Similarly for $M_{11}(m_1,m_2)$ given by Eq. (\ref{strn-eq27}), in the 
asymptotic limit only the terms with $\nu_1=i$ in $X_{11}$ and $\nu_2=j$ in 
$Y_{11}$ will contribute. Then, simplifying using Eqs. (\ref{eq-a1asymp}) and 
(\ref{eq-a2asymp}) we have
\be
M_{11}(m_1,m_2) \stackrel{\mbox{asymp}}{\longrightarrow} 
\resizebox{0.7\hsize}{!}{$V^2_\co 
\dis\sum_{i+j=k} V^2_H(i,j)\,\dis\binom{N_1-m_1-i}{k_0}
\dis\binom{m_2-j}{k_0}\,T(N_1,m_1,i)\,T(N_2,m_2,j)\;.$}
\label{asym12-2}
\ee
Turning to fourth order moments, firstly for $\overline{\lan H^4(k) \ran^{m_1, 
m_2}}$, using Eqs. (\ref{strn-eq23}) and (\ref{eq-tfasymp}), we get
\be
\barr{l}
\overline{\lan H^4(k) \ran^{m_1,m_2}} \stackrel{\mbox{asymp}}{\longrightarrow} 
2\,\l[\; \overline{\lan H^2(k) \ran^{m_1,m_2}} \;\r]^2 + \dis\sum_{i_1+j_1=k} 
\;\dis\sum_{i_2 + j_2 =k} V^2_H(i_1,j_1) V^2_H(i_2,j_2) \\
\\
\times\; F(N_1,m_1,i_1,i_2)\; F(N_2,m_2,j_1,j_2)\;.
\earr \label{asym12-3}
\ee
This formula follows from the fact that in Eq. (\ref{strn-eq23}), in the asymptotic 
limit only terms with $\nu_1=i$ in $X$ and $\nu_2=j$ in $Y$ will contribute.
Equation (\ref{asym12-3}) also gives $M_{40}(m_1,m_2)$ and 
$M_{04}(m_1,m_2)$ via Eq. (\ref{strn-eq24}). 

The first non-trivial fourth order moment is $M_{31}(m_1,m_2)$ and it is 
given by Eq. (\ref{strn-eq28}). As only terms with $\nu_1=i_1$  
in $X_{11}$ and $\nu_2=j_1$ in $Y_{11}$ in Eq. (\ref{strn-eq28}) will 
contribute in the asymptotic limit, we have
\be
\barr{l}
M_{31}(m_1,m_2) = 2\,\overline{\lan H^2(k) \ran^{m_1,m_2}}\;M_{11}(m_1,m_2)
\\ \\
+ V^2_\co \dis\sum_{i_1+j_1=k}\; \dis\sum_{i_2+j_2=k} V^2_H(i_1,j_1) 
V^2_H(i_2,j_2) \l\{ \Lambda^{i_1}(N_1,m_1,i_2) A_1(N_1,m_1,i_1,k_0) \r.
\\ \\
\l. + \;\Lambda^{j_1}(N_2,m_2,j_2) A_2(N_2,m_2,j_1,k_0) \r\} \;.
\earr \label{asym12-4}
\ee
Substituting Eqs. (\ref{eq-a1asymp}) and (\ref{eq-a2asymp}) for $A_1$ and $A_2$
respectively and Eq. (\ref{reduced}) for $\Lambda$ will give,
\be
\barr{l}
M_{31}(m_1,m_2) \stackrel{\mbox{asymp}}{\longrightarrow} 
2\,\overline{\lan H^2(k) \ran^{m_1,m_2}}\;M_{11}(m_1,m_2) \\ \\
+ V^2_\co \dis\sum_{i_1+j_1=k} \dis\sum_{i_2+j_2=k} V^2_H(i_1,j_1) 
V^2_H(i_2,j_2) \dis\binom{N_1-m_1-i_1}{k_0}\,\dis\binom{m_2-j_1}{k_0}
\\ \\
\times\;F(N_1,m_1,i_1,i_2)\,F(N_2,m_2,j_1,j_2) \;.
\earr \label{asym12-5}
\ee
Moment $M_{13}(m_1,m_2)$ is given by Eq. (\ref{strn-eq29}) and in the asymptotic
limit only the terms with $\nu_1=i_2$ in $X_{11}$ and $\nu_2=j_2$ in $Y_{11}$ will
survive. Then,
\be
\barr{l}
M_{13}(m_1,m_2) = 2\,\overline{\lan H^2(k)\ran^{m_1+k_0,m_2-k_0}}\;M_{11}(m_1,m_2)
\\
\\
+ V^2_\co \dis\sum_{i_1+j_1=k} \dis\sum_{i_2+j_2=k} V^2_H(i_1,j_1) 
V^2_H(i_2,j_2) \;\Lambda^{i_2}(N_1,m_1+k_0,i_1) A_1(N_1,m_1,i_2,k_0) 
\\ \\
\times \;\Lambda^{j_2}(N_2,m_2-k_0,j_1) A_2(N_2,m_2,j_2,k_0)\;.
\earr \label{asym12-6}
\ee
Substituting Eqs. (\ref{eq-a1asymp}) and (\ref{eq-a2asymp}) for $A_1$ and $A_2$
respectively and Eq. (\ref{reduced}) for $\Lambda$ will give,
\be
\barr{l}
M_{13}(m_1,m_2) \stackrel{\mbox{asymp}}{\longrightarrow} 
2\,\overline{\lan H^2(k) \ran^{m_1+k_0,m_2-k_0}}\;M_{11}(m_1,m_2) \\
\\
+ V^2_\co \dis\sum_{i_1+j_1=k} \;\;\dis\sum_{i_2+j_2=k} V^2_H(i_1,j_1) \;
V^2_H(i_2,j_2)
\dis\binom{m_1+k_0-i_2}{i_1}\\ \\
\times \;\dis\binom{N_1-m_1-k_0+i_1-i_2}{i_1}\,
\dis\binom{m_1}{i_2}\,\,\dis\binom{N_1}{i_2}\;\dis\binom{N_1-m_1-i_2}{k_0} \\
\\
\times\;\dis\binom{m_2-k_0-j_2}{j_1}\,\dis\binom{N_2-m_2+k_0+j_1-j_2}{j_1}\,
\dis\binom{m_2}{j_2}\,\dis\binom{N_2}{j_2}\;\dis\binom{m_2-j_2}{k_0}\;.
\earr \label{asym12-7}
\ee
Finally, the most complicated moment $M_{22}(m_1,m_2)$ is given by Eq.
(\ref{strn-eq30}) and it has three terms. The first term $M^{(1)}_{22}(m_1,m_2)$
is simple,
\be
M^{(1)}_{22}(m_1,m_2) =\overline{\lan \cod(k_0) \co(k_0) \ran^{m_1,m_2}}\;
\overline{\lan H^2(k) \ran^{m_1,m_2}}\;\overline{\lan H^2(k) 
\ran^{m_1+k_0,m_2-k_0}}\;.
\label{asym12-8}
\ee
The second term $M^{(2)}_{22}(m_1,m_2)$ is given by Eq. (\ref{strn-eq30}) 
with functions $X_{22:a}$ and $Y_{22:a}$ given by Eqs. (\ref{strn-eqnn1}) and 
Eq. (\ref{strn-eqnn2}) respectively. In the asymptotic limit,
only the terms with $\nu_1=i_1$ and $\nu_2=i_2$ will survive in Eq.
(\ref{strn-eqnn1}).  Similarly, only the terms with $\nu_1=j_1$ and $\nu_2=j_2$
will survive in Eq. (\ref{strn-eqnn2}). These will give
\be
\barr{l}
M^{(2)}_{22}(m_1,m_2) = V^2_\co \dis\sum_{i_1+j_1=k} \dis\sum_{i_2+j_2=k} 
V^2_H(i_1,j_1) \, V^2_H(i_2,j_2) \\
\\
\times\;{\dis\binom{N_1}{k_0}}^{-1} \dis\binom{N_1}{m_1} 
{\dis\binom{N_1-k_0}{m_1}}^{-1}
A_1(N_1,m_1,i_1,k_0) \, A_1(N_1,m_1,i_2,k_0) \\
\\
\times\;{\dis\binom{N_2}{k_0}}^{-1} \dis\binom{N_2}{m_2} 
{\dis\binom{N_2-k_0}{m_2-k_0}}^{-1}
A_2(N_2,m_2,j_1,k_0) \, A_2(N_2,m_2,j_2,k_0)\;.
\earr \label{asym12-9}
\ee  
Substituting Eqs. (\ref{eq-a1asymp}) and (\ref{eq-a2asymp}) for $A_1$ and $A_2$
respectively and further simplifications using the assumption that $m_1 >> k,k_0$ 
and $m_2 >> k,k_0$ will give,
\be
\barr{l}
M^{(2)}_{22}(m_1,m_2)  \stackrel{\mbox{asymp}}{\longrightarrow}  
V^2_\co \dis\sum_{i_1+j_1=k} \dis\sum_{i_2+j_2=k} 
V^2_H(i_1,j_1) V^2_H(i_2,j_2)\;  \dis\binom{m_2-j_1-j_2}{k_0}
\\  \\
\times\;\dis\binom{N_1-m_1-i_1-i_2}{k_0}\;
T(N_1,m_1,i_1)\;T(N_1,m_1,i_2)\;T(N_2,m_2,j_1)\;T(N_2,m_2,j_2)\;.
\earr \label{asym12-10}
\ee
Lastly, the third term $M^{(3)}_{22}(m_1,m_2)$ of $M_{22}(m_1,m_2)$ is given by Eq.
(\ref{strn-eq30}) along with Eq. (\ref{strn-eq31}). In the asymptotic limit
only the terms with $\nu_1=i_1$, $\nu_2=i_2$ and $\nu=i_1+i_2$ will survive in
$X_{22:b}$ given by Eq. (\ref{strn-eq31}). Similarly, only terms with $\nu_1=j_1$, 
$\nu_2=j_2$ and $\nu=j_1+j_2$ will survive in $Y_{22:b}$. Then,
\be
\barr{l}
M^{(3)}_{22}(m_1,m_2) = V^2_\co \dis\binom{N_1}{m_1}^{-1} 
\dis\binom{N_2}{m_2}^{-1} \dis\binom{N_1}{k_0}^{1/2}\dis\binom{N_2}{k_0}^{1/2}\;
\dis\binom{N_1-k_0}{m_1} \, \dis\binom{N_2-k_0}{m_2-k_0} \\ \\
\times \dis\sum_{i_1+j_1=k} \dis\sum_{i_2+j_2=k} 
V^2_H(i_1,j_1) V^2_H(i_2,j_2) 
\;  \l\{d(N_1 : i_1+i_2) \; d(N_2 : j_1+j_2)\r\}^{1/2} \\ \\
\resizebox{0.98\hsize}{!}{$\times \;\lan m_1+k_0 \mid\mid 
\l[B^{i_1}(i_1) B^{i_2}(i_2)\r]^{i_1+i_2} \mid\mid m_1+k_0\ran \;
\lan m_1 \mid\mid \l[B^{i_1}(i_1) B^{i_2}(i_2)\r]^{i_1+i_2} \mid\mid 
m_1\ran$}
\\ \\
\times \;\l|U(f_{m_1+k_0}\,\overline{f_{m_1}}\,f_{m_1+k_0}\,f_{m_1}\,;\,
f_{k_0}\,i_1+i_2)\r|\; 
\lan m_2 \mid\mid \l[B^{j_1}(j_1) B^{j_2}(j_2)\r]^{j_1+j_2} \mid\mid 
m_2\ran\;
\\ \\
\resizebox{0.98\hsize}{!}{$\times\; \lan m_2-k_0 \mid\mid \l[B^{j_1}(j_1) 
B^{j_2}(j_2)\r]^{j_1+j_2} \mid\mid m_2-k_0\ran
\; \l|U(f_{m_2}\,\overline{f_{m_2-k_0}}\,f_{m_2}\,f_{m_2-k_0}\,;\,
f_{k_0}\,j_1+j_2)\r|\;.$}
\earr \label{asym12-11}
\ee
Now, simplifying the reduced matrix elements as in Section 4 along with the
$U$-coefficients using Eqs. (\ref{eq-a1asymp}) and (\ref{eq-a2asymp}) will give
the final result,
\be
\barr{l}
M^{(3)}_{22}(m_1,m_2) \stackrel{\mbox{asymp}}{\longrightarrow} 
V^2_\co \dis\sum_{i_1+j_1=k} \dis\sum_{i_2+j_2=k} 
V^2_H(i_1,j_1) V^2_H(i_2,j_2) \\
\\
\times\;\dis\binom{N_1-m_1-i_1-i_2}{k_0}\,\dis\binom{m_2-j_1-j_2}{k_0}\,
F(N_1,m_1,i_1,i_2)\,F(N_2,m_2,j_1,j_2) \;.
\earr \label{asym12-12}
\ee
Combining Eqs. (\ref{asym12-8}), (\ref{asym12-10}) and (\ref{asym12-12}) we 
have,
\be
\barr{l}
M_{22}(m_1,m_2) \stackrel{\mbox{asymp}}{\longrightarrow} 
\overline{\lan \cod(k_0) \co(k_0) \ran^{m_1,m_2}}\;
\overline{\lan H^2(k) \ran^{m_1,m_2}}\;\overline{\lan H^2(k) 
\ran^{m_1+k_0,m_2-k_0}} 
\\ \\
+ V^2_\co \dis\sum_{i_1+j_1=k} \;\dis\sum_{i_2+j_2=k} 
V^2_H(i_1,j_1) V^2_H(i_2,j_2) \l\{ \dis\binom{N_1-m_1-i_1-i_2}{k_0} \r.
\\ \\
\times\; \l.
\dis\binom{m_2-j_1-j_2}{k_0} \,
T(N_1,m_1,i_1)\;T(N_1,m_1,i_2)\;T(N_2,m_2,j_1)\;T(N_2,m_2,j_2) \r. 
\\ \\ \l.
+ \; \dis\binom{N_1-m_1-i_1-i_2}{k_0}\,\dis\binom{m_2-j_1-j_2}{k_0}\,
F(N_1,m_1,i_1,i_2)\,F(N_2,m_2,j_1,j_2) \r\}\;.
\earr \label{asym12-13}
\ee
The asymptotic formulas given by Eqs. (\ref{asym12-1}),  (\ref{asym12-2}),
(\ref{asym12-3}), (\ref{asym12-5}), (\ref{asym12-7}) and (\ref{asym12-13}) are
identical to those obtained using the asymptotic theory of Mon and 
French as derived in detail in \cite{Ma-arx} for EGOE. This agreement gives a 
good check of the exact formulas derived in Section 5. They 
also show (as the asymptotic results should be valid for any $k$ and $k_0$) that 
the term $(-1)^{\phi(\dots) + \phi(\dots)}\, U(\dots)$ in Eqs. (\ref{strn-eq27}) and 
(\ref{strn-eq31}) will be $\l|U(\dots)\r|$, i.e. the phase of the $U$-coefficient
[see Eq. (\ref{snt-8})] cancels with the phase factor $(-1)^{\phi(\dots) + 
\phi(\dots)}$ in these equations. Rewriting Eq. (\ref{snt-8}) in a form that
extends easily to boson systems, we have for $U^2$,
\be
\l[U(f_m,\, \overline{f_p},\, f_m,\, f_p\, ;\,f_{m-p}\,\nu)\r]^2 = 
\dis\frac{{\binom{N+1}{\nu}}^2 \binom{m-\nu}{p-\nu} \binom{N-\nu -p}{m-p}\;
(N-2\nu +1)}{{\binom{N-m+p}{p}}^2 \binom{N}{m-p}\;(N+1)}\;.
\label{snt-9}
\ee
We will discuss extension of Eq. (\ref{snt-9}) to boson systems in Section 8.

All the formulas given above simplify further in the dilute limit (or
strict asymptotic limit) 
defined by $N_1 \rightarrow \infty$, $N_2 \rightarrow \infty$, $m_1 \rightarrow
\infty$, $m_2 \rightarrow \infty$, $m_1/N_1 \rightarrow 0$ and $m_2/N_2
\rightarrow 0$ and $k$ and $k_0$ fixed. Also, assuming that $V^2_H(i,j)=V^2_H$
independent of $i$ and $j$, the reduced moments and cumulants will
be independent of $V^2_\co$ and $V^2_H$. With these, the dilute limit
formulas for $\xi$ and $k_{rs}$ with $r+s=4$ are obtained using Eqs.
(\ref{eq.normom}), (\ref{eq.cumu}), (\ref{strn-eq20}), (\ref{asym12-1}),  
(\ref{asym12-2}), (\ref{asym12-3}), (\ref{asym12-5}), (\ref{asym12-7}) and 
(\ref{asym12-13}). The results are as follows. Firstly, it is easy to see that
$\wtM_{P0}=M_{P0}/M_{00}=\overline{\lan H^P(k)\ran^{m_1,m_2}}$ and 
$\wtM_{0P}=M_{0P}/M_{00} =\overline{\lan H^P(k)\ran^{m_1+k_0,m_2-k_0}}$; $P=2,4$.
Using Eqs. (\ref{asym12-1}), (\ref{asym12-2}) , (\ref{asym12-3}) and
(\ref{strn-eq20}) we have asymptotic formulas for bivariate cumulants 
$\xi$, $k_{40}$ and $k_{04}$,
\be
\barr{l}
\xi(m_1,m_2) = \dis\frac{M_{11}(m_1,m_2)}{M_{00}(m_1,m_2)\,\l[\wtM_{20}(m_1,m_2)
\wtM_{02}(m_1,m_2)\r]^{1/2}} \\
\\
\stackrel{\mbox{asymp}}{\longrightarrow} 
\l[\dis\binom{m_2}{k_0}\,\l\{
\dis\sum_{i_1+j_1=k} T(N_1,m_1,i_1)\,T(N_2,m_2,j_1) \r.\r.\\
\\
\times\;\l.\l.\dis\sum_{i_2+j_2=k}
T(N_1,m_1+k_0,i_2)\,T(N_2,m_2-k_0,j_2)\r\}^{1/2}\r]^{-1} \\
\\
\times\;\dis\sum_{i+j=k}
\binom{m_2-j}{k_0}\,T(N_1,m_1,i)\,T(N_2,m_2,j)\;, \\
\\
k_{40}(m_1,m_2) = \dis\frac{\wtM_{40}(m_1,m_2)}{\l[\wtM_{20}(m_1,m_2)\r]^2} 
\;-3\\
\\
\stackrel{\mbox{asymp}}{\longrightarrow} 
\dis\frac{\dis\sum_{i_1+j_1=k}\, \dis\sum_{i_2+j_2=k}
F(N_1,m_1,i_1,i_2)\,F(N_2,m_2,j_1,j_2)}{\l[
\dis\sum_{i+j=k} T(N_1,m_1,i)\,T(N_2,m_2,j)\r]^2}\,-1\;, \\
\\
k_{04}(m_1,m_2) = \dis\frac{\wtM_{04}(m_1,m_2)}{\l[\wtM_{02}(m_1,m_2)\r]^2}
\;-3\\
\\
\stackrel{\mbox{asymp}}{\longrightarrow} 
\dis\frac{\dis\sum_{i_1+j_1=k}\, \dis\sum_{i_2+j_2=k}
F(N_1,m_1+k_0,i_1,i_2)\,F(N_2,m_2-k_0,j_1,j_2)}{\l[
\dis\sum_{i+j=k} T(N_1,m_1+k_0,i)\,T(N_2,m_2-k_0,j)\r]^2}\,-1\;. 
\earr \label{asym12-14}
\ee
Note that the functions $T$ and $F$ are given by Eq. (\ref{eq-tfasymp}).
Similarly, Eqs. (\ref{asym12-5}) and (\ref{asym12-7}) will give the formulas 
for $k_{31}$ and $k_{13}$ respectively,
\be  
\barr{l}
k_{31}(m_1,m_2) = \dis\frac{\wtM_{31}(m_1,m_2)}{\l[\wtM_{20}(m_1,m_2)\r]^{3/2}\;
\l[\wtM_{02}(m_1,m_2)\r]^{1/2}} \;-3\,\xi(m_1,m_2) \\
\\
\stackrel{\mbox{asymp}}{\longrightarrow}\;-\xi(m_1,m_2)  + \l\{
\dis\binom{m_2}{k_0}\,\l[
\dis\sum_{i_1+j_1=k} T(N_1,m_1,i_1)\,T(N_2,m_2,j_1)\r]^{3/2} \r. \\
\\
\times\; \l. \l[\dis\sum_{i_2+j_2=k} T(N_1,m_1+k_0,i_2)\,T(N_2,m_2-k_0,j_2)
\r]^{1/2}\r\}^{-1} \\
\\
\times\; \dis\sum_{i_1+j_1=k}\,
\dis\sum_{i_2+j_2=k} \binom{m_2-j_1}{k_0}\,F(N_1,m_1,i_1,i_2)\,
F(N_2,m_2,j_1,j_2)\;,\\
\\
k_{13}(m_1,m_2) = \dis\frac{\wtM_{13}(m_1,m_2)}{\l[\wtM_{20}(m_1,m_2)\r]^{1/2}\;
\l[\wtM_{02}(m_1,m_2)\r]^{3/2}} \;-3\,\xi(m_1,m_2) \\
\\
\stackrel{\mbox{asymp}}{\longrightarrow}\;-\xi(m_1,m_2) + \l\{ 
\dis\binom{m_2}{k_0}\,\l[\dis\sum_{i_1+j_1=k} T(N_1,m_1,i_1)\,T(N_2,m_2,j_1)
\r]^{1/2} \r. \\
\\
\times\;\l. \l[\dis\sum_{i_2+j_2=k} T(N_1,m_1+k_0,i_2)\,T(N_2,m_2-k_0,j_2)
\r]^{3/2}\r\}^{-1} \\
\\
\times\;\dis\sum_{i_1+j_1=k}\,\dis\sum_{i_2+j_2=k} \dis\binom{m_2-j_2}{k_0}\,
T(N_1,m_1,i_2)\,T(N_2,m_2,j_2) \\
\\
\times\; \dis\binom{N_1}{i_1}\,\dis\binom{m_1+k_0-i_2}{i_1}\,\dis\binom{N_2}{j_1}\,
\dis\binom{m_2-k_0-j_2}{j_1}\;.
\earr \label{asym12-15}
\ee
Finally, using (\ref{asym12-13}) we have,
\be
\barr{l}
k_{22}(m_1,m_2) = \dis\frac{\wtM_{22}(m_1,m_2)}{\l[\wtM_{20}(m_1,m_2)\;
\wtM_{02}(m_1,m_2)\r]} \;-2 \, \xi^2(m_1,m_2) \;-1 \\ \\
\stackrel{\mbox{asymp}}{\longrightarrow} -2\,\xi^2(m_1,m_2) + 
\l\{ \dis\binom{m_2}{k_0}\,
\dis\sum_{i_1+j_1=k} T(N_1,m_1,i_1)\,T(N_2,m_2,j_1) \r.
\\ \\
\l. \times \;
\dis\sum_{i_2+j_2=k} T(N_1,m_1+k_0,i_2)\,T(N_2,m_2-k_0,j_2)\r\}^{-1} \\
\\
\times\;\dis\sum_{i_1+j_1=k}\,\dis\sum_{i_2+j_2=k}
\binom{m_2-j_1-j_2}{k_0}\;\l\{
T(N_1,m_1,i_1)\;T(N_1,m_1,i_2) \r.
\\ \\ \l.
\times \;T(N_2,m_2,j_1)\;T(N_2,m_2,j_2)
+ F(N_1,m_1,i_1,i_2)\,F(N_2,m_2,j_1,j_2) \r\}\;.
\earr \label{asym12-16}
\ee
Numerical results obtained using Eqs. (\ref{asym12-14}) - (\ref{asym12-16})
are shown in Tables 2 for $k_0=2$ and $k=2$ and similarly in Table 3 for
$k_0=1$ and $k=2$. It is seen that in general $|k_{PQ}(m_1,m_2)| \leq 0.3$ implying
that the strength densities are close to a bivariate Gaussian. However, as seen
from Tables 2 and 3, it is necessary to add the corrections due to $k_{PQ}$.
Also, expanding $k_{PQ}$ in powers of  $1/m_1$ and $1/m_2$
using Mathematica, it is seen that all the  $k_{PQ}$ with $P+Q =4$ behave as,
for $k=2$ and $k_0=2$,
\be
k_{PQ} = - \dis\frac{4}{m_1} + O\l( \dis\frac{1}{m_1^2} \r)
+ O\l( \dis\frac{m_2^2}{m_1^3} \r) + \ldots \;.
\label{asym12-fin}
\ee
Therefore, for $m_1 >> 1$ and $m_2 << m_1^{3/2}$, the transition 
strength density approaches bivariate Gaussian form in general. 
It is important to recall that the strong dependence on $m_1$ in 
Eq. (\ref{asym12-fin}) is due to the
nature of the operator $\co$ i.e., $\co(k_\co) \l| m_1, m_2 \ran = \l| m_1 +
k_\co, m_2 - k_\co \ran$. Thus, we conclude that bivariate Gaussian form with
corrections (see Appendix A for the form with Edgeworth corrections) due to 
$k_{PQ}$, $P+Q=4$ will form a good approximation for transition strength 
densities generated by beta and double beta decay type operators.

\section{Lower-order moments of transition strength densities:
results for particle removal operators}

Particle removal (or addition) operators are of great interest in nuclear
physics. For example one particle (proton or neutron) removal from a target 
nucleus gives information about the single particle levels in the target and
similarly, two-particle removal gives information about pairing force
\cite{Sch-12,Fr-07}. Let us begin with $m$ spinless fermions in $N$ sp states
and a particle removal operator $\co$ that removes $k_0$  number of particles
when acting on a $m$ fermion state. Then the general form of $\co$ is,
\be
\co = \dis\sum_{\alpha_0} V_{\alpha_0}\;A_{\alpha_0}(k_0) \;.
\label{snt-1}
\ee
Here, $A_{\alpha_0}(k_0)$ is a $k_0$ particle annihilation operator and
$\alpha_0$ are indices for a $k_0$ particle state. Note that $A_{\alpha_0}(k_0)$
transforms as $\{\overline{f_{k_0}}\} = \{1^{N-k_0}\}$ with respect to $U(N)$
and $A^\dagger_{\alpha_0}(k_0)$  transforms as $\{f_{k_0}\}$. It important to
recognize that the $\co$ matrices will be rectangular matrices connecting $m$
particle states to $m-k_0$ particle states. In the defining space, the matrix
will be a $1 \times d_0$ matrix with matrix elements given by $V_{\alpha_0}$.
Note that $\alpha_0$ takes $d_0$ values and $d_0=\binom{N}{k_0}$. We will
represent the $\co$ matrix in the defining space by GUE implying that the 
defining space matrix elements $V_{\alpha_0}$ are zero centered independent
Gaussian random variables. Also, the $V_{\alpha_0}$ are assumed to be
independent of the $V_{ij}(k)$ variables in Eq. (\ref{strn-eq2})  and therefore
also independent of the $W$ variables in Eq. (\ref{strn-eq5})] with variance
satisfying 
\be
\overline{V_{\alpha} V^{\dagger}_\beta} =  V^2_\co\;\delta_{\alpha
\beta} \;.
\label{snt-2}
\ee
In many particle spaces the $\co$ matrices will be $d_1 \times d_2$ matrices
connecting $d_1=\binom{N}{m}$ number of $m$-particle states to
$d_2=\binom{N}{m-k_0}$ number of $(m-k_0)$-particle states. Fig. 4 gives an
example for the $H$ and $\co$ matrices in the defining space and in the $m$
particle spaces. Using Eqs. (\ref{snt-1}) and (\ref{snt-2}), we have 
\be
\overline{\lan \cod \co\ran^m} = V^2_\co\;\binom{m}{k_0}\;,\;\;\;
\overline{\lan \co \cod\ran^m} = V^2_\co\;\binom{N-m}{k_0}\;.
\label{snt-3}
\ee
Similarly, Eq. (\ref{sum-hh}) gives the relations,
\be
\overline{\lan \cod \co H^p\ran^m} = \overline{\lan \cod \co\ran^m}\;\; 
\overline{\lan H^p\ran^m}\;,\;\;\;
\overline{\lan \cod H^p \co\ran^m} = \overline{\lan \cod \co\ran^m}
\;\; \overline{\lan H^p\ran^{m-k_0}}\;.
\label{snt-4}
\ee
We will also make use of Eq. (\ref{strn-eq20z}) given before.
Following the procedure used in Section 2, it is possible to derive formulas for
the lower order bivariate moments of the transition strength densities generated by $\co$ 
defined by Eq. (\ref{snt-1}). Just as in Sections 3 and 5, we will consider 
the bivariate moments 
\be
M_{PQ} = \overline{\lan \cod H^Q \co H^P\ran^m}
\label{snt-6}
\ee
with $P+Q=2$ and $4$ (the $P+Q=3$ moments are zero as we are using independent
EGUE representations for $\co$ and $H$ matrices in many particle spaces).

\subsection{Exact formulas for the bivariate moments}

Firstly, Eqs. (\ref{snt-4}) gives,
\be
\barr{l}
M_{20} = \overline{\lan \cod \co \ran^m}\;\;\overline{\lan H^2\ran^m}\;,\;\;\;
M_{02} = \overline{\lan \cod \co \ran^m}\;\;\overline{\lan H^2\ran^{m-k_0}}\;, 
\\
M_{40} = \overline{\lan \cod \co \ran^m}\;\;\overline{\lan H^4\ran^m}\;,\;\;\;
M_{04} = \overline{\lan \cod \co \ran^m}\;\;\overline{\lan H^4\ran^{m-k_0}}\;.
\earr \label{snt-7}
\ee
Now, Eq. (\ref{snt-3}) along with Eqs. (\ref{strn-eq7a}) and (\ref{strn-eq7b}) 
will give the formulas for $M_{20}$, $M_{02}$, $M_{40}$ and $M_{04}$.
Formula for the first non-trivial moment $M_{11}=\overline{\lan \cod H \co
H\ran^m}$ is derived by introducing complete set of states between $\cod$ and
$H$, $H$ and $\co$ and $\co$ and $H$ in the trace giving, 
\be 
\barr{l} 
M_{11}(m) = \overline{\lan \cod H \co H\ran^m} = \\
\\
{\dis\binom{N}{m}}^{-1}\;  
\dis\sum_{v_1, v_2, v_3, v_4} \overline{\lan m, v_1 \mid \cod \mid m-k_0, 
v_2\ran \lan m-k_0, v_3 \mid \co \mid  m, v_4\ran} \\
\\ 
\times\;\overline{\lan m-k_0, v_2 \mid H \mid m-k_0, v_3\ran \lan  m,
v_4 \mid H \mid m, v_1\ran}\;.  
\earr \label{snt-8a} 
\ee 
Using Eq. (\ref{snt-1}) and applying Eq. (\ref{snt-2}) along with
Eqs. (\ref{strn-eq5}) - (\ref{reduced}) and the Wigner-Eckart theorem  
will give, 
\be 
\barr{l} 
M_{11}(m) = V_{\co}^2 V_H^2\; {\dis\binom{N}{m}}^{-1}\; \dis\binom{N-k_0}{m-k_0}
\;\\
\\
\dis\sum_{\nu=0}^k \l[\Lambda^{\nu}(N,m-k_0,m-k_0-k)\; \Lambda^{\nu}
(N,m,m-k)\r]^{1/2} \\
\\
\times\; \dis\sum_{v_1, v_2, v_3, v_4;\,\alpha;\,\omega_\nu} 
C^{f_{k_0} , \alpha}_{f_m v_1\,,\;\overline{f_{m-k_0}} \overline{v_2}}\,
C^{\overline{f_{k_0}} , \overline{\alpha}}_{f_{m-k_0} v_3\,,\;\overline{f_{m}} 
\overline{v_4}}\,
C^{\nu , \omega_{\nu}}_{f_{m-k_0} v_2\,,\;\overline{f_{m-k_0}} 
\overline{v_3}}\,
C^{\nu , \omega_{\nu}}_{f_m v_4\,,\;\overline{f_m} 
\overline{v_1}}\;.
\earr 
\label{snt-8b} 
\ee 
Simplifying the four C-G coefficients will give finally,
\be
\barr{l}
M_{11}(m) = V_{\co}^2 V_H^2\; {\dis\binom{N}{m}}^{-1}\; 
\dis\binom{N-k_0}{m-k_0}\;
\dis\sum_{\nu=0}^k Z_{11}(N, m, k_0, k, \nu)\;;\\
\\
Z_{11}(N, m, k_0, k, \nu) = \l[\binom{N}{k_0}\,d(N : \nu)\,
\Lambda^{\nu}(N,m,m-k)\,\Lambda^{\nu}(N,m-k_0,m-k_0-k)\r]^{1/2} \\
\\
\times \l|U(f_{m}\,\overline{f_{m-k_0}}\,
f_{m}\,f_{m-k_0}\,;\,f_{k_0}\,\nu)\r|\;.
\earr \label{snt-8c}
\ee
Here, $|U|$ appears as discussed just after Eq. (\ref{asym12-13}) and Eq.
(\ref{snt-9}) gives the formula for $U^2$.

Turning to the fourth order moments, we need $M_{13}$, $M_{31}$ and $M_{22}$.
As $\cod \neq \co$, here $M_{13} \neq M_{31}$ [similarly $M_{40} \neq M_{04}$ 
and $M_{20} \neq M_{02}$ as seen from Eq. (\ref{snt-7})]. Following the 
procedure used for deriving the formula for $M_{11}(m)$, we have for 
$M_{31}(m)$
\be
\barr{l}
M_{31}(m) = \overline{\lan \cod H \co H^3 \ran^{m}} = 
2\;\overline{\lan H^2\ran^{m}}\;M_{11}(m) \\
\\
+\; V^2_\co\;V^2_H\;\dis\binom{N}{m}^{-1} \;\dis\sum_{v_1, v_2, 
v_3, v_4, v_5, v_6: \alpha , \nu_1, \omega_{\nu_1}, \nu_2, \omega_{\nu_2}}  \\
\\
\lan m, v_1 \mid A^{\dagger}_{\alpha}(k_0) \mid m-k_0, v_2 \ran \lan m-k_0, 
v_3 \mid A_{\alpha}(k_0)\mid m, v_4\ran \\
\\
\times \lan m-k_0, v_2 \mid B^{\nu_1, \omega_{\nu_1}}(k) \mid m-k_0, 
v_3\ran \lan m, v_5 \mid B^{\nu_1, \omega_{\nu_1}}(k) \mid m, v_6\ran \\
\\
\times\; \lan m, v_4 \mid B^{\nu_2, \omega_{\nu_2}}(k) \mid m, 
v_5\ran \lan m, v_6 \mid B^{\nu_2, \omega_{\nu_2}}(k) \mid m, v_1\ran\;.
\earr \label{snt-10}
\ee
Now, applying the Wigner-Eckart theorem, using the results in Section 2 and
simplifying the resulting C-G coefficients will give,
\be
\barr{l}
M_{31}(m) = \overline{\lan \cod H \co H^3 \ran^{m}} = 
2\;\overline{\lan H^2\ran^{m}}\;M_{11}(m) \\
\\
+ V^2_\co \, V^2_H\;\dis\binom{N}{m}^{-1}\; \dis\binom{N-k_0}{m-k_0}\; 
\dis\sum_{\nu=0}^{min(k,m-k)} \Lambda^{\nu}(N,m,k)\;
Z_{11}(N, m, k_0, k, \nu)\;.
\earr \label{snt-11}
\ee
The function $Z_{11}$ is defined in Eq. (\ref{snt-8c}). Following the same 
procedure as above, the formula for $M_{13}$ is,
\be
\barr{l}
M_{13}(m) = \overline{\lan \cod H^3 \co H \ran^{m}} = 
2\;\overline{\lan H^2\ran^{m-k_0}}\;M_{11}(m) \\
\\
+ V^2_\co\,V^2_H\;  \binom{N}{m}^{-1}
\binom{N-k_0}{m-k_0}\;  \dis\sum_{\nu=0}^{min(k,m-k_0-k)} 
\Lambda^{\nu}(N, m-k_0, k)\;Z_{11}(N, m, k_0, k, \nu)\;.
\earr \label{snt-12}
\ee   
Formula for $M_{22}$ follows from the formula given in Section 5 for
$M_{22}(m_1,m_2)$ by using the $m_2$ part appropriately. The final result 
(with $\rho$ a multiplicity label) is
\be
\barr{l}
M_{22}(m) = \overline{\lan \cod H^2 \co H^2\ran^m} =
\overline{\lan \cod \co\ran^{m}}\;\overline{\lan H^2\ran^{m}}\;
\overline{\lan H^2\ran^{m-k_0}} \\
\\
+ V^2_\co \;V^2_H \,\l\{\dis\binom{N}{m}\, \dis\binom{N}{k_0}\r\}^{-1}
\dis\binom{N-k_0}{m-k_0}\;\l\{\dis\sum_{\nu=0}^k Z_{11}(N, m, k_0, k, \nu)\r\}^2 
\\
\\
+ V^2_\co \;V^2_H\, \dis\binom{N}{m}^{-1}\,\dis\binom{N-k_0}{m-k_0}\;
\dis\sum_{\nu_1=0}^{k}
\dis\sum_{\nu_2=0}^{k} \dis\sum_{\nu=0}^{2k} 
\dis\sqrt{\binom{N}{k_0}\;d(N : \nu)} \\
\\
\times \dis\sum_\rho \lan m \mid\mid \l[B^{\nu_1}(k) B^{\nu_2}(k)\r]^{
\nu:\rho} \mid\mid m\ran\;\lan m-k_0 \mid\mid \l[B^{\nu_1}(k) 
B^{\nu_2}(k)\r]^{\nu:\rho} \mid\mid m-k_0\ran \\
\\
\times\; (-1)^{\phi(f_{m},\overline{f_{m-k_0}}, k_0) + \phi(f_{m-k_0},
\overline{f_{m-k_0}}, \nu)}\;
U(f_{m}\,\overline{f_{m-k_0}}\,f_{m}\,f_{m-k_0}\,;\,
f_{k_0}\,\nu)\;.
\earr \label{snt-13}
\ee
The moments $M_{PQ}$ can be converted into reduced (scale free) cumulants
$k_{PQ}$ that gives information about the shape of the bivariate transition 
strength density. For our purpose the first non-trivial cumulants are the
fourth order cumulants and they are defined by Eq. (\ref{eq.cumu}).
The $k_{PQ}$, $P+Q=4$ follow from Eqs. (\ref{strn-eq7a}), (\ref{strn-eq7b}),
(\ref{snt-3}), (\ref{snt-7}),(\ref{snt-8c}), (\ref{snt-11}), (\ref{snt-12})
and (\ref{snt-13}).
Numerical results for some typical values of $(N,m,k,k_0)$ are shown in
Table 4. These results show that in general $|k_{PQ}| \lazz 0.3$ indicating
that the bivariate strength density will be close to a bivariate Gaussian. For
further confirming this result, we will derive asymptotic results for
$k_{PQ}$. 

\subsection{Asymptotic formulas for bivariate moments and approach to bivariate 
Gaussian form}

Here we will consider the asymptotic limit defined by $N \rightarrow \infty$
with $m$, $k$ and $k_0$ fixed and $k$, $k_0 << m$. Note that in the dilute limit
(or true asymptotic limit) we  also have $m \rightarrow \infty$ and  $m/N
\rightarrow 0$ with $k$ and $k_0$ fixed. Firstly, from Section 6 it is easy to
see that in the asymptotic limit: (i)
${\binom{N}{m}}^{-1}\binom{N-k_0}{m-k_0}\;Z_{11}(N,m,k_0,k,k) \rightarrow  
\binom{m}{k}\,\binom{N}{k}\,\binom{m-k}{k_0}$; (ii)
$\Lambda^0(N,m,k) \rightarrow \binom{m}{k}\,\binom{N}{k}$; (iii)
${\binom{N}{m}}^{-1}\,\Lambda^k(N,m,m-k)\,\Lambda^k(N,m,k)\,d(N:k) 
\rightarrow \binom{m}{k}\,\binom{m-k}{k}
\,{\binom{N}{k}}^2$.
Starting with $\xi$, it should be clear that in the asymptotic limit only the 
term with $\nu=k$ in Eq. (\ref{snt-8c}) will survive. Then, applying (i) and
(ii) above we have 
\be
\xi(m) = \dis\frac{M_{11}(m)}{M_{00}(m)\,\l[\wtM_{20}(m)
\wtM_{02}(m)\r]^{1/2}} 
\stackrel{\mbox{asymp}}{\longrightarrow} \dis\frac{\binom{m-k}{k_0}\; 
{\binom{m}{k}}^{1/2}}{\binom{m}{k_0}\;{\binom{m-k_0}{k}}^{1/2}}\;. 
\label{asymp-snt2}
\ee
Similarly, for $k_{40}$ and $k_{04}$ only the terms with $\nu=k$ in Eq. 
(\ref{strn-eq7b}) will survive and then applying (ii) and (iii) above will give,
\be
\barr{l}
k_{40}(m) = \dis\frac{\wtM_{40}(m)}{\l[\wtM_{20}(m)\r]^2} \;-3
\stackrel{\mbox{asymp}}{\longrightarrow} \dis\frac{\binom{m-k}{k}}{
\binom{m}{k}} -1\;,\\
\\
k_{04}(m) = \dis\frac{\wtM_{04}(m)}{\l[\wtM_{02}(m)\r]^2} \;-3
\stackrel{\mbox{asymp}}{\longrightarrow} \dis\frac{\binom{m-k_0-k}{k}}{
\binom{m-k_0}{k}} -1\;.
\earr \label{asymp-snt3}
\ee
For $M_{31}$, the first term in Eq. (\ref{snt-11}) is trivial and in the sum in
the second term only the $\nu=k$ term will survive in the asymptotic limit. Now,
applying (i)-(iii) above will give the result for $k_{31}(m)$,
\be
k_{31}(m) \stackrel{\mbox{asymp}}{\longrightarrow} \dis\frac{\binom{m-k}{k} 
\binom{m-k}{k_0}}{\binom{m}{k_0}\;\dis\sqrt{\binom{m}{k}\,\binom{m-k_0}{k}}} 
-\xi(m)= \xi(m)\,k_{40}(m)\;.
\label{asymp-snt4}
\ee
Similarly $k_{13}(m)$ is given by,
\be
k_{13}(m) \stackrel{\mbox{asymp}}{\longrightarrow} \dis\frac{\binom{m-k_0-k}{k}
\binom{m-k}{k_0}\;{\binom{m}{k}}^{1/2}}{\binom{m}{k_0}\;{\binom{m-k_0}{k}
}^{3/2}} -\xi(m) = \xi(m)\,k_{04}(m)\;.
\label{asymp-snt5}
\ee
Finally, in $M_{22}$ only the third term in Eq. (\ref{snt-13}) is complicated. 
This is simplified using its relation, valid in the asymptotic limit, 
to $\xi(m)$ as described in Section 4. Following this we have for $k_{22}$,
\be
\barr{l}
k_{22}(m) \stackrel{\mbox{asymp}}{\longrightarrow} -2\,[\xi(m)]^2 +
\dis\frac{\binom{m}{k}\,{\binom{m-k}{k_0}}^2}{\binom{m-k_0}{k}\,
{\binom{m}{k_0}}^2}
+\dis\frac{\binom{m-2k}{k_0}\,\binom{m-k}{k}}{\binom{m}{k_0}\,\binom{m-k_0}{k}}
\\
\\
\approx -2\,[\xi(m)]^2 + \dis\frac{\binom{m-2k}{k_0}}{\binom{m}{k_0}\,
\binom{m-k_0}{k}}\l[\binom{m}{k} + \binom{m-k}{k}\r]\;.
\earr \label{asymp-snt6}
\ee
In the dilute limit with $m \rightarrow \infty$ and $m/N \rightarrow 0$ and
expanding the binomials in Eqs. (\ref{asymp-snt2}) to (\ref{asymp-snt6}), 
it is seen that to order $1/m$ the cumulants $k_{rs}$, $r+s=4$ will be $-k^2/m$
(independent of $k_0$) and the correlation coefficient $\xi(m) \rightarrow 1- 
(k k_0)/2m$. Thus, the cumulants will tend to zero giving bivariate Gaussian
form. However, as $\xi \rightarrow 1$ as $m \rightarrow \infty$, in practice it
is necessary to add the $k_{rs}$, $r+s=4$ corrections to the bivariate Gaussian.

Let us add that the results given in this Section extend easily to particle
addition operators $\co^\dg=\sum_\alpha V_\alpha \,  A^\dagger_{\alpha}(k_0)$,
acting on a $m$-particle state generating $m+k_0$ particle states, by using the
results in Section 5 for the $m_1$ part appropriately. 

In addition to the three fermionic systems considered so far, for example in 
nuclear physics applications in particular it is also important to consider
explicitly the parity symmetry. For this situation, we need to consider two
orbits one with $+$ve and other with $-$ve parity (for two types of fermions, we
will have four orbits). By combining the formulation given in \cite{MKP} for
embedded ensembles with parity with the formulation in Sections 3 and 5, it is
possible to derive formulas for the bivariate moments over spaces with fixed-$m$
[or fixed-$(m_1,m_2)$] and parity. Results for all these extensions will be
discussed elsewhere.

\section{Lower-order moments of transition strength densities: Results 
for boson systems}

For $m$ spinless bosons in $N$ sp states with a general $k$-body Hamiltonian, 
we have BEGUE($k$) [`B' here stands for bosonic]. The embedding algebra for this
system is again $U(N)$. As $m$ boson states should be symmetric under interchange
of any two bosons, the  irrep $f_m =\{m\}$; the totally symmetric irrep of
$U(N)$. Similarly, $\overline{f_m}=\{m^{N-1}\}$. For $H$ a BEGUE($k$) and the
transition operator an independent BEGUE($k_0$), the results of Section 3
translate to those of BEGUE($k$) by applying the well known $N
\rightarrow -N$ symmetry, i.e. in the fermion results replace $N$ by $-N$ and
then take the absolute value of the final result; see \cite{Ko-05,kota} for
discussion on this property and \cite{Ag-02} for explicit derivation of the
formulas for boson systems for the moments of the one and two-point functions in
eigenvalues without using the $N \rightarrow -N$ symmetry. Firstly, it is easy
to see that the $m$ boson space dimension $D_B(N,m)=\binom{N+m-1}{m}$ follows 
from the $m$ fermion space dimension $D_F(N,m)=\binom{N}{m}$ by replacing $N$ 
by $-N$ in $\binom{N}{m}$ and taking the absolute value giving in general,
\be
\binom{N+r}{s} \stackrel{N \rightarrow -N}{\longrightarrow} 
\dis\binom{N-r+s-1}{s}\;.
\label{eq.bose1}
\ee
More strikingly, as pointed out in \cite{Ko-05},
\be
\barr{rcl}
\Lambda^\nu(N,m,k) \stackrel{bosons}{\longrightarrow}  \Lambda^\nu_B(N,m,k) 
& = & 
\l|\dis\binom{m-\nu}{k}\,\dis\binom{-N-m+k-\nu}{k}\r| \\
\\
& = & \dis\binom{m-\nu}{k}\,\dis\binom{N+m+\nu-1}{k} \;.
\earr \label{eq.bose2}
\ee
Here we have used the formula given by Eq. (\ref{reduced}) for fermion systems.
Moreover, for bosons the irreps $\nu$ for a $k$-body operator take the values
$\nu=0,1,\ldots,k$ as it is for fermions but for the fact that they correspond 
to the Young tableaux $\{2\nu,\nu^{N-2}\}$. Also, the $N \rightarrow -N$ 
symmetry and Eq. (\ref{eq-dnu}) will give 
\be
d_B(N:\nu) ={\binom{N+\nu-1}{\nu}}^2-{\binom{N+\nu-2}{\nu-1}}^2\;.
\label{eq.bose3}
\ee
Using Eqs. (\ref{eq.bose1}), (\ref{eq.bose2}) and (\ref{eq.bose3}), it is
possible to write the formulas for $M_{rs}(m)$, $r+s=2,4$, that correspond
to the Eqs. (\ref{strn-eq7a}), (\ref{strn-eq7b}), (\ref{strn-eq10}),
(\ref{strn-eq12}) and (\ref{strn-eq13}). As an example, the bivariate
correlation coefficient for the system considered in Section 3 but for bosons
is given by,
\be
\resizebox{0.95\hsize}{!}{$
\xi(m) = \dis\frac{M_{11}(m)}{\dis\sqrt{M_{20}(m)\, M_{02}(m)}} = 
\dis\frac{\dis\sum_{\nu=0}^{min(t,m-k)}\,
\Lambda_B^{\nu}(N,m,k)\, \Lambda_B^{\nu}(N,m,m-t)\,d_B(N:\nu)}{
\dis\binom{N+m-1}{m}\;\Lambda_B^0(N,m,t)\,\Lambda_B^0(N,m,k)}\;.$}
\label{strn-boson}
\ee
Again, it can be verified that Eq. (\ref{strn-boson}) has correctly the  $k
\leftrightarrow t$ symmetry. Let us add that for boson systems, asymptotic
results correspond to the dense limit defined by  $m \rightarrow \infty$, $N
\rightarrow \infty$,  $m/N \rightarrow \infty$ and $k, \;k_0$ fixed. 

Two species boson systems are also important in quantum physics \cite{CKM,kota}
and for these systems a situation similar to the one we have considered in
Section 5 is possible. Again, it is possible to apply the  $N \rightarrow -N$
symmetry but now for both $N_1$ and $N_2$ using Eq. (\ref{eq.bose1}).   Then,
$\binom{N_1+r}{s}$ will change to $\binom{N_1-r+s-1}{s}$ and $\binom{N_2+t}{u}$
changes to $\binom{N_2-t+u-1}{u}$. Similarly $\Lambda^{\nu}(N_1,m_1,k)$,
$d(N_1:\nu)$, $\Lambda^{\nu}(N_2,m_2,k)$ and $d(N_2:\nu)$ will change to the
bosonic $\Lambda_B$ and $d_B$ via Eqs. (\ref{eq.bose2}) and (\ref{eq.bose3}).
Also, for example for $k$-particle boson creation and annihilation operators
$A^\dagger_\alpha(k)$ and $A_\alpha(k)$, it is easy to prove that
\be
\lan m \mid\mid A^\dagger(k) \mid\mid m-k\ran\,\lan m-k \mid\mid A(k) 
\mid\mid m\ran = \binom{N+m-1}{m-k}\;.
\label{eq.bose4}
\ee
This follows from the relation $\lan \sum_\alpha A^\dagger_\alpha(k) A_\alpha(k)
\ran^m = \binom{m}{k}$, the Wigner-Eckart theorem and the sum rules for C-G
coefficients. More importantly, Eq. (\ref{eq.bose4}) follows also from the
fermionic Eq. (\ref{strn-eq20z}) by applying $N \rightarrow -N$ symmetry. 

Going further, as seen from Sections 3, 5 and 7, the bivariate moments contain
not only the  functions $\Lambda^\nu(N,m,k)$ and $d(N:\nu)$ but also $U(N)$
Racah (or $U-$)  coefficients [(see Eqs. (\ref{strn-eq13}) and
(\ref{strn-eq27})]. In principle the $N \rightarrow -N$ law applies also to the
Racah coefficients by translating the irreps appropriately. For example, the 
fermionic  $U$-coefficient given by Eq. (\ref{snt-9}) can be translated, using
the $N \rightarrow -N$ law, for boson systems and the result is
\be
\resizebox{0.98\hsize}{!}{$U_B^2(f_m,\, \overline{f_p},\, f_m,\, f_p\, ;\,
f_{m-p},\,\nu) = 
\dis\frac{{\dis\binom{N+\nu-2}{\nu}}^2 \dis\binom{m-\nu}{p-\nu} 
\dis\binom{N+m+\nu -1}{m-p}\;(N+2\nu -1)}{{\dis\binom{N+m-1}{p}}^2 
\dis\binom{N+m-p-1}{m-p}\;(N-1)}\;.$}
\label{eq.bose6}
\ee
Note that in Eq. (\ref{eq.bose6}), $f_r=\{r\}$, $\overline{f_s}=\{s^{N-1}\}$ 
and $\nu=\{2^\nu ,\nu^{N-2}\}$. It is easy to verify Eq. (\ref{eq.bose6}) for
$\nu=0$. Using Eq. (\ref{eq.bose6}), it is possible to deal with boson systems
that are similar to the fermionic systems considered in Sections 5 and 7. For
example, Eqs. (\ref{eq.bose6}) and (\ref{snt-8c}) will give the formula for
the bivariate correlation coefficient for particle removal operator for bosons. 
Full details of the formulas for the bivariate moments for boson systems and
their asymptotic structure will be presented elsewhere.

\section{Conclusions and future outlook}

Embedded random matrix ensembles were used with success in the past to 
understand the form of the eigenvalue density of finite quantum systems
\cite{Mo-75} and now there are good applications of the Gaussian form found  for
these systems \cite{KH,Senk1,Senk2}. Similarly, these ensembles are shown to
provide the basis for the theory for expectation values of operators
\cite{Ko-03,KH}. In the present paper the focus is turned to transition 
strengths. 

Employing embedded Gaussian unitary ensembles of random matrices, for the first
time we have derived exact group theoretical formulas for the second and fourth
order bivariate moments of the transition strength densities. Explicit results
for a spinless many fermion system with $k$-body Hamiltonian and the transition
operator a $k_0$-body operator are presented in Sections 3. Similarly, results
for a system with two types of spinless fermions and beta decay (and double beta
decay) type operators are presented in Section 5. In addition, results for a
particle removal operator are presented in Section 7.1. The corresponding
asymptotic results presented in Sections 4, 6 and 7.2  respectively and the
numerical results from the exact formulas for the bivariate fourth order
cumulants as presented in Tables 1-4 clearly show that the smoothed transition
densities can be very well approximated by Edgeworth  corrected bivariate
Gaussian. Also, values of the bivariate correlation coefficient ($\xi$) shown in
Tables 1-4 clearly confirm that the strength distribution will be a narrow
distribution (unlike for a GOE). Briefly discussed in Section 8 are some results
for boson systems. The formulation and results given in this paper hold for a
general $k$-body Hamiltonian (similarly for the body rank of the transition
operators considered).  However, it should be  emphasized that two-body
Hamiltonians, i.e. EGUE(2)s , are in general more relevant for systems such as
nuclei and atoms. Clearly, the work presented in the paper represents major
progress in random matrix theory for smoothed transition strengths after the
paper by French et al in 1988 \cite{FKPT}.

One gap in the present results is that there is not yet a formula  (or a good
procedure) available for the $U(N)$ Racah coefficients of the type,
$$
U(f_m,\, \nu_1,\,f_m,\,\nu_2\,:\;f_m,\,\nu)
$$
where $f_m=\{1^m\}$ for fermions (and $\{m\}$ for bosons) and $\mu=\{2^\mu,
1^{N-2\mu}\}$ (for bosons $\mu=\{2^\mu, \mu^{N-2}\}$). These coefficients are 
needed for the $M_{22}$ moment.  In future, it is also important to consider
fermions systems with spin as these are of direct interest in mesoscopic systems
\cite{Al-00}. Embedding algebra for the ensembles with spin for the fermions is
$U(\Omega) \otimes SU(2)$ with $SU(2)$ generating spin and $\Omega=N/2$
\cite{Ko-07}. It is also possible to consider ensembles with more general
$U(\Omega) \otimes SU(r)$ embedding \cite{Ma-12}; the ensembles with $r=1$, $2$
and $4$ are important for fermionic systems and $r=1$, $2$ and $3$ for bosonic
systems (for example $r=3$ is appropriate for spinor BEC). The ensembles with
$U(\Omega) \otimes SU(r)$ embedding  \cite{Ko-07,Ma-12,kota} are studied so far
only for one and two-point functions in eigenvalues.  Even for these, the group
theoretical results are incomplete \cite{Ko-07,Ma-12} as the Racah coefficients
needed, for example for the fourth moment of the eigenvalue density, are not yet
available. Another important extension is to consider ensembles with a
mean-field one-body term which is more realistic for systems such as atomic
nuclei and atoms.

In summary, in this paper a detailed analytical study of transition strengths 
has been carried out, going beyond the results in \cite{FKPT}, using embedded
random matrix ensembles and established clearly that the form of the transition 
strength densities for isolated finite interacting fermion systems will be {\it
generically a bivariate Gaussian}  with fourth order cumulant corrections. Fig.
5 shows a bivariate Gaussian with and without Edgeworth corrections. 
We expect that in the near future
the bivariate Gaussian form will be used in practical calculations of transition
strengths (see \cite{KM} for an attempt in the past) just as it is being done at
present in a systematic manner by the Michigan  group \cite{Senk1,Senk2} for
nuclear level densities with good success using  the Gaussian form given by the
embedded ensembles.

\flushleft{\bf Acknowledgments} \\

Thanks are due to Prof. H.A. Weidenm\"{u}ller, for many useful discussions. 
M.V. acknowledges financial support from CONACyT Project No. 154586 
and PAPIIT-UNAM Project No. IG101113. M.V. is DGAPA/UNAM postdoctoral 
fellow.  

\renewcommand{\theequation}{A-\arabic{equation}}
\setcounter{equation}{0}  
\section*{APPENDIX A}  

\noindent Given a bivariate distributions $\rho(x,y)$, its integral over $y[x]$
the marginal density $\rho_2(y)$[$\rho_1(x)$]. The centroids and
variances of these are the marginal centroids and  variances and say that
they are  ($\epsilon_1$, $\sigma_1^2$) for $\rho_1(x)$ and similarly 
($\epsilon_2$, $\sigma_2^2$) for $\rho_2(y)$. Note that the $\wtM_{20}$ and
$\wtM_{02}$ in Section 4 are $\sigma_1^2$ and $\sigma_2^2$ respectively. 
Bivariate Gaussian in terms of the standardized variables
$\wx=(x-\epsilon_1)/\sigma_1$ and $\wy=(y-\epsilon_2)/\sigma_2$ is given by,
\be
\eta_\cg(\wx , \wy) = \dis\frac {1}{2\pi \sqrt{(1 - \xi^2)}} \;
\exp \l\{- \dis\frac {\wx^2 - 2\xi \wx \wy + \wy^2}{2(1 - \xi^2)} \r\}\;.
\label{eq.ew6}
\ee
Here $\xi$ is the correlation coefficient. Thus, a bivariate Gaussian is
defined by the five variables $(\epsilon_1,\epsilon_2,\sigma_1,\sigma_2,\zeta)$.
Now, the Edgeworth (ED) corrected bivariate Gaussian including $k_{rs}$ up to
$r+s=4$ is given by
{\footnotesize
\be
\barr{rcl}
\eta_{biv-ED}(\wx,\wy) &=&
\l\{ 1 + \l({\dis\frac{k_{30}}{ 6}} He_{30}(\wx,\wy)
+ {\dis\frac{k_{21}}{2}} He_{21}(\wx,\wy)
\r . \r . \\ \\
& + & \l.\l.  {\dis\frac{k_{12}}{2}} He_{12}(\wx,\wy) +
{\dis\frac{k_{03}}{6}} He_{03}(\wx,\wy) \r) \r. \\ \\
& + & \l. \l(\l\{ {\dis\frac{k_{40}}{24}} He_{40}
(\wx,\wy) +
{\dis\frac{k_{31}}{6}} He_{31}(\wx,\wy)
\r.\r.\r. \\ \\
& + & \l.\l.\l.  {\dis\frac{k_{22}}{4}} He_{22}(\wx,\wy)
+ {\dis\frac{k_{13}}{6}} He_{13}(\wx,\wy) +
{\dis\frac{k_{04}}{24}}
He_{04}(\wx,\wy) \r\} \r.\r.\\ \\
& + & \l.\l. \l\{ {\dis\frac{k_{30}^2}{72}} He_{60}(\wx,
\wy)
+ {\dis\frac{k_{30} k_{21}}{12}} He_{51}(\wx,\wy)
\r.\r.\r. 
\\ \\
& + & \l.\l.\l. \l[{\dis\frac{k_{21}^2}{8}} + {\dis\frac{k_{30} k_{12}}{12}}\r]
He_{42}(\wx,\wy) \r.\r.\r. \\ \\
& + & \l.\l.\l.
\l[{\dis\frac{k_{30}k_{03}}{36}} + {\dis\frac{k_{12}k_{21}}{4}}\r]
He_{33}(\wx,\wy) \r.\r.\r. \\ \\
& + & \l.\l.\l.  \l[{\dis\frac{k_{12}^2}{8}} + {\dis\frac{k_{21}k_{03}}{12}}\r]
He_{24}(\wx,\wy) +
{\dis\frac{k_{12} k_{03}}{12}} He_{15}(\wx,\wy)
\r.\r.\r. \\ \\
& + & \l.\l.\l. {\dis\frac{k_{03}^2}{72}}
He_{06}(\wx,\wy) \r\} \r)
\r\} \eta_{\cal G}(\wx,\wy) \;.
\earr \label{eq.ew9}
\ee}
Note that for EGUE and BEGUE, by definition the $k_{rs}=0$ for $r+s=3$. However
in practical applications, these will be non-zero though expected to be small in
magnitude.  The bivariate  Hermite polynomials $He_{m_1m_2}(\wx,\wy)$ in Eq.
(\ref{eq.ew9}) satisfy the recursion relation,
\be
\barr{rcl}
(1 - \xi^2)He_{m_1+1,m_2}(\wx, \wy) & = & (\wx - \xi \wy) He_{m_1,m_2}(
\wx,\wy) \\ \\
& - & m_1 He_{m_1-1,m_2}(\wx,\wy) + m_2 \xi He_{m_1,m_2-1}(\wx,\wy) \;.
\earr \label{ew.11}
\ee
This can be solved using $He_{00}(\wx,\wy) =  1$, $He_{10}(\wx,\wy) =
\l(\wx - \xi \wy\r)/\l(1 - \xi^2\r)$ and $He_{01}(\wx,\wy) =
\l(\wy - \xi \wx\r)/\l(1 - \xi^2\r)$.

\flushleft{\bf References} \\

\newpage

\begin{figure}[ht] 

\begin{center}
\includegraphics[width=5in,height=5.95in]{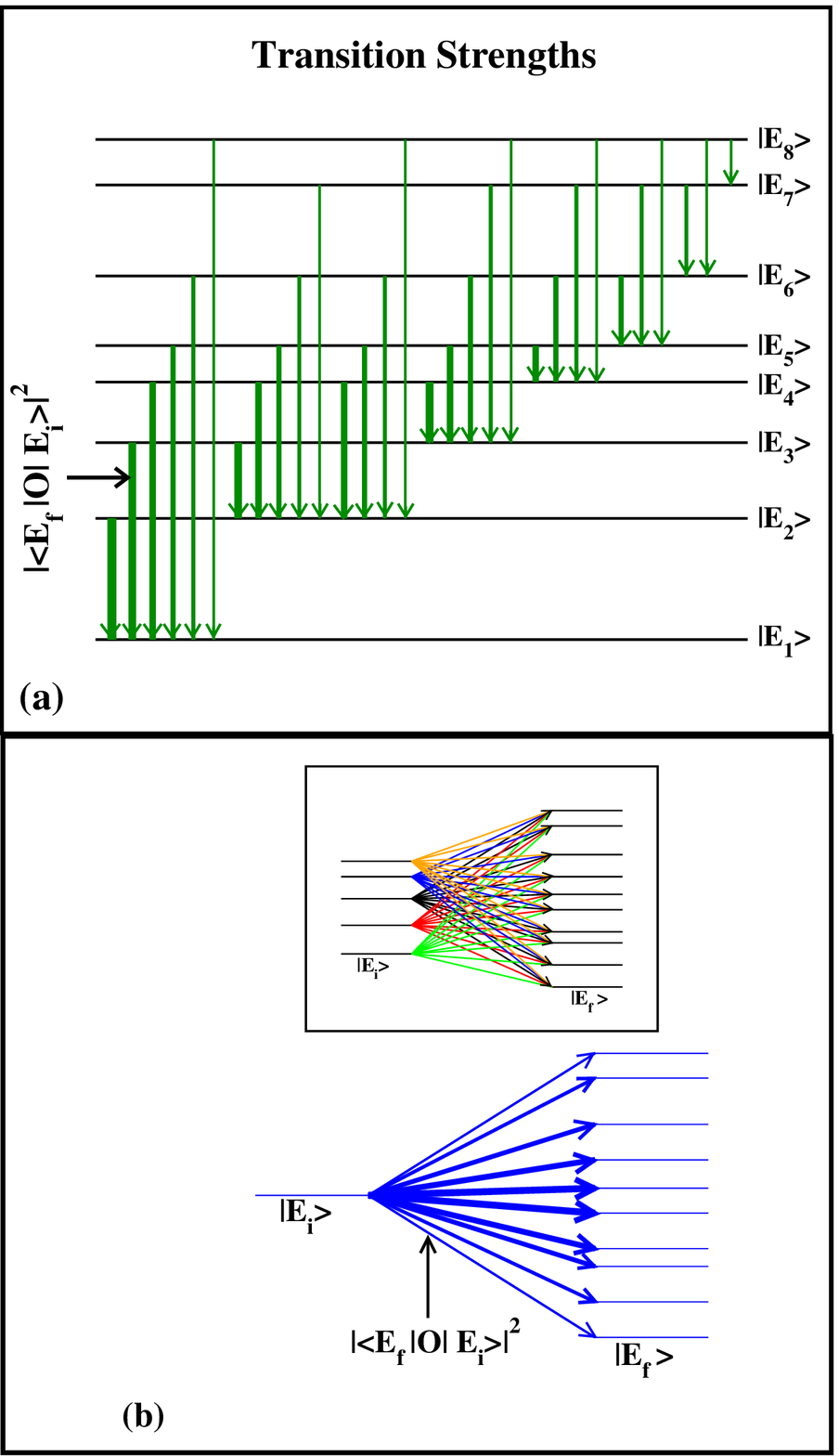} 
\end{center}
\caption{Schematic figure showing transition strengths. (a) Transition 
strengths for transitions induced by an operator ${\cal O}_1$, from levels with
energies (eigenvalues of the Hamiltonian) $E_i$ of a system `$a$' to levels with
energies $E_f$ of the same system `$a$'. The strengths $\l|\lan a, E_f \mid
{\cal O}_1 \mid a, E_i\ran\r|^2$ are proportional to the widths of the lines in the
figure. (b) Transition strengths for transitions induced by an operator ${\cal
O}_2$, from one particular level with energy (eigenvalue of the Hamiltonian)
$E_i$ of a system `$a$' to levels with energies $E_f$ of another system
`$b$'. The strengths $\l|\lan b, E_f \mid {\cal O}_1 \mid a, E_i\ran\r|^2$ are
proportional to the widths of the lines in the figure. In general, transitions from
several levels of the system `$a$' to the levels of system `$b$' are possible as
shown in the inset figure.} 
\end{figure}

\newpage

\begin{figure}[ht]
\begin{center}
\includegraphics[width=5.25in,height=6.75in]{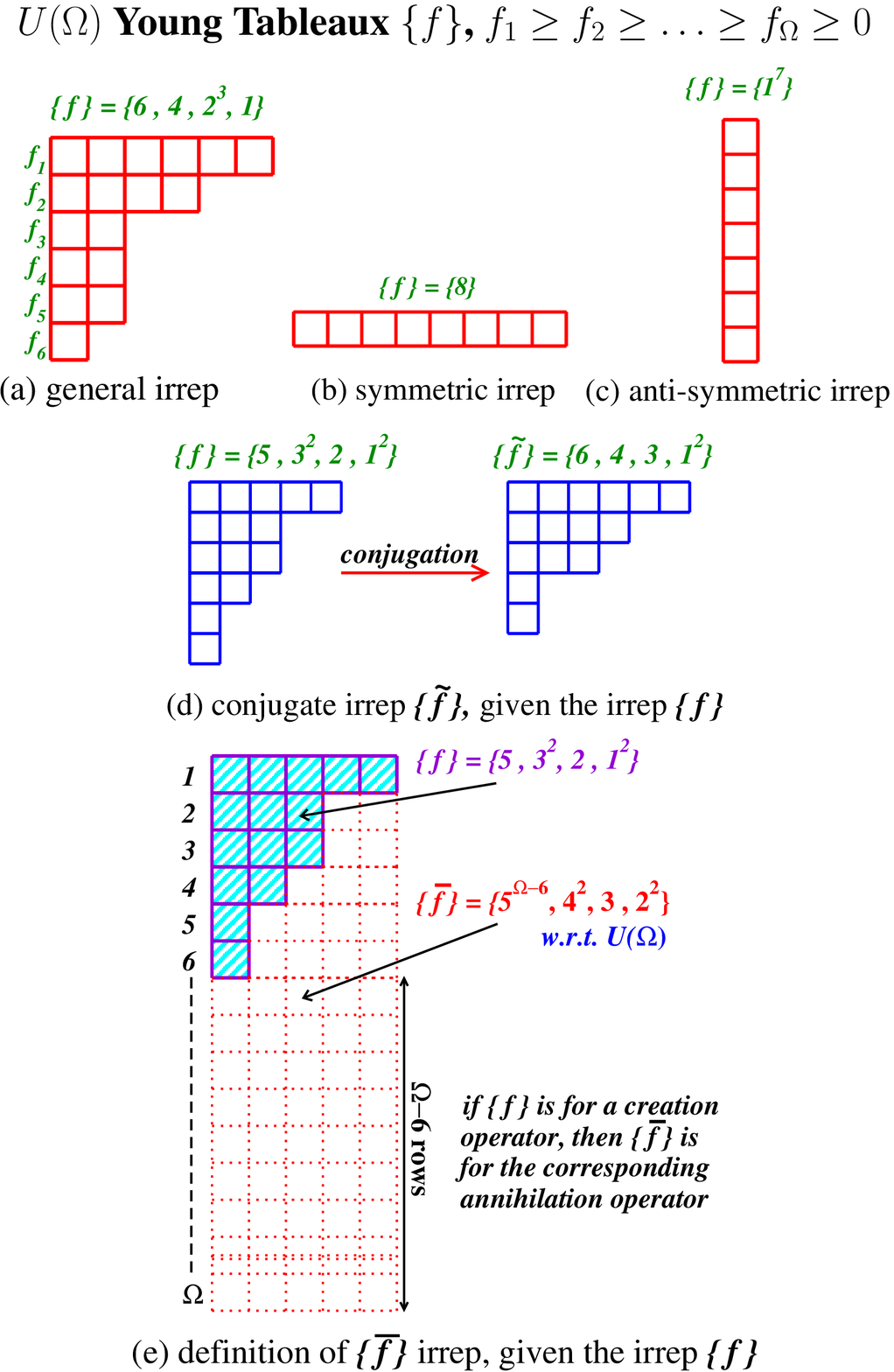}
\end{center}
\caption{Young tableaux representation of the irreps of $U(\Omega)$.  Shown are
examples of: (a) a general irrep $\{f\}$; (b) symmetric irrep $\{m\}$; (c)
antisymmetric irrep $\{1^m\}$; (d) conjugate irrep $\{\tilde{f}\}$ that
corresponds to a given $\{f\}$; (e) irrep $\{\overline{f}\}$ that corresponds to
a given  $\{f\}$. Note the importance of $\Omega$ in defining
$\{\overline{f}\}$.}
\end{figure}

\newpage

\begin{figure}[ht]
\begin{center}
\includegraphics[width=5.5in]{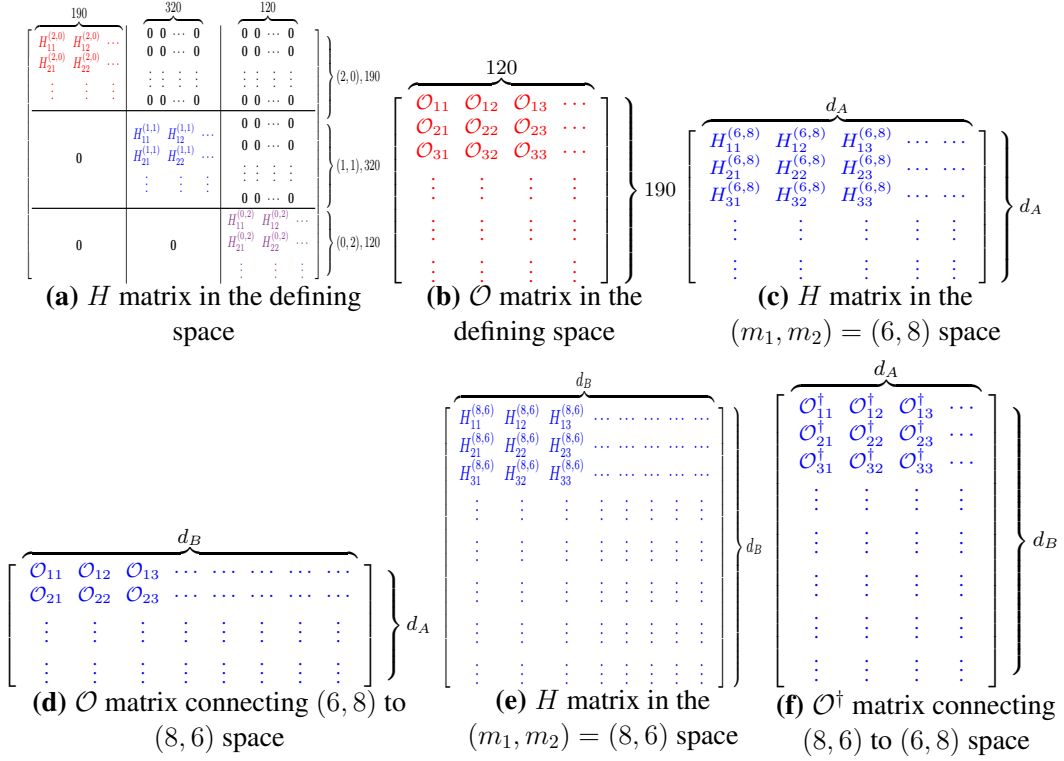}
\end{center}
\caption{Schematic diagram showing the matrix representations of the Hamiltonian
($H$) and the transition operator ($\co$) in the respective defining spaces and
in the many particle spaces for the system considered in Section 5 with
$N_1=20$, $N_2=16$, $k=2$ and $k_0=2$. (a) $H$ matrix in the defining space.
Note that the matrix is in block diagonal form with 3 blocks and they correspond
to $(m_1,m_2)=(2,0)$, $(1,1)$ and $(0,2)$ as shown in the figure. The matrix
dimensions of each block are also shown in the figure. (b) $\co$ matrix in the
defining space. Note that $\co$ is a rectangular matrix connecting 
$(m_1,m_2)_i=(0,2)$ to $(m_1,m_2)_f=(2,0)$. The number of rows and columns in the
matrix are shown in the figure. (c) $H$ matrix $H_i$ in the initial
$(m_1,m_2)=(6,8)$ space with matrix dimension $d_A=498841200$. (d) $\co$ matrix
connecting states in $(m_1,m_2)_i =(6,8)$ space with the states in $(m_1,m_2)_f
=(8,6)$ space with matrix elements $\lan (m_1,m_2)_f=(8,6) \beta \mid \co \mid 
(m_1,m_2)_i=(6,8), \alpha\ran$. Note that the matrix is a $d_B \times d_A$
rectangular matrix with $d_A$ given in (c) and $d_B=1008767760$. (d) $H$ matrix
$H_f$ in the final $(8,6)$ space. (e) same as (d) but for the $\cod$ matrix. It
is useful to note that the bivariate moments $M_{PQ}$ of the transition 
strength density are given by $(d_A)^{-1}\; \mbox{\bf Tr}\l[\cod (H_f)^Q \co (H_i)^P\r]$
where $\mbox{\bf Tr}$ stands for the matrix trace.}
\end{figure}

\newpage

\begin{figure}[ht]
\begin{center}
\includegraphics[width=5.5in]{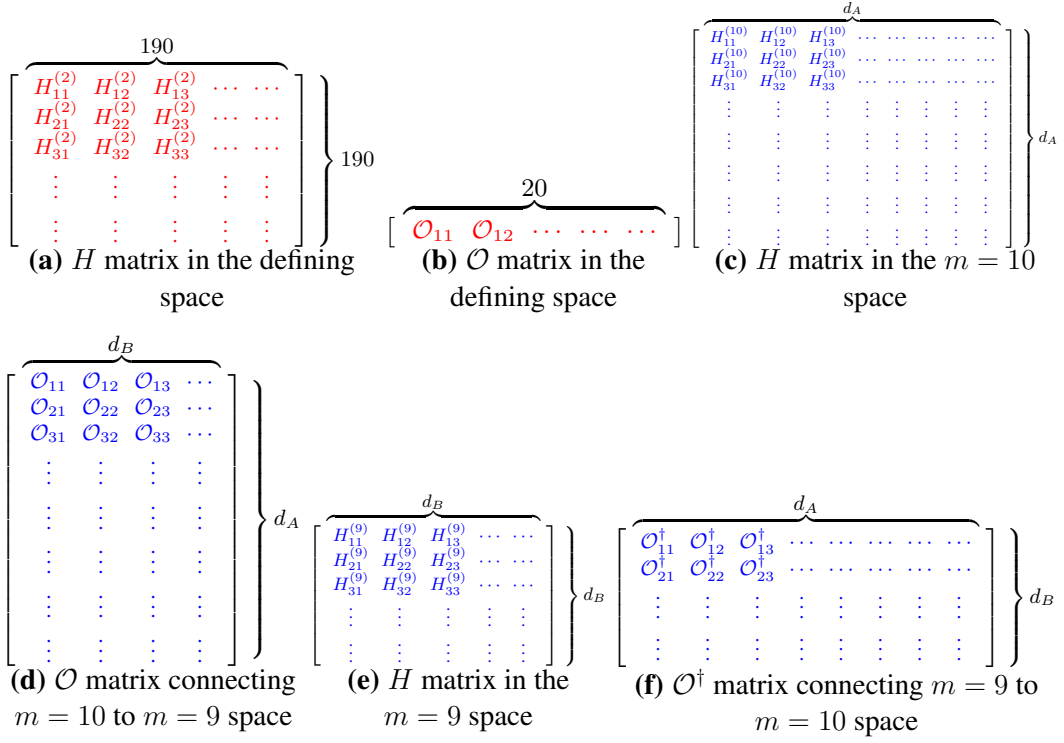}
\end{center}
\caption{Schematic diagram showing the matrix representations of the Hamiltonian
($H$) and the transition operator ($\co$) in the respective defining spaces and
in the many particle spaces for the system considered in Section 7 with $N=20$,
$k=2$ and $k_0=1$. (a) $H$ matrix in the defining space. (b) $\co$ matrix in the
defining space. (c) $H$ matrix $H_i$ in the initial $m=10$ space with matrix
dimension $d_A=184756$. (d) $\co$ matrix connecting states in $m_i=10$ space
with the states in $m_f =9$ space with matrix elements $\lan m=9, \beta \mid \co
\mid m=10, \alpha\ran$. Note that the matrix is a $d_B \times d_A$ rectangular
matrix with $d_A$ given in (c) and $d_B=167960$. (e) $H$ matrix $H_f$ in the
final $m=9$ space. (f) same as (d) but for the $\cod$ matrix. See Fig. 3 for
further details.}
\end{figure}

\newpage

\begin{figure}[ht]
\begin{center}
    \subfigure[t][{Bivariate Gaussian}]
    {
        \includegraphics[width=4.25in]{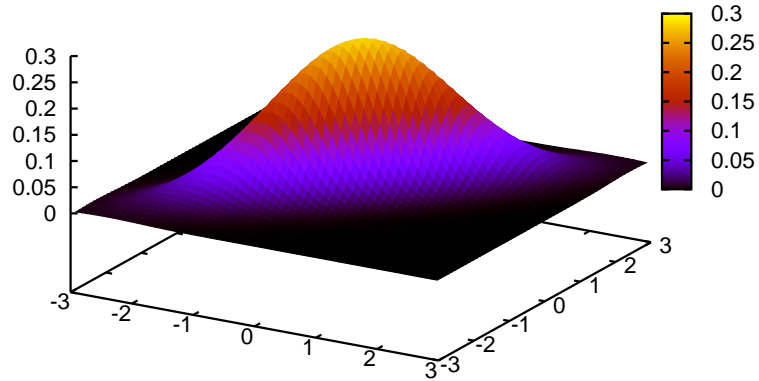}
        \label{fig5a}
    }
    \\
    \subfigure[t][{Bivariate Gaussian with ED corrections}]
    {
        \includegraphics[width=4.25in]{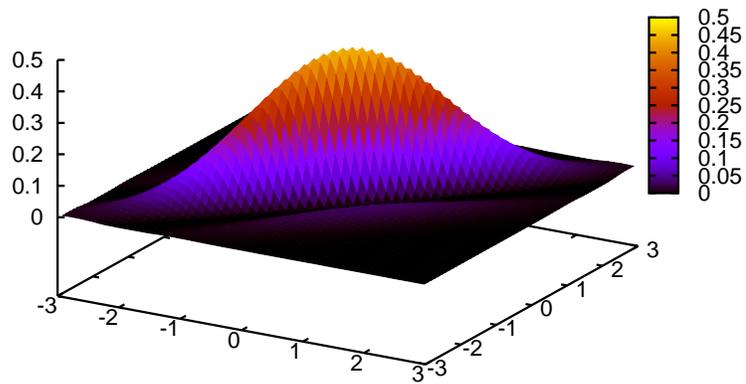}
        \label{fig5b}
    }
\end{center}
\caption{Schematic diagram showing (a) normalized bivariate Gaussian and
(b) normalized bivariate Gaussian with Edgeworth (ED) corrections (see Appendix A). 
The fourth order cumulants used are $\xi=0.83$, $k_{40}=-0.18$, $k_{04}=-0.17$,
$k_{31}=-0.15$, $k_{13}=-0.14$, and $k_{22}=-0.03$ corresponding to $(N_1,m_1) = 
(44,10)$ and $(N_2,m_2)=(58,20)$ in Table 2.}
\end{figure}

\newpage
\begin{table}[ht]

\caption{Bivariate correlation coefficient ($\xi$) and fourth order bivariate
cumulants $k_{rs}$ ($=k_{sr}$) with $r \geq s$ and $r+s=4$ for various values of
number of sp state ($N$), number of fermions ($m$), Hamiltonian body rank ($k$)
and the body rank ($k_0$) of the transition operator. Results are obtained using
the formulas given in Section 3. Note that for the $M_{22}$ that is needed for 
$k_{22}$, we have used Eq. (\ref{strn-eq13}) with the third term replaced by the
corresponding asymptotic formula given by Eq. (\ref{asymp11}) as the
$U$-coefficient needed for the finite-$(N,m)$ formula is not available. The
finite $N$ and $m$ results are compared with the asymptotic limit results (these
are given in the brackets). See text for further details.}

\begin{center}
\begin{tabular}{cccccccc}
\hline 
$N$ & $m$ & $k$ & $k_0$ & $\xi$ & $k_{40}$ & $k_{31}$ & $k_{22}$ \\
\hline \\
$20$ & $10$ & $2$ & $1$ & $0.68(0.8)$ & $-0.54(-0.38)$ & $-0.36(-0.3)$ & 
$-0.09(-0.27)$ \\
$30$ & & & & $0.73(0.8)$ & $-0.48(-0.38)$ & $-0.35(-0.3)$ & 
$-0.16(-0.27)$ \\
$50$ & & & & $0.76(0.8)$ & $-0.43(-0.38)$ & $-0.33(-0.3)$ & 
$-0.21(-0.27)$ \\
$80$ & & & & $0.78(0.8)$ & $-0.41(-0.38)$ & $-0.32(-0.3)$ & 
$-0.23(-0.27)$ \\
$40$ & $8$ & $2$ & $1$ & $0.71(0.75)$ & $-0.52(-0.46)$ & $-0.37(-0.35)$ & 
$-0.23(-0.3)$ \\
& $12$ & & & $0.78(0.83)$ & $-0.4(-0.32)$ & $-0.31(-0.27)$ & 
$-0.15(-0.24)$ \\
& $15$ & & & $0.81(0.87)$ & $-0.36(-0.26)$ & $-0.29(-0.22)$ & 
$-0.1(-0.21)$ \\
& $20$ & & & $0.82(0.9)$ & $-0.32(-0.2)$ & $-0.27(-0.18)$ & 
$-0.03(-0.17)$ \\
$60$ & $8$ & $2$ & $1$ & $0.72(0.75)$ & $-0.5(-0.46)$ & $-0.36(-0.35)$ & 
$-0.26(-0.3)$ \\
& $15$ & & & $0.83(0.87)$ & $-0.32(-0.26)$ & $-0.26(-0.22)$ & 
$-0.15(-0.21)$ \\
& $20$ & & & $0.86(0.9)$ & $-0.27(-0.2)$ & $-0.23(-0.18)$ & 
$-0.09(-0.17)$ \\
& $30$ & & & $0.88(0.93)$ & $-0.23(-0.13)$ & $-0.2(-0.12)$ & 
$-0.02(-0.12)$ \\
$24$ & $8$ & $2$ & $1$ & $0.67(0.75)$ & $-0.56(-0.46)$ & $-0.38(-0.35)$ & 
$-0.18(-0.3)$ \\
& & $2$ & $2$ & $0.44(0.54)$ & $-0.56(-0.46)$ & $-0.24(-0.25)$ & 
$-0.07(-0.17)$ \\
& & $3$ & $1$ & $0.54(0.63)$ & $-0.88(-0.82)$ & $-0.47(-0.51)$ & 
$-0.24(-0.35)$ \\
& & $3$ & $2$ & $0.27(0.36)$ & $-0.88(-0.82)$ & $-0.23(-0.29)$ & 
$-0.06(-0.12)$ \\
& & $4$ & $1$ & $0.41(0.5)$ & $-0.99(-0.99)$ & $-0.41(-0.49)$ & 
$-0.16(-0.25)$ \\
$40$ & $12$ & $1$ & $1$ & $0.89(0.92)$ & $-0.11(-0.08)$ & $-0.1(-0.08)$ & 
$-0.02(-0.08)$ \\
& & $2$ & $1$ & $0.78(0.83)$ & $-0.4(-0.32)$ & $-0.31(-0.27)$ & 
$-0.15(-0.24)$ \\
& & $3$ & $1$ & $0.68(0.75)$ & $-0.71(-0.62)$ & $-0.48(-0.46)$ & 
$-0.27(-0.37)$ \\
& & $3$ & $2$ & $0.45(0.55)$ & $-0.71(-0.62)$ & $-0.32(-0.34)$ & 
$-0.12(-0.21)$ \\
& & $4$ & $1$ & $0.59(0.67)$ & $-0.91(-0.86)$ & $-0.53(-0.57)$ & 
$-0.3(-0.4)$ \\
& & $4$ & $2$ & $0.33(0.42)$ & $-0.91(-0.86)$ & $-0.3(-0.36)$ & 
$-0.1(-0.17)$ \\
& & $5$ & $1$ & $0.5(0.58)$ & $-0.98(-0.97)$ & $-0.49(-0.57)$ & 
$-0.24(-0.34)$ \\
& & $5$ & $2$ & $0.24(0.32)$ & $-0.98(-0.97)$ & $-0.23(-0.31)$ & 
$-0.05(-0.1)$ \\
\hline
\end{tabular}
\end{center}
\label{tab1}
\end{table}

\newpage
\begin{sidewaystable}[hb]
\caption{Bivariate correlation coefficient ($\xi$) and fourth order bivariate
cumulants $k_{rs}$ with $r \geq s$ and $r+s=4$ for the system considered in
Section 5 for double beta decay type transition operators with $k_0=2$.  Results
are given for various values of number of sp state $N_1$ and $N_2$ and number of
fermions $m_1$ and $m_2$ in these sp states respectively. The Hamiltonian body
rank ($k$) is chosen to be $k=2$. Results are obtained using the formulas given
in Section 5. Note that for the $M_{22}$ that is needed for
$k_{22}$ we have used Eq. (\ref{strn-eq30}) with the
third term replaced by the corresponding asymptotic formula given by Eq.
(\ref{asym12-12}) as the $U$-coefficient needed for the finite-$(N,m)$ formula
is not available. The finite $N$ and $m$ results are compared with the
asymptotic limit results (these are given in the brackets). See text for further
details.}

\begin{center} 
\begin{tabular}{cccccccccc} 
\hline  $N_1$ & $m_1$ & $N_2$ & $m_2$ & $\xi$ &
$k_{40}$ & $k_{04}$ & $k_{31}$ & $k_{13}$ & $k_{22}$ \\ 
\hline \\
20 & 8 & 20 & 8 & $0.66(0.76)$ & $-0.34(-0.24)$ & $-0.35(-0.24)$ & 
$-0.22(-0.18)$ & $-0.23(-0.19)$ & $-0.01(-0.18)$ \\
20 & 10 & 20 & 10 & $0.68(0.81)$ & $-0.31(-0.19)$ & $-0.32(-0.20)$ & 
$-0.21(-0.16)$ & $-0.22(-0.16)$ & $0.05(-0.16)$ \\
32 & 10 & 32 & 10 & $0.74(0.81)$ & $-0.26(-0.19)$ & $-0.26(-0.20)$ & 
$-0.19(-0.16)$ & $-0.19(-0.16)$ & $-0.04(-0.16)$ \\
32 & 12 & 32 & 12 & $0.77(0.84)$ & $-0.23(-0.16)$ & $-0.24(-0.16)$ &
$-0.18(-0.14)$ & $-0.18(-0.14)$ & $-0.01(-0.14)$ \\
32 & 16 & 32 & 16 & $0.78(0.88)$ & $-0.21(-0.12)$ & $-0.22(-0.12)$ &
$-0.17(-0.11)$ & $-0.17(-0.11)$ & $0.06(-0.11)$ \\
32 & 10 & 44 & 8 & $0.73(0.78)$ & $-0.27(-0.22)$ & $-0.29(-0.23)$ &
$-0.20(-0.17)$ & $-0.21(-0.18)$ & $-0.07(-0.18)$ \\
32 & 10 & 44 & 15 & $0.79(0.85)$ & $-0.21(-0.16)$ & $-0.22(-0.16)$ &
$-0.17(-0.13)$ & $-0.17(-0.13)$ & $-0.03(-0.13)$ \\
32 & 10 & 44 & 20 &  $0.80(0.87)$ & $-0.20(-0.13)$ & $-0.20(-0.13)$ &
$-0.16(-0.11)$ & $-0.16(-0.11)$ & $0.01(-0.11)$ \\
32 & 12 & 44 & 20 & $0.81(0.88)$ & $-0.19(-0.12)$ & $-0.19(-0.12)$ &
$-0.16(-0.11)$ & $-0.16(-0.11)$ & $0.02(-0.11)$ \\
32 & 16 & 44 & 20 & $0.81(0.89)$ & $-0.19(-0.11)$ & $-0.19(-0.11)$ &
$-0.15(-0.10)$ & $-0.15(-0.10)$ & $0.04(-0.10)$ \\
44 & 10 & 58 & 20 & $0.83(0 .87)$ & $-0.18(-0.13)$ & $-0.17(-0.13)$ &
$-0.15(-0.11)$ & $-0.14(-0.11)$ & $-0.03(-0.11)$ \\
44 & 15 & 58 & 20 & $0.84(0.89)$ & $-0.16(-0.11)$ & $-0.16(-0.11)$ &
$-0.13(-0.10)$ & $-0.14(-0.10)$ & $-0.01(-0.10)$ \\
44 & 20 & 58 & 20 & $0.85(0.90)$ & $-0.15(-0.10)$ & $-0.16(-0.10)$ &
$-0.13(-0.09)$ & $-0.13(-0.09)$ & $0.01(-0.09)$ \\
44 & 10 & 58 & 24 & $0.83(0.88)$ & $-0.17(-0.12)$ & $-0.16(-0.12)$ &
$-0.14(-0.10)$ & $-0.14(-0.10)$ & $-0.01(-0.10)$ \\
44 & 10 & 58 & 28 & $0.84(0.90)$ & $-0.16(-0.10)$ & $-0.16(-0.10)$ &
$-0.14(-0.09)$ & $-0.13(-0.09)$ & $0.01(-0.09)$ \\
\hline 
\end{tabular} 
\end{center} 
\label{tab2} 
\end{sidewaystable}

\newpage
\begin{sidewaystable}[hb]
\caption{Bivariate correlation coefficient ($\xi$) and fourth order bivariate
cumulants $k_{rs}$ with $r \geq s$ and $r+s=4$ for the system considered in
Section 5 for beta decay type transition operators with $k_0=1$.  Results
are given for various values of number of sp state $N_1$ and $N_2$ and number of
fermions $m_1$ and $m_2$ in these sp states respectively. The Hamiltonian body
rank ($k$) is chosen to be $k=2$. Results are obtained using the formulas given
in Section 5. Note that for the $M_{22}$ that is needed for
$k_{22}$ we have used Eq. (\ref{strn-eq30}) with the
third term replaced by the corresponding asymptotic formula given by Eq.
(\ref{asym12-12}) as the $U$-coefficient needed for the finite-$(N,m)$ formula
is not available. The finite $N$ and $m$ results are compared with the
asymptotic limit results (these are given in the brackets). See text for further
details.}

\begin{center} 
\begin{tabular}{cccccccccc} 
\hline  $N_1$ & $m_1$ & $N_2$ & $m_2$ & $\xi$ & $k_{40}$ & $k_{04}$ & $k_{31}$ 
& $k_{13}$ & $k_{22}$ \\ 
\hline \\
20 & 8 & 30 & 8 & $0.83(0.88)$ & $-0.32(-0.24)$ & $-0.33(-0.25)$ & 
$-0.26(-0.21)$ & $-0.27(-0.22)$ & $-0.12(-0.22)$ \\
20 & 8 & 30 & 10 & $0.84(0.89)$ & $-0.30(-0.22)$ & $-0.30(-0.22)$ &
$-0.25(-0.19)$ & $-0.25(-0.20)$ & $-0.10(-0.20)$ \\
20 & 10 & 30 & 10 & $0.84(0.90)$ & $-0.29(-0.20)$ & $-0.30(-0.20)$ &
$-0.24(-0.18)$ & $-0.25(-0.18)$ & $-0.07(-0.18)$ \\
20 & 10 & 30 & 15 & $0.86(0.92)$ & $-0.26(-0.16)$ & $-0.26(-0.16)$ &
$-0.22(-0.14)$ & $-0.23(-0.15)$ & $-0.03(-0.15)$ \\
36 & 8 & 36 & 8 & $0.85(0.88)$ & $-0.29(-0.24)$ & $-0.29(-0.24)$ &
$-0.24(-0.21)$ & $-0.25(-0.21)$ & $-0.15(-0.21)$ \\
36 & 12 & 36 & 12 & $0.88(0.92)$ & $-0.22(-0.16)$ & $-0.22(-0.16)$ &
$-0.20(-0.15)$ & $-0.20(-0.15)$ & $-0.08(-0.15)$ \\
36 & 16 & 36 & 16 & $0.90(0.94)$ & $-0.20(-0.12)$ & $-0.20(-0.12)$ &
$-0.18(-0.12)$ & $-0.18(-0.12)$ & $-0.03(-0.12)$ \\
36 & 18 & 36 & 18 & $0.90(0.94)$ & $-0.19(-0.11)$ & $-0.19(-0.11)$ &
$-0.17(-0.10)$ & $-0.17(-0.10)$ & $-0.01(-0.10)$ \\
36 & 8 & 36 & 10 & $0.86(0.89)$ & $-0.27(-0.22)$ & $-0.27(-0.21)$ &
$-0.23(-0.19)$ & $-0.23(-0.19)$ & $-0.13(-0.19)$ \\
36 & 8 & 36 & 12 & $0.87(0.90)$ & $-0.25(-0.20)$ & $-0.25(-0.19)$ &
$-0.22(-0.18)$ & $-0.22(-0.17)$ & $-0.11(-0.17)$ \\
36 & 8 & 36 & 16 & $0.88(0.92)$ & $-0.23(-0.16)$ & $-0.23(-0.16)$ &
$-0.21(-0.15)$ & $-0.20(-0.15)$ & $-0.08(-0.15)$ \\
44 & 8 & 44 & 8 & $0.85(0.88)$ & $-0.28(-0.24)$ & $-0.28(-0.24)$ &
$-0.24(-0.21)$ & $-0.24(-0.21)$ & $-0.16(-0.21)$ \\
44 & 12 & 44 & 12 & $0.89(0.92)$ & $-0.21(-0.16)$ & $-0.21(-0.16)$ &
$-0.19(-0.15)$ & $-0.19(-0.15)$ & $-0.09(-0.15)$ \\
44 & 20 & 44 & 20 & $0.91(0.95)$ & $-0.16(-0.10)$ & $-0.16(-0.10)$ &
$-0.15(-0.09)$ & $-0.15(-0.09)$ & $-0.02(-0.09)$ \\
\hline
\end{tabular}
\end{center}
\label{tab3}
\end{sidewaystable}

\begin{table}[ht]

\caption{Bivariate correlation coefficient ($\xi$) and fourth order bivariate
cumulants $k_{rs}$ with $r+s=4$ for various values of number of
sp states ($N$), number of fermions ($m$), Hamiltonian body rank ($k$) and the
rank ($k_0$) of the particle removal transition operator. Results are obtained 
using the formulas given in Section 7.1. Note that for the $M_{22}$ that is needed
for  $k_{22}$, we have used Eq. (\ref{snt-13}) with the third term replaced by
the corresponding asymptotic formula given by Eq. (\ref{asymp-snt6}) as a
formula for the reduced matrix elements in Eq. (\ref{snt-13}) is not
available.} 

\begin{center}
\begin{tabular}{cccccccccc}
\\
\\
\hline 
$N$ & $m$ & $k$ & $k_0$ & $\xi$ & $k_{40}$ & $k_{04}$ & $k_{31}$ & 
$k_{13}$ & $k_{22}$ \\
\hline \\
$20$ & $10$ & $2$ & $1$ & $0.82$ & $-0.54$ & $-0.55$ & $-0.44$ & $-0.45$ &
$-0.21$ \\
$30$ & $10$ & $2$ & $1$&  $0.85$ & $-0.48$ & $-0.50$ & $-0.41$ & $-0.43$ &
$-0.26$ \\
$60$ & $10$ & $2$ & $1$ & $0.88$ & $-0.42$ & $-0.46$ & $-0.37$ & $-0.40$ &
$-0.30$ \\
$80$ & $10$ & $2$ & $1$ & $0.88$ & $-0.41$ & $-0.45$ & $-0.36$ & $-0.39$ &
$-0.31$ \\
$50$ & $12$ & $2$ & $1$ & $0.89$ & $-0.38$ & $-0.40$ & $-.034$ & $-0.36$ &
$-0.25$ \\
& $15$ & $2$ & $1$ & $0.91$ & $-0.33$ & $-0.35$ & $-0.30$ & $-0.31$ & $-0.19$ 
\\
& $20$ & $2$ & $1$ & $0.92$ & $-0.29$ & $-0.29$ & $-0.26$ & $-0.27$ & $-0.13$
\\
& $25$ & $2$ & $1$ & $0.92$ & $-0.27$ & $-0.27$ & $-0.25$ & $-0.25$ & $-0.08$ \\
$24$ & $8$ & $2$ & $1$ & $0.82$ & $-0.56$ & $-0.61$ & $-0.46$ & $-0.49$ &
$-0.31$ \\
& & $2$ & $2$ & $0.66$ & $-0.56$ & $-0.67$ & $-0.37$ & $-0.43$ & $-0.22$ \\
$40$ & $15$ & $2$ & $1$ & $0.90$ & $-0.36$ & $-0.37$ & $-0.32$ & $-0.33$ &
$-0.18$ \\
& & $2$ & $2$ & $0.80$ & $-0.36$ & $-0.38$ & $-0.29$ & $-0.31$ & $-0.12$ \\
$60$ & $20$ & $2$ & $1$ & $0.93$ & $-0.27$ & $-0.27$ & $-0.25$ & $-0.25$ &
$-0.14$ \\
& & $3$ & $1$ & $0.89$ & $-0.51$ & $-0.53$ & $-0.46$ & $-0.47$ & $-0.30$ \\
& & $3$ & $2$ & $0.79$ & $-0.51$ & $-0.54$ & $-0.40$ & $-0.43$ & $-0.22$ \\
\hline
\end{tabular}
\end{center}
\label{tab4}
\end{table}

\end{document}